\documentclass{article}
\usepackage{authblk}
\usepackage[margin=1in]{geometry}
\usepackage[square,numbers]{natbib}
\usepackage{mathtools}
\usepackage{amssymb}
\usepackage{bm}
\usepackage{quantum}
\usepackage{xcolor}
\usepackage[justification=justified]{caption}
\usepackage{subcaption}	
\usepackage{tikz}
\usetikzlibrary{calc, shapes, arrows, backgrounds, decorations.markings}
\usepackage{tikz-3dplot}
\usepackage{graphicx}
\usepackage{fancybox}
\usepackage[hidelinks]{hyperref}
\usepackage{cleveref}	

\crefname{subsection}{subsection}{subsections}
\crefname{subsubsection}{subsubsection}{subsubsections}

\tikzset{
	gate/.style = {draw, fill=white, rectangle, minimum height = 1.5em, minimum width = 1.5em, fill opacity=1, text height = 0.9em, inner sep=0.1em, align=center, text centered, text depth=0.5em},
	transform shape
	}
	
\def\beamsplitterpm(#1,#2){
	\draw (#1,#2) -- ++(1,-1);
	\draw (#1+1,#2) -- (#1,#2-1);
	\draw[very thick] (#1,#2-0.5) -- (#1+1,#2-0.5);
	\node (BS1a) at (#1+0.5,#2-0.1) {\scriptsize +};
	\node (BS1b) at (#1+0.5,#2-0.9){\scriptsize -};
}
\def\beamsplittermp(#1,#2){
	\draw (#1,#2) -- ++(1,-1);
	\draw (#1+1,#2) -- (#1,#2-1);
	\draw[very thick] (#1,#2-0.5) -- (#1+1,#2-0.5);
	\node (BS1a) at (#1+0.5,#2-0.1) {\scriptsize -};
	\node (BS1b) at (#1+0.5,#2-0.9){\scriptsize +};
}

\def\beamsplittertune(#1,#2,#3){
	\fill[white] (#1,#2+0.1) rectangle (#1+1,#2-1.1);
	\draw (#1,#2) -- ++(1,-1);
	\draw (#1+1,#2) -- (#1,#2-1);
	\draw[very thick] (#1,#2-0.5) -- (#1+1,#2-0.5);
	\node[right] (BS1a) at (#1+1.01,#2-0.5) {\scriptsize #3};
}

\def\meas(#1,#2,#3){
	\draw (#1,#2-0.3) -- (#1,#2+0.3);
	\draw[fill=white] (#1,#2-0.3) to[out=0, in=0, distance=2em] (#1,#2+0.3);
	\node (meas) at (#1+0.2,#2){#3};
}

\def\cplus(#1,#2,#3,#4,#5){
	\draw (#1,#2) -- (#3,#4 \ifnum#2<#4{+}\else{-}\fi 0.25);
	\draw (#3,#4) circle (0.25);
	\draw[fill=black] (#1,#2) circle (0.065);
	\node at (#3,#4 \ifnum#2<#4{+}\else{-}\fi 0.4){\centering\scriptsize $#5$};
}

\def\drawwires(#1,#2){
	\foreach \i in {1,...,#1}
		\draw (0,\i-1) -- (#2,\i-1);
}

\def\drawwiresat(#1,#2,#3){
	\begin{scope}[shift={(#1)}]
	\drawwires(#2,#3);
	\end{scope}
}

\def\cwire(#1,#2){
	\draw [line width=0.3em] (#1) -- (#2);
	\draw [ultra thick, white] (#1) -- (#2);
}

\def\cvswap(#1,#2){
	\draw (#1) -- (#2);
	\draw ($(#1)-(0.2,0.2)$) -- ($(#1)+(0.2,0.2)$);
	\draw ($(#1)-(0.2,-0.2)$) -- ($(#1)+(0.2,-0.2)$);
	\draw ($(#2)-(0.2,0.2)$) -- ($(#2)+(0.2,0.2)$);
	\draw ($(#2)-(0.2,-0.2)$) -- ($(#2)+(0.2,-0.2)$);
}

\newcommand*{\ArcAngle}{60}%
\newcommand*{\ArcRadius}{2.0}%
\def\VR{\kern-\arraycolsep\strut\vrule &\kern-\arraycolsep}
\def\vr{\kern-\arraycolsep & \kern-\arraycolsep}

\renewcommand{\vec}[1]{\bm{#1}}
\newcommand{\EP}{\varepsilon}

\begin{document}
\title{Spacetime replication of continuous variable quantum information}
\date{June 28, 2016}
\author[1,3]{Patrick Hayden}
\author[1]{Sepehr Nezami}
\author[1]{Grant Salton}
\affil[1]{Stanford Institute for Theoretical Physics, Stanford University, Stanford, CA 94305}
\author[2,3]{Barry C.\ Sanders}
\affil[2]{Institute for Quantum Science and Technology, University of Calgary, Alberta, Canada T2N 1N4}
\affil[3]{Program in Quantum Information Science, Canadian Institute for Advanced Research, Toronto, Ontario M5G 1Z8, Canada}
\maketitle
\begin{abstract}
The theory of relativity requires that no information travel faster than light, whereas the unitarity of quantum mechanics ensures that quantum information cannot be cloned.  These conditions provide the basic constraints that appear in information replication tasks, which formalize aspects of the behavior of information in relativistic quantum mechanics.  In this article, we provide continuous variable (CV) strategies for spacetime quantum information replication that are directly amenable to optical or mechanical implementation.  We use a new class of homologically-constructed CV quantum error correcting codes to provide efficient solutions for the general case of information replication.  As compared to schemes encoding qubits, our CV solution requires half as many shares per encoded system.  We also provide an optimized five-mode strategy for replicating quantum information in a particular configuration of four spacetime regions designed not to be reducible to previously performed experiments. For this optimized strategy, we provide detailed encoding and decoding procedures using standard optical apparatus and calculate the recovery fidelity when finite squeezing is used.  As such we provide a scheme for experimentally realizing quantum information replication using quantum optics.
\end{abstract}

\section{Introduction}
The no-cloning theorem~\cite{Wooters1982} is one of the simplest and most powerful observations in quantum mechanics.  The fact that quantum information cannot be copied is the ultimate foundation for the security of quantum key distribution~\cite{bennett1984quantum} and plays a central role in the theory of quantum error correction~\cite{gottesman1997stabilizer}. The no-cloning theorem has also been used as a powerful tool for studying the consistency of proposals in quantum gravity~\cite{susskind1993stretched,hayden2007black}. Likewise, a static version of the no-cloning theorem known as the monogamy of entanglement is at the core of a recent controversy over whether spacetime ends at the horizon of a black hole, a potentially drastic conflict with Einstein's equivalence principle that is nonetheless difficult to refute~\cite{almheiri2013black}. 

In light of the confusion that has arisen regarding the replication of information in spacetimes with subtle causal structures such as evaporating black holes,~\citet{Hayden2012} studied the much simpler problem of replicating quantum information in multiple regions of Minkowski spacetime. The theory of relativity requires that information
cannot travel faster than light,
whereas the unitarity of quantum mechanics ensures that quantum information cannot be cloned. In Minkowski spacetime, it turns out that those trivial constraints are the \emph{only} restrictions on replicating quantum information. Provided those two conditions are met, information can in principle be encoded into the state of multiple particles which are then sent along causal curves, with the net result that the quantum information is available for decoding inside each of the specified regions.

The demonstration in~\cite{Hayden2012} used a codeword-stabilized quantum code~\cite{Cross2008} to replicate the quantum information of a single qubit. If the replication were to be done across~$N$ regions of spacetime, $N(N-1)$ qubits would be required, most naturally with one particle for each qubit. While impressive experimental progress on quantum error correction has been made, demonstrating spacetime quantum information replication would require the particles to propagate in space. For that reason, it is natural to use light as the carrier of quantum information and to encode not just a qubit but the full state of a mode of the electromagnetic field.
To that end, in this article we develop continuous variable (CV) stabilizer codes suitable for use in laboratory demonstrations of spacetime quantum information replication. Our codes protect against pure bosonic erasure in ways specifically tailored for the task of information replication.
We will show that in general only $N(N-1)/2$ modes are required to replicate the information in a single mode across~$N$ spacetime regions. 

A related laboratory demonstration was conducted by~\citet{Niset2008}, encoding two logical modes into four physical modes such that the logical state could be recovered with access to any three out of the four modes. The motivation in that setting was simply to demonstrate CV quantum error correction, so the spacetime configuration was not particularly interesting.  Similar work has been done in the area of CV quantum secret sharing on both the theoretical~\cite{Tyc2002,Lance2003,Tyc2003} and experimental~\cite{Lance2004,Lance2005} fronts.  
The simplest example of information replication that does not reduce to previously demonstrated quantum secret sharing schemes from a quantum error correction perspective involves information replicated across four spacetime regions. For that special case, we provide an optimized five-mode code along with detailed encoding and decoding procedures using standard optical apparatus. In addition,
we calculate the recovery fidelity when finite squeezing~\cite{LK87} is used.

The article is organized as follows.  In \cref{RepSection} we describe the problem of replicating quantum information in spacetime. \Cref{CVSection} then begins by explaining the role of quantum error correction in the solution of~\cite{Hayden2012} and briefly reviewing CV quantum  error correction.  With the CV framework available, we describe the general CV error correcting codes for spacetime information replication.  In \cref{FiveModeCodeSection} we formulate a specific CV error correcting code that should be experimentally feasible, and we provide an experimental proposal in \cref{sec:proposed}.  We end with a summary of our findings and a call for action from experimentalists in \cref{ConclusionsSection}.
%
\section{Replicating quantum information}
\label{RepSection}
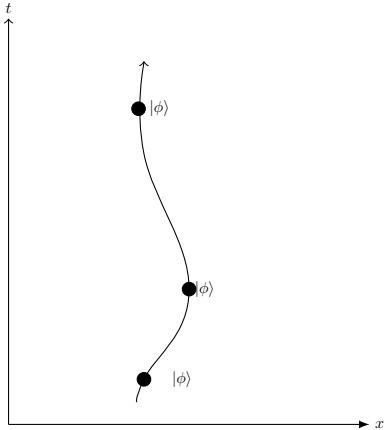
\begin{figure}[!t]
	\centering
		\begin{tikzpicture}[scale=0.6]
			\draw[->] (-2,-1) -- (-2,8) node[above] {$t$};
			\draw[-latex] (-2,-1) -> (6,-1) node[right]{$x$};
			\draw plot [smooth, tension=0.75] coordinates { (0.85,-0.5)(1,0)(2,2)(1,5)(1,7) }[->];
			\node[right] at (2,2) {$\ket{\phi}$};
			\draw[fill=black] (2,2) circle (0.15);
			\node[right] at (1,6) {$\ket{\phi}$};
			\draw[fill=black] (0.88,6) circle (0.15);
			\node[right] at (1.5,0) {$\ket{\phi}$};
			\draw[fill=black] (1,0) circle (0.15);
		\end{tikzpicture}
	\caption{\label{worldline}Example worldline of a quantum state $\ket\phi$.  In this example, the state moves through spacetime with trivial time evolution.  The state is necessarily replicated in time as it persists from one time to another.  It cannot, however, be transported along a spacelike trajectory (due to special relativity) or be replicated on a given spatial slice (due to the no-cloning theorem).}
\end{figure}

Consider a single particle moving through spacetime along the worldline shown in \cref{worldline}.
Obviously, the quantum information consisting of the internal state of that particle exists at every point on the world line or, equivalently, at every instant of (proper or other) time. 
Thus, whereas the no-cloning theorem prohibits the replication of quantum information at spacelike separations, it is absolutely trivial to replicate quantum information at timelike separations. Indeed, the unitarity of quantum mechanics and, in particular, its microscopic reversibility requires quantum information to be ``replicated'' in time.  Taking this broader spacetime view then suggests a natural question: when can a collection of regions in spacetime each contain a copy of the same quantum information? Relativistic causality prohibits superluminal signalling and no-cloning excludes replicating quantum information across two spacelike separated regions. What other constraints might exist? ~\citet{Hayden2012} answered this question in the case when each region is a causal diamond in spacetime. 

Given two spacetime events (four-vectors)~$y_j$ and $z_j$ with $z_j$ in the future of~$y_j$, we define the \emph{causal diamond}~$D_j$ to be the intersection of the future light cone of~$y_j$ and the past light cone of $z_j$.    
We say that a piece of quantum information can be \emph{replicated} across a given set of causal diamonds if the information can, in principle, be localized at some point in each causal diamond. In other words, we could, in principle, produce the state in any one of the diamonds, although doing so might prevent us from producing the state in another diamond.  In essence, the information exists in each of the diamonds \emph{as potential}. (This loose terminology will be given a precise operational definition suitable for experimental demonstration below.) 

\begin{figure}
	\centering
		\begin{tikzpicture}[scale=0.6]
			\draw[->] (-3.5,-0.5) -- (-3.5,0.5) node[above] {$t$};
			\draw[-latex] (-3.5,-0.5) -> (-2.5,-0.5) node[right]{$x$};
			\node (a) at (-3.5,6.5) {a)};
			\node (p) at (-2,6.5) {Possible};
			\coordinate (s) at (0cm, 0cm);
			\draw[fill=yellow] (s) circle (0.15cm);
			\node[below right] at (s) {$s$};
			\node[label={[yshift=-0.6cm]$D_1$}] (y1) at (0, 1){};
			\node[label={[yshift=-0.6cm]$D_2$}] (y2) at (1, 3){};
			\node[label={[yshift=-0.6cm]$D_3$}] (y3) at (-1, 4){};
			\fill[blue, opacity=0.3, rotate=45, draw=black] (y1) rectangle ($(y1)+(0.5,2)$);
			\fill[blue, opacity=0.3, rotate=45, draw=black] (y2) rectangle ($(y2)+(2.5,1)$);
			\fill[blue, opacity=0.3, rotate=45, draw=black] (y3) rectangle ($(y3)+(1,1)$);
			\draw (-4,-1) rectangle (4,7);
		\end{tikzpicture}
		\quad
		\begin{tikzpicture}[scale=0.6]
			\draw[->] (-3.5,-0.5) -- (-3.5,0.5) node[above] {$t$};
			\draw[-latex] (-3.5,-0.5) -> (-2.5,-0.5) node[right]{$x$};
			\node (a) at (-3.5,6.5) {b)};
			\node (p) at (-2,6.5) {Possible};
			\coordinate (s) at (0cm, 0cm);
			\draw[fill=yellow] (s) circle (0.15cm);
			\node[below right] at (s) {$s$};
			\node[label={[yshift=-0.6cm]$D_1$}] (y1) at (0, 1){};
			\node[label={[yshift=-0.6cm]$D_2$}] (y2) at (1, 1.5){};
			\node[label={[yshift=-0.6cm]$D_3$}] (y3) at (-1, 3){};
			\fill[blue, opacity=0.3, rotate=45, draw=black] (y1) rectangle ($(y1)+(4,3)$);
			\fill[blue, opacity=0.3, rotate=45, draw=black] (y2) rectangle ($(y2)+(3,3)$);
			\fill[blue, opacity=0.3, rotate=45, draw=black] (y3) rectangle ($(y3)+(2,2)$);
			\draw (-4,-1) rectangle (4,7);
		\end{tikzpicture}
		\quad
		\begin{tikzpicture}[scale=0.6]
			\draw[->] (-3.5,-0.5) -- (-3.5,0.5) node[above] {$t$};
			\draw[-latex] (-3.5,-0.5) -> (-2.5,-0.5) node[right]{$x$};
			\node (a) at (-3.5,6.5) {c)};
			\node (p) at (-1.5,6.5) {Impossible};
			\coordinate (s) at (0cm, 0cm);
			\draw[fill=yellow] (s) circle (0.15cm);
			\node[below right] at (s) {$s$};
			\node[label={[yshift=-0.6cm]$D_1$}] (y1) at (0, 1){};
			\node[label={[yshift=-0.6cm]$D_2$}] (y2) at (-1.5, 3.5){};
			\node[label={[yshift=-0.6cm]$D_3$}] (y3) at (1.5, 3.5){};
			\fill[blue, rotate=45, draw=black, opacity=0.3] (y1) rectangle ($(y1)+(2,2)$);
			\fill[blue, rotate=45, draw=black, opacity=0.3] (y2) rectangle ($(y2)+(1.5,1.5)$);
			\fill[blue, rotate=45, draw=black, opacity=0.3] (y3) rectangle ($(y3)+(1.5,1.5)$);
			\draw (-4,-1) rectangle (4,7);
		\end{tikzpicture}
	\caption{Examples of information replication tasks.  In example (a), quantum information localized at point~$s$ \emph{can} be replicated in all three causal diamonds by simply propagating it along a timelike curve passing through all three diamonds.  Similarly, in example (b) we can replicate quantum information in all three diamonds by simply moving the information to a point in the overlap area.  However, in example (c), information replication is \emph{not} possible, as no causal relationship exists between diamonds $D_2$ and $D_3$.\label{replicationexamples}}
\end{figure}
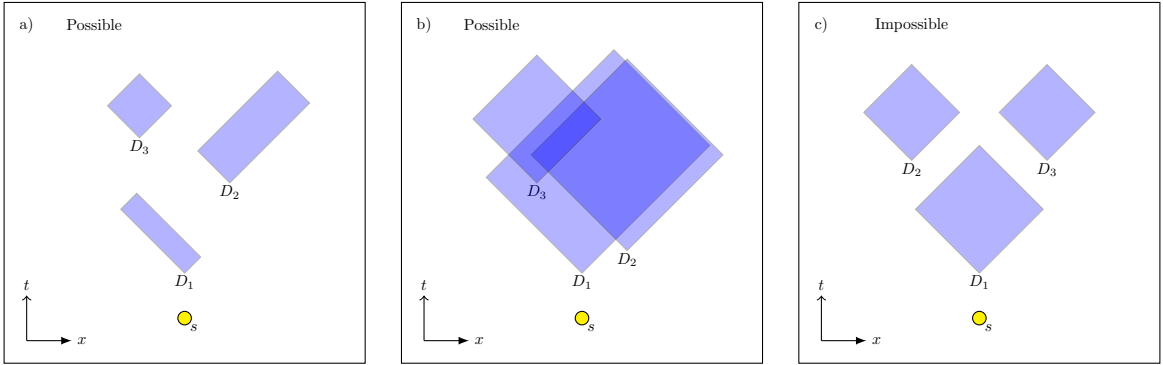

Two causal diamonds are \emph{causally related} if there exists a causal curve from at least one point in one of the diamonds to at least one point in the other diamond.  Intuitively, this means it is possible to send a message from one diamond to the other, without specifying in which direction.  Assuming that the information is initially localized at a starting point~$s$,~\citet{Hayden2012} proved that quantum information can be replicated in a set of causal diamonds if and only if
\begin{quote}
(a) there is a causal curve from~$s$ to a point in each diamond. \\
(b) for each pair $(j,k)$, the diamonds~$D_j$ and $D_k$ are causally related. 
\end{quote}
Because the statement is if and only if, whenever these two conditions are met, there exists some way to replicate the same quantum information across all of the causal diamonds. Note, however, that the conditions encode only the most elementary and obvious constraints. If information could be replicated in a pair of diamonds that were not causally related, that would constitute a violation of the no-cloning theorem. Likewise, the requirement that there be a causal curve from~$s$ to each diamond is the most basic expression of the impossibility of superluminal signalling. Some simple examples of collections of spacetime regions for which replication is either obviously possible or impossible are show in \cref{replicationexamples}.


Whereas the abstract notion of information replication across spacetime regions is conceptually appealing, its operational significance is not immediately obvious. An equivalent, operational notion called \emph{summoning} was introduced in pioneering work by Kent~\cite{Kent2011,Kent2012,adlam2015quantum}. Kent's motivation was to find simple quantum information theoretic tasks for which both relativistic causality and quantum mechanical unitarity played important roles. Replication of quantum information across a pair of causal diamonds is obviously one such example.

A summoning task is defined by a spacetime starting point~$s$,
a collection of request points $\{y_j\}$ and a collection of reveal points $\{z_j\}$.
The task is an adversarial game played by a referee against a player Alice:
the referee supplies a quantum system in a state $\ket{\phi}$ unknown to Alice at~$s$ and has agents stationed at the various spacetime points~$y_j$ and $z_j$.
Alice may have agents and apparatus arranged as she likes in spacetime. One and only one of the referee's agents, located at a request point~$y_j$, will request the replicated quantum information from Alice. In that case, Alice, her agents, and apparatus must arrange for the quantum information to appear at $z_j$. That is, they must supply a system at spacetime point $z_j$ in the same quantum state $\ket{\phi}$, up to isomorphism. The referee can test that the state is the same as the one supplied using a suitable measurement. If Alice and her agents pass the test 100\% of the time,
then they succeed at the summoning task.

If Alice succeeds at summoning then,
for each $j$, the quantum information must be in the past of $z_j$ because the information ultimately appears there. Because the task is adversarial and Alice does not know in advance at which of the~$y_j$ the request will occur, her plan must work for each of them. Likewise, the information must be in the future of~$y_j$ because that is the set of spacetime points that can be affected by a request at~$y_j$. Thus, succeeding at summoning implies that the information is replicated across each and every causal diamond. Our experimental proposal is,
therefore, one for a demonstration of nontrivial quantum information summoning in spacetime.

%
\section{Continuous variable codes}\label{CVSection}

The strategy Hayden and May used to prove the aforementioned replication theorem relied heavily
on employing quantum error correcting codes~\cite{Hayden2012}. The same approach is applicable to the replication of CV quantum information but the specific codes they invented are not. As such, our first major task is to design appropriate CV codes.

Before doing so, let us briefly review how the codes will be used to replicate quantum information across spacetime regions. Given a set of~$N$ causal diamonds in spacetime, construct the undirected graph of causal relationships between all the causal diamonds.  If each pair of diamonds is causally related (that is, condition (b) above holds), then the graph of causal relations is necessarily a complete graph with $n=\binom{N}{2}$ edges.

Now suppose we have a quantum error correcting code with one share for each of those~$n$ edges and that, for each vertex of the graph, the encoded state can be recovered using only the shares associated to edges subtending that vertex. In that case, Alice could take the system at~$s$ supplied by the referee and encode it into~$n$ particles using the error correcting code. Each particle would then be propagated along a causal curve with a segment corresponding to its associated edge in the complete graph. (If the edge represented the existence of a causal curve from~$D_i$ to~$D_j$, the particle would be sent through~$D_i$ to~$D_j$.)

By capturing all the particles passing through any given causal diamond and then acting on them with a suitable decoding operation, it will be possible to recover the encoded quantum information $\ket{\phi}$ using the properties of the quantum error correcting code. Thus, if a request is made at~$y_j$, it will be possible to exhibit the referee's state at $z_j$. See \cref{FourRegionGraph}(b) for an example with $N=4$.

The task at hand is to design a CV quantum error correcting code with $n=\binom{N}{2}$ shares with the error correcting property described in the previous paragraph. 
Whereas Hayden and May used codeword-stabilized quantum codes~\cite{Hayden2012},
we will exhibit a new class of CV stabilizer codes.  As mentioned earlier, CV codes have the advantage of being directly applicable to bosonic modes, which are ubiquitous in quantum optics and quantum field theory.  We will first describe the general CV codes for~$N$ spacetime regions, and then provide an optimized five-mode code for information replication in a specific four region example.  We will then translate the encoding and decoding of this quantum error correcting code into the language of quantum optics, describing how to demonstrate spacetime information replication on an optical bench.

In the CV stabilizer formalism,
the discrete Pauli group is replaced by the Heisenberg-Weyl group~\cite{BSBN02,Barnes2004}, and an arbitrary unitary in this group (acting on~$n$ bosonic modes) is fully specified (up to phase) by two real vectors of length~$n$: $\vec{s},\vec{t}\in\RR^n$. In other words, there is a $2n$-dimensional, real vector space that provides an equivalent description of the group.  The (unitary) group elements correspond to phase space displacements of the~$n$ modes, and they take the form
\begin{equation}
	U(\vec{s},\vec{t})=\exp(i(\vec{s}\cdot \vec{X}+\vec{t}\cdot \vec{P})),
\end{equation}
where $\vec{X}=(X_1,\ldots,X_n)^\top$ and $\vec{P}=(P_1,\ldots,P_n)^\top$ are vectors of the~$n$ quadrature operators $X_i$ and $P_i$ acting on each of the modes. Two of these unitaries commute up to a phase:
\begin{equation}
	U(\vec{s}_1,\vec{t}_1)U(\vec{s}_2,\vec{t}_2)
		=e^{i \omega [(\vec{s}_1,\vec{t}_1),(\vec{s}_2,\vec{t}_2)]}U(\vec{s}_2,\vec{t}_2)U(\vec{s}_1,\vec{t}_1),
\end{equation}
where $\omega$ is a symplectic form on the real vector space.  In particular,
\begin{equation}
\omega [(\vec{s}_1,\vec{t}_1),(\vec{s}_2,\vec{t}_2)]=\vec{s}_1\cdot\vec{t}_2-\vec{t}_1\cdot\vec{s}_2.
\end{equation}
With this inner product, the $2n$-dimensional vector space is a symplectic space.

In addition to single-mode unitaries, we will also need to be able to implement controlled, two-mode unitaries.  Of particular interest is the continuous variable analog of the qubit CNOT gate, usually called the controlled-addition, controlled-shift, or CPLUS operator.  For our purposes, we focus on a slight generalization of the usual operator to allow for a ``gain'' -- a multiplicative constant.  To connect with experimental quantum optics, we refer to this gate as a ``QND'' gate (for quantum non-demolition). In particular,
\begin{equation}\label{cplusdefn}
\text{QND}_c\ket{x,y}=\ket{x+cy,y}\quad\leftrightarrow\quad\raisebox{-1.25em}{
	\begin{tikzpicture}[scale=0.6, every node/.style={inner sep=2pt}]
		\node [left](mode1start) at (0,0) {$\ket x$};
		\node [left](mode2start) at (0,-1) {$\ket y$};
		\node [right](mode2end) at (2,-1) {$\ket{y}$};
		\node [right](mode1end) at (2,0) {$\ket{x+cy}$};
		\draw (mode1start) -- (mode1end);
		\draw (mode2start) -- (mode2end);
		\cplus(1,-1,1,0,c);
	\end{tikzpicture}}.
\end{equation}

Our goal is to replicate continuous variable quantum information.  We will show that (for an allowed replication task) we can succeed by encoding one bosonic mode into~$\binom{N}{2}$ modes.  The structure of the code will be such that we can correct for erasure of a known subset of the modes.  We model erasure of a mode as arbitrary displacement in phase space.

To construct a CV stabilizer code we need~$\binom{N}{2}-1$ stabilizer generators.  Since the unitary group is actually a Lie group, we are searching for elements of the associated Lie algebra that correspond to the infinitesimal stabilizer generators.  In particular, we're searching for an abelian Lie subalgebra, $\mathfrak{s}$, so that the (abelian) subgroup of the full Heisenberg-Weyl group formed by exponentiating $\mathfrak{s}$ is the stabilizer group.  The infinitesimal stabilizer generators are described by real vectors $\vec{v}^m=(\vec{s}^m,\vec{t}^m)\in\RR^{2n}$, where the index $m$ runs over the list of generators.  The generators are then operators of the form $\vec{s}^m\cdot\vec{X}+\vec{t}^m\cdot\vec{P}$.  The codespace is  defined to be the set of all states $\ket{c}$ for which $a\ket{c}=0$ for all $a\in\mathfrak{s}$.  In other words, the codespace is the intersection of the kernels of all the stabilizer generators.

Following the code construction technique described above, we associate one mode of our code to each edge of a complete graph with vertices labeled $1$ through~$N$ indicating causal relationships between the causal diamonds in the replication problem. Hayden and May~\cite{Hayden2012} attached \emph{two} qubits to each edge of the graph, 
whereas we associate \emph{one} mode.  An example of such a graph is shown in \cref{FourRegionGraph}(a) for the case of $N=4$ spacetime regions (6 modes).  Each edge/mode can be uniquely specified by an ordered pair $(a,b)$, written simply as $ab$ with $a<b$.  This is just a notational convention we choose to simplify the presentation.

The code that we will propose is in fact a CSS code~\cite{CSS1996}, which means it has the simplifying property that all of the stabilizer generators either consist of pure~$X$ displacements or pure~$P$ displacements.  In other words, a generator $g$ for a pure~$X$ stabilizer has the form
\begin{equation}
g= \vec{v}\cdot\vec{X}.
\end{equation}
Similarly, a generator $h$ for a pure~$P$ displacement is of the form
\begin{equation}
h=\vec{w}\cdot\vec{P}.
\end{equation}
With this construction the condition that the stabilizer group be abelian is equivalent to an orthogonality condition on the vectors specifying the stabilizer generators:
\begin{equation}\label{orthcon}
 \vec{v}\cdot\vec{w}=0
\end{equation}
for all pure~$X$ and pure~$P$ generators specified by~$\vec{v}$ and~$\vec{w}$, respectively. 
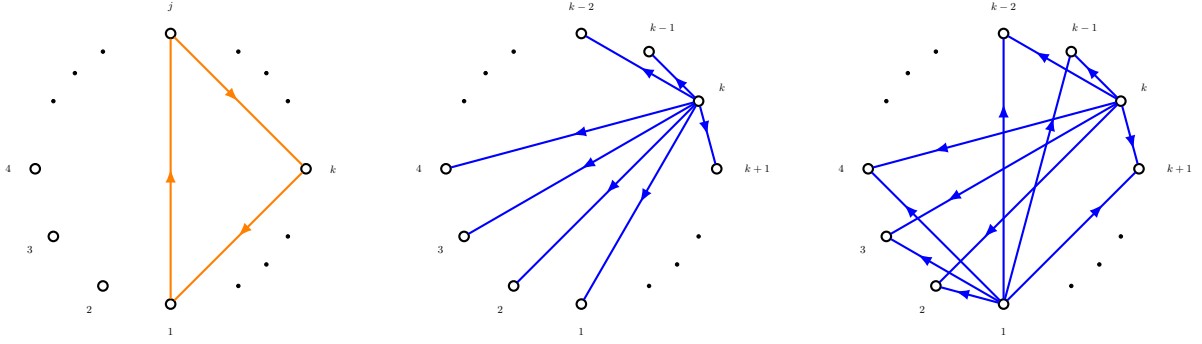
\begin{figure*}\centering
   \begin{subfigure}[t]{0.32\linewidth}\centering
		\begin{tikzpicture}[scale=0.45]
			\node[draw, circle, thick, inner sep=0pt,minimum size=8pt] (A) at (270:4){};
			\node at (270:4.8){\small{$1$}};
			\node[draw, circle, thick, inner sep=0pt,minimum size=8pt] at (240:4){};
			\node at (240:4.8){\small{$2$}};
			\node[draw, circle, thick, inner sep=0pt,minimum size=8pt] at (210:4){};
			\node at (210:4.8){\small{$3$}};
			\node[draw, circle, thick, inner sep=0pt,minimum size=8pt] at (180:4){};
			\node at (180:4.8){\small{$4$}};

			\draw[fill=black] (120:4) circle (0.05);
			\draw[fill=black] (135:4) circle (0.05);
			\draw[fill=black] (150:4) circle (0.05);
			
			\node[draw, circle, thick, inner sep=0pt,minimum size=8pt] (J) at (90:4){};
			\node at (90:4.8){\small{$j$}};
			
			\draw[fill=black] (30:4) circle (0.05);
			\draw[fill=black] (45:4) circle (0.05);
			\draw[fill=black] (60:4) circle (0.05);
			
			\node[draw, circle, thick, inner sep=0pt,minimum size=8pt] (K) at (0:4){};
			\node at (00:4.8){\small{$k$}};
			
			\draw[fill=black] (-30:4) circle (0.05);
			\draw[fill=black] (-45:4) circle (0.05);
			\draw[fill=black] (-60:4) circle (0.05);
			
			\begin{scope}[thick, orange, decoration={markings,mark=at position 0.5 with {\arrow{latex}}}]
				\draw[postaction={decorate}] (A) -- (J);
				\draw[postaction={decorate}] (J) -- (K);
				\draw[postaction={decorate}] (K) -- (A);
			\end{scope}
		\end{tikzpicture}
		\caption{\label{closedLoops} Pure~$X$ stabilizer $g_{jk}$ described by the vector $\vec{v}_{jk}$.  Pure~$X$ stabilizer generators are directed, closed loops of length three containing vertex 1.}
   \end{subfigure}
   \hfill
   \begin{subfigure}[t]{0.32\linewidth}\centering
		\begin{tikzpicture}[scale=0.45]
			\node[draw, circle, thick, inner sep=0pt,minimum size=8pt] (A) at (270:4){};
			\node at (270:4.8){\small{$1$}};
			\node[draw, circle, thick, inner sep=0pt,minimum size=8pt] (B) at (240:4){};
			\node at (240:4.8){\small{$2$}};
			\node[draw, circle, thick, inner sep=0pt,minimum size=8pt] (C) at (210:4){};
			\node at (210:4.8){\small{$3$}};
			\node[draw, circle, thick, inner sep=0pt,minimum size=8pt] (D) at (180:4){};
			\node at (180:4.8){\small{$4$}};

			\draw[fill=black] (120:4) circle (0.05);
			\draw[fill=black] (135:4) circle (0.05);
			\draw[fill=black] (150:4) circle (0.05);
			
			\node[draw, circle, thick, inner sep=0pt,minimum size=8pt] (K2) at (90:4){};
			\node at (90:4.8){\small{$k-2$}};
			\node[draw, circle, thick, inner sep=0pt,minimum size=8pt] (K1) at (60:4){};
			\node at (60:4.8){\small{$k-1$}};
			\node[draw, circle, thick, inner sep=0pt,minimum size=8pt] (K) at (30:4){};
			\node at (30:4.8){\small{$k$}};
			\node[draw, circle, thick, inner sep=0pt,minimum size=8pt] (K11) at (0:4){};
			\node at (0:5.2){\small{$k+1$}};
			
			\draw[fill=black] (-30:4) circle (0.05);
			\draw[fill=black] (-45:4) circle (0.05);
			\draw[fill=black] (-60:4) circle (0.05);
			
			\begin{scope}[thick, blue, decoration={markings,mark=at position 0.5 with {\arrow{latex}}}]
				\draw[postaction={decorate}] (K) -- (A);
				\draw[postaction={decorate}] (K) -- (B);
				\draw[postaction={decorate}] (K) -- (C);
				\draw[postaction={decorate}] (K) -- (D);
				\draw[postaction={decorate}] (K) -- (K2);
				\draw[postaction={decorate}] (K) -- (K1);
				\draw[postaction={decorate}] (K) -- (K11);
			\end{scope}
		\end{tikzpicture}
		\caption{\label{Ai} Graphical representation of the star vector $\vec{A}_k$.}
   \end{subfigure}
   \hfill
   \begin{subfigure}[t]{0.32\linewidth}\centering
		\begin{tikzpicture}[scale=0.45]
			\node[draw, circle, thick, inner sep=0pt,minimum size=8pt] (A) at (270:4){};
			\node at (270:4.8){\small{$1$}};
			\node[draw, circle, thick, inner sep=0pt,minimum size=8pt] (B) at (240:4){};
			\node at (240:4.8){\small{$2$}};
			\node[draw, circle, thick, inner sep=0pt,minimum size=8pt] (C) at (210:4){};
			\node at (210:4.8){\small{$3$}};
			\node[draw, circle, thick, inner sep=0pt,minimum size=8pt] (D) at (180:4){};
			\node at (180:4.8){\small{$4$}};

			\draw[fill=black] (120:4) circle (0.05);
			\draw[fill=black] (135:4) circle (0.05);
			\draw[fill=black] (150:4) circle (0.05);
			
			\node[draw, circle, thick, inner sep=0pt,minimum size=8pt] (K2) at (90:4){};
			\node at (90:4.8){\small{$k-2$}};
			\node[draw, circle, thick, inner sep=0pt,minimum size=8pt] (K1) at (60:4){};
			\node at (60:4.8){\small{$k-1$}};
			\node[draw, circle, thick, inner sep=0pt,minimum size=8pt] (K) at (30:4){};
			\node at (30:4.8){\small{$k$}};
			\node[draw, circle, thick, inner sep=0pt,minimum size=8pt] (K11) at (0:4){};
			\node at (0:5.2){\small{$k+1$}};
			
			\draw[fill=black] (-30:4) circle (0.05);
			\draw[fill=black] (-45:4) circle (0.05);
			\draw[fill=black] (-60:4) circle (0.05);
			
			\begin{scope}[thick, blue, decoration={markings,mark=at position 0.75 with {\arrow{latex}}}]
				\draw[postaction={decorate}] (K) -- (B);
				\draw[postaction={decorate}] (K) -- (C);
				\draw[postaction={decorate}] (K) -- (D);
				\draw[postaction={decorate}] (K) -- (K2);
				\draw[postaction={decorate}] (K) -- (K1);
				\draw[postaction={decorate}] (K) -- (K11);
				
				\draw[postaction={decorate}] (A) -- (B);
				\draw[postaction={decorate}] (A) -- (C);
				\draw[postaction={decorate}] (A) -- (D);
				\draw[postaction={decorate}] (A) -- (K2);
				\draw[postaction={decorate}] (A) -- (K1);
				\draw[postaction={decorate}] (A) -- (K11);
			\end{scope}
		\end{tikzpicture}
		\caption{\label{PStab} Pure~$P$ stabilizer generator $h_k$ described by the vector $\vec{w}_k=\vec{A}_1+\vec{A}_k$.}
   \end{subfigure}
   \caption{\label{stabfigures}Graphical representations of pure~$X$ stabilizers ($\vec{v}_{ij}$), star vectors ($\vec{A}_k$), and pure~$P$ stabilizers ($\vec{w}_k$).}
\end{figure*}

Using the formalism described above, we now define the CV codes used for information replication. The initial motivation for the codes presented herein is based on simplicial homology of an $n$-dimensional sphere. However, in this paper we present a graph theoretic description of the code and defer discussion of the homological motivation to \cref{Homology}.

Recall that any pure~$X$ or~$P$ stabilizer generator has an equivalent description in terms of a real vector~$\vec{v}$ or~$\vec{w}$ of length~$\binom{N}{2}$. We define the standard basis vectors for this~$\binom{N}{2}$-dimensional vector space to be $\vec{e}_{jk}$ ($1 \leq j <k \leq N$), such that $\vec{e}_{jk}$ contains only one non-zero element corresponding to the mode living on the edge connecting vertices $j$ and $k$ of the complete causal graph.  This entry is normalized to unity.  We also take the convention $\vec{e}_{jk}=-\vec{e}_{kj}$. With these definitions, we are in a position to describe the stabilizer generators in terms of the basis $\{\vec{e}_{jk}\}$.

\textbf{Pure~$X$ stabilizer generators:  }
In our CSS code, there are $\binom{N-1}{2}$ pure~$X$ stabilizer generators described by the vectors $\vec{v}_{jk} = \vec{e}_{1j} + \vec{e}_{jk} + \vec{e}_{k1}$ ($2 \leq j < k \leq N$)\footnote{The promotion of vertex $1$ to a special vertex is arbitrary and taken without loss of generality.}. These pure~$X$ stabilizers only generate displacements in the~$P$ quadrature.  In the graph picture, pure~$X$ generators are directed, closed triangles containing vertex $1$. \Cref{stabfigures}(a) shows an example of a pure~$X$ stabilizer in our code, and the corresponding unitaries are given by $g_{jk}=\exp(i \vec{v}_{jk}\cdot\vec{X})$, $(2\leq j < k \leq N)$.

\textbf{Pure~$P$ Stabilizer generators:  }
As a useful intermediate step in the definition of pure~$P$ stabilizer generators, we first define \emph{star vectors}, $\vec{A}_j=\sum_{k\neq j}{\vec{e}_{jk}}$.  A graphical representation of such a vector is shown in \cref{stabfigures}(b).  In terms of these star vectors, the corresponding vectors for all $N-2$ pure~$P$ generators are given by $\vec{w}_k=\vec{A}_1 + \vec{A}_k,\quad (2\leq k\leq N-1)$.  These pure~$P$ stabilizer generators are depicted graphically in \cref{stabfigures}(c), and the unitaries are of the form $h_k=\exp(i \vec{w}_k\cdot\vec{P})$, $(2\leq k\leq N-1)$.

Having defined both pure~$X$ and pure~$P$ stabilizer generators, the full Lie algebra is the space spanned by all stabilizer generators.  The stabilizer group is then formed by exponentiating the stabilizer Lie subalgebra.

It can be checked that the generators of~$\mathfrak{s}$ are independent (see \cref{ErrorCorrectionProof}) and that the total number of generators is $\binom{N-1}{2} + N-2 = \binom{N}{2} -1$, as desired. The orthogonality condition \cref{orthcon} becomes $\vec{v}_{jk}\cdot(\vec{A}_1+\vec{A}_l)=0$. To prove this, it suffices to show $\vec{v}_{jk}\cdot\vec{A}_{l}=0$. We have
\begin{equation}
\vec{v}_{jk}\cdot\vec{A}_l=(\vec{e}_{1j} + \vec{e}_{jk} + \vec{e}_{k1})\cdot\left(\sum_{m\neq l}{\vec{e}_{lm}}\right).
\end{equation}
Thus, if the vertex $l$ of the star $\vec{A}_l$ is not equal to $1$, $j$, or $k$, then the two vectors are trivially orthogonal as they share no common edges. On the other hand, if $l$ \emph{is} equal to one of the vertices of the triangle $\vec{v}_{jk}$ (say, for example vertex $j$), then the sum of the only non-zero terms in the inner product of $\vec{v}_{lk}\cdot\vec{A}_l$ is
\begin{equation}
\vec{e}_{1l}\cdot\vec{e}_{l1}+ \vec{e}_{lk}\cdot\vec{e}_{lk} = -1+1=0.
\end{equation}
Thus, we have a valid set of stabilizer generators.

Having defined the stabilizer generators, we claim that our code is a valid error correcting code for the error model described above.

\emph{Theorem 1:} \begin{minipage}[t]{0.75\textwidth}The stabilizer code generated by $\mathfrak{s}$ can correct erasure errors on all modes not adjacent to any vertex.
\end{minipage}

The proof can be found in \cref{ErrorCorrectionProof}.

\subsection{Example: four regions, six modes}\label{6modeEx}

To construct nontrivial examples of information replication, one needs at least three spacetime regions in 2+1 dimensions. (In one spatial dimension, it is sufficient to encode information into a single particle and then propagate the particle along a causal trajectory.) There is an interesting configuration of three regions in 2 or more spatial dimensions that was analyzed in~\citet{Hayden2012},
known as the causal merry-go-round. The underlying code in that case has three shares and can correct for the erasure of any single share. As there has already been an experimental demonstration of the corresponding CV quantum error correcting code~\cite{Lance2005}, in this subsection we investigate the next simplest example, which involves four regions.

\begin{figure}[!t]
	\centering
		\begin{tikzpicture}[scale=1]		
			\coordinate (s) at (-1,-3,1.25);
					
			\coordinate (P4) at (1.25,-2.5,1.25);
			
			\coordinate (CA) at (0.5,-0.25,0);
			\coordinate (CB) at (3,-0.25,0);
			\coordinate (CC) at (2.5,0,2.5);
			\coordinate (CD) at (0,0,2.5);
			
			\coordinate (QD) at (0.5,2.25,0); 
			\coordinate (QB) at (2.75,1.2,1.25);
			\coordinate (QA) at (3,2.25,0);
			\coordinate (QC) at (1.25,1.25,2.5);

			\draw[dashed, thin] (CA) -- (CB) -- (CC) -- (CD) -- cycle;
			\draw[dashed, thin] (0.5,2.25,0) -- (3,2.25,0) -- (2.5,2.5,2.5) -- (0,2.5,2.5) -- cycle;
			
			\draw[dashed, thin] (0,0,2.5) -- (0,2.5,2.5);
			\draw[dashed, thin] (2.5,0,2.5) -- (2.5,2.5,2.5);
			
			\node [below] at (CA) {$y_3$};
			\node [below right] at (CB) {$y_1$};
			\node [below right] at (CC) {$y_2$};
			\node [below left] at (CD) {$y_4$};
			
			\draw[red][-triangle 45] (CC) -- (QB);
			\draw[red][-triangle 45] (CB) -- (QA);
			\draw[red][-triangle 45] (CA) -- (QD);
			\draw[red][-triangle 45] (CD) -- (QC);
			\draw[red][-triangle 45] (CB) -- (QD);
			\draw[red][-triangle 45] (CC) -- (QA);
			
			\draw[very thick] (CA) -- (QA);
			\draw[very thick] (CB) -- (QB);
			\draw[ultra thick] (CC) -- (QC);
			\draw[ultra thick] (CD) -- (QD);
			
			\draw[fill=black] (CA) circle (0.15);
			\draw[fill=black] (CB) circle (0.15);
			\draw[fill=black] (CC) circle (0.15);
			\draw[fill=black] (CD) circle (0.15);
			
			\draw[fill=blue]  (QA) circle (0.15);
			\draw[fill=blue]  (QB) circle (0.15);
			\draw[fill=blue]  (QC) circle (0.15);
			\draw[fill=blue]  (QD) circle (0.15);
			
			\node [above right] at (QA) {$z_3$};
			\node [right] at (QB) {$z_1$};
			\node [above] at (QC) {$z_2$};
			\node [above left] at (QD) {$z_4$};
			
			\draw[fill=yellow] (P4) circle (0.15);
			\node [below right] at (P4) {$s$};	
		\end{tikzpicture}
	\centering
	\caption{A configuration of four spacetime regions which allows information replication. The vertical direction is time and the four causal diamonds are drawn in black. In this example, the timelike separated points $y_j$ and $z_j$ have different spatial coordinates, making the diamonds look like long, thin lightlike line segments. The red rays are also lightlike and depict causal curves between various causal diamonds. Note the causal curve from $y_2$ to $z_3$ passing through $z_1$.  This curve passes through three of the four causal diamonds and allows us to reduce the number of shares required.\label{4regions}}
\end{figure}
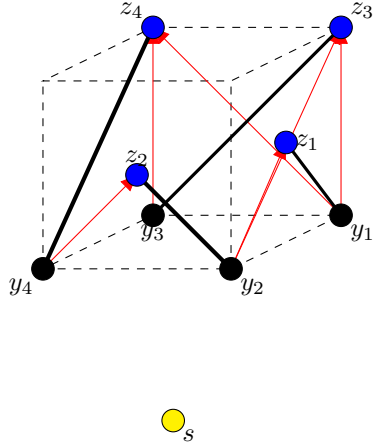

Many configurations of four regions are reducible to fewer regions using ad hoc arguments, but the configuration shown in \cref{4regions}, also studied in~\cite{Hayden2012}, is not.

The graph of causal relations and our error model are are shown in \cref{FourRegionGraph}.    Using our general CV code construction, we require six modes to complete the task.  The generator matrix for this code, which is a matrix whose rows contain the real vectors that define the stabilizer unitaries, is given by
\begin{equation}
\bordermatrix{
&1 & 2  & 3  & 4 & 5 & 6 &\VR 1 & 2 & 3 & 4  & 5 & 6\cr
&1 & -1 & 0  & 1 & 0 & 0 &\VR 0 & 0 & 0 & 0  & 0 & 0\cr
&0 & 1  & -1 & 0 & 0 & 1 &\VR 0 & 0 & 0 & 0  & 0 & 0\cr
&1 & 0  & -1 & 0 & 1 & 0 &\VR 0 & 0 & 0 & 0  & 0 & 0\cr
&0 & 0  & 0  & 0 & 0 & 0 &\VR 0 & 1 & 1 & 1  & 1 & 0\cr
&0 & 0  & 0  & 0 & 0 & 0 &\VR 1 & 0 & 1 & -1 & 0 & 1
},
\end{equation}
where the labels $1$ to $6$ above the matrix represent the six modes, equivalently labeled as $e_{12}$, $e_{13}$, $e_{14}$, $e_{23}$, $e_{24}$, $e_{34}$.  The stabilizer generators for this example are shown in \cref{6modegraphs}.
\begin{figure}[!h]
	\centering
		\begin{tikzpicture}[scale=0.6]
			\node (t') at (-1,0) {a)};

			\node[label=left:\small{$2$}] (A) at (0,0) {};
			\node[label=right:\small{$3$}] (B) at (7,0) {};
			\node[label=right:\small{$4$}] (C) at (7,-7) {};
			\node[label=left:\small{$1$}] (D) at (0,-7) {};
			
			\begin{scope}[decoration={markings,mark=at position 0.5 with {\arrow{latex}}}]
			\draw[postaction={decorate}] (A) -- (B);
			\draw[postaction={decorate}] (B) -- (C);
			\draw[postaction={decorate}] (D) -- (C);
			\draw[postaction={decorate}] (D) -- (A);
			\end{scope}
			\begin{scope}[decoration={markings,mark=at position 0.6 with {\arrow{latex}}}]
			\draw[postaction={decorate}] (D) -- (B);
			\draw[postaction={decorate}] (A) -- (C);
			\end{scope}
				
			\draw[fill=white] (A) circle (0.2);
			\draw[fill=white] (B) circle (0.2);
			\draw[fill=white] (C) circle (0.2);
			\draw[fill=white] (D) circle (0.2);
			
		\end{tikzpicture}
		\qquad\qquad
		\begin{tikzpicture}[scale=0.6]
			\node (t') at (-1,-10) {b)};

			\node[label=left:\small{$2$}] (P) at (0,-10) {};
			\node[label=right:\small{$3$}] (Q) at (7,-10) {};
			\node[label=right:\small{$4$}] (R) at (7,-17) {};
			\node[label=left:\small{$1$}] (W) at (0,-17) {};
			
			\coordinate (PQ) at (3.5,-10);
			\coordinate (QR) at (7, -13.5);
			\coordinate (RW) at (3.5, -17);
			\coordinate (WP) at (0, -13.5);
			\coordinate (PR) at (1.75, -11.75);
			\coordinate (WQ) at (5.25, -11.75);
			
			\draw (P) circle (0.2);
			\draw (Q) circle (0.2);
			\draw (R) circle (0.2);
			\draw (W) circle (0.2);
			
			\draw(0.2,-10) -- (6.8,-10);
			\draw[dashed] (7,-10.2) -- (7,-16.8);
			\draw[dashed] (6.8,-17) -- (3.7,-17);
			\draw[dashed] (3.7,-17) -- (0.2,-17);
			\draw (0,-16.8) -- (0,-10.2);
			\draw (6.86,-16.86) -- (0.14,-10.14);
			\draw[dashed] (6.86,-10.14) -- (0.14,-16.86);
			
			\pgfmathsetmacro{\XValueArc}{\ArcRadius*cos(\ArcAngle)}%
			\pgfmathsetmacro{\YValueArc}{\ArcRadius*sin(\ArcAngle)}%
			\pgfmathsetmacro{\XValueLabel}{\ArcRadius*cos(\ArcAngle/2)}%
			\pgfmathsetmacro{\YValueLabel}{\ArcRadius*sin(\ArcAngle/2)}%
			
			\draw[purple] ($(-0.5,-9.5)+(1.7,0)$) arc (0:-90:1.7);
			\draw[purple,dashed] ($(-0.5,-17.5)+(1.7,0)$) arc (0:90:1.7);
			\draw[purple,dashed] ($(7.5,-9.5) -(1.7,0)$)  arc (0:90:-1.7);
			\draw[purple,dashed] ($(7.5,-17.5) -(1.7,0)$) arc (0:-90:-1.7);
		\end{tikzpicture}
	\caption{\label{FourRegionGraph} (a) The directed graph of causal relations for the configuration shown in \cref{4regions}.  Our six shares live on the edges of this graph. Directionality is chosen to be from the vertex with the smaller label to the vertex with the larger label.  (b) The error model: dashed lines correspond to unavailable shares of the code that we lose, solid lines represent recovered shares, and arcs cover the shares sufficient to reproduce the state at the encompassed vertex. We want to recover the encoded state at any given vertex whenever we have access to the shares on the three adjacent edges (i.e., those crossed by the arc at that vertex).  We model erasure of shares as random displacements on modes that we do not use to recover.}
\end{figure}
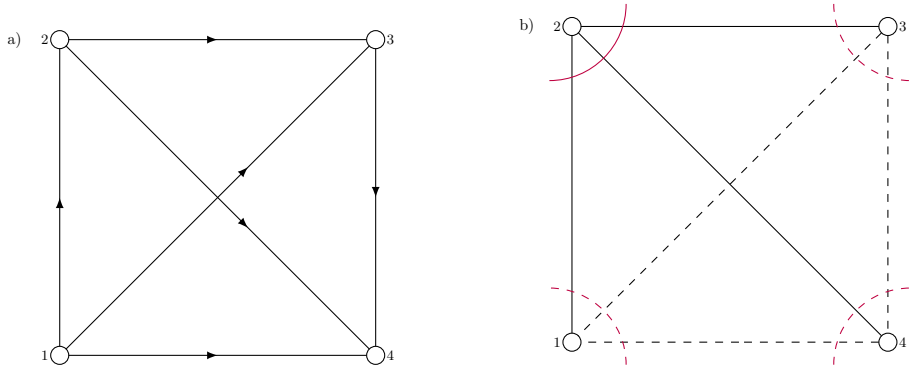
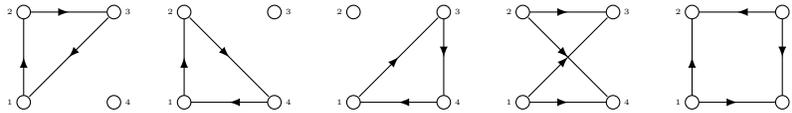
\begin{figure}[!htb]
	\centering
	\begin{tikzpicture}[scale=0.6]

		\node[label=left:\tiny{$1$}] (A) at (0,0) {};
		\node[label=left:\tiny{$2$}] (B) at (0,2) {};
		\node[label=right:\tiny{$3$}] (C) at (2,2) {};
		\node[label=right:\tiny{$4$}] (D) at (2,0) {};
		
		\begin{scope}[decoration={markings,mark=at position 0.5 with {\arrow{latex}}}]
		\draw[postaction={decorate}] (A) -- (B);
		\draw[postaction={decorate}] (B) -- (C);
		\draw[postaction={decorate}] (C) -- (A);
		\end{scope}
		
		\draw[fill=white] (A) circle (0.15cm);
		\draw[fill=white] (B) circle (0.15cm);
		\draw[fill=white] (C) circle (0.15cm);
		\draw[fill=white] (D) circle (0.15cm);
	\end{tikzpicture}\quad
	\begin{tikzpicture}[scale=0.6]
		\node[label=left:\tiny{$1$}] (A) at (0,0) {};
		\node[label=left:\tiny{$2$}] (B) at (0,2) {};
		\node[label=right:\tiny{$3$}] (C) at (2,2) {};
		\node[label=right:\tiny{$4$}] (D) at (2,0) {};
		
		\begin{scope}[decoration={markings,mark=at position 0.5 with {\arrow{latex}}}]
		\draw[postaction={decorate}] (A) -- (B);
		\draw[postaction={decorate}] (B) -- (D);
		\draw[postaction={decorate}] (D) -- (A);
		\end{scope}
		
		\draw[fill=white] (A) circle (0.15cm);
		\draw[fill=white] (B) circle (0.15cm);
		\draw[fill=white] (C) circle (0.15cm);
		\draw[fill=white] (D) circle (0.15cm);
	\end{tikzpicture}
	\quad
	\begin{tikzpicture}[scale=0.6]
		\node[label=left:\tiny{$1$}] (A) at (0,0) {};
		\node[label=left:\tiny{$2$}] (B) at (0,2) {};
		\node[label=right:\tiny{$3$}] (C) at (2,2) {};
		\node[label=right:\tiny{$4$}] (D) at (2,0) {};
		
		\begin{scope}[decoration={markings,mark=at position 0.5 with {\arrow{latex}}}]
		\draw[postaction={decorate}] (A) -- (C);
		\draw[postaction={decorate}] (C) -- (D);
		\draw[postaction={decorate}] (D) -- (A);
		\end{scope}
		
		\draw[fill=white] (A) circle (0.15cm);
		\draw[fill=white] (B) circle (0.15cm);
		\draw[fill=white] (C) circle (0.15cm);
		\draw[fill=white] (D) circle (0.15cm);
	\end{tikzpicture}
	\quad
	\begin{tikzpicture}[scale=0.6]
		\node[label=left:\tiny{$1$}] (A) at (0,0) {};
		\node[label=left:\tiny{$2$}] (B) at (0,2) {};
		\node[label=right:\tiny{$3$}] (C) at (2,2) {};
		\node[label=right:\tiny{$4$}] (D) at (2,0) {};
		
		\begin{scope}[decoration={markings,mark=at position 0.5 with {\arrow{latex}}}]
		\draw[postaction={decorate}] (A) -- (C);
		\draw[postaction={decorate}] (B) -- (C);
		\draw[postaction={decorate}] (B) -- (D);
		\draw[postaction={decorate}] (A) -- (D);
		\end{scope}
		
		\draw[fill=white] (A) circle (0.15cm);
		\draw[fill=white] (B) circle (0.15cm);
		\draw[fill=white] (C) circle (0.15cm);
		\draw[fill=white] (D) circle (0.15cm);
	\end{tikzpicture}
	\quad
	\begin{tikzpicture}[scale=0.6]
		\node[label=left:\tiny{$1$}] (A) at (0,0) {};
		\node[label=left:\tiny{$2$}] (B) at (0,2) {};
		\node[label=right:\tiny{$3$}] (C) at (2,2) {};
		\node[label=right:\tiny{$4$}] (D) at (2,0) {};
		
		\begin{scope}[decoration={markings,mark=at position 0.5 with {\arrow{latex}}}]
		\draw[postaction={decorate}] (A) -- (B);
		\draw[postaction={decorate}] (C) -- (B);
		\draw[postaction={decorate}] (C) -- (D);
		\draw[postaction={decorate}] (A) -- (D);
		\end{scope}
		
		\draw[fill=white] (A) circle (0.15cm);
		\draw[fill=white] (B) circle (0.15cm);
		\draw[fill=white] (C) circle (0.15cm);
		\draw[fill=white] (D) circle (0.15cm);
	\end{tikzpicture}
\caption{\label{6modegraphs}The five graph state stabilizer generators of our six mode code for four spacetime regions.}
\end{figure}
%
\section{A five-mode code for four spacetime regions}\label{FiveModeCodeSection}
Having developed a general framework for replicating information in spacetime using $N>3$ modes, we now propose an experimentally feasible demonstration of information replication for the configuration of $N=4$ spacetime regions in \cref{4regions}.  The general method as depicted in \cref{6modeEx} uses six modes to complete the task.  As experimental feasibility is a priority, we would like to reduce the number of required modes.
As such, we will deviate from the codes presented above and employ a \emph{different} optimized five-mode CV CSS code.
The improvement comes about by exploiting the presence of a causal curve passing through three of the four regions (i.e., regions 1, 2, and 3).  The existence of this curve allows us to use five shares instead of six, as one of the shares can be physically transported along this causal curve through three instead of just two causal diamonds.  This causal curve is clear in \cref{4regions} as it starts at point $y_2$, passes through point $z_1$, and ends at point $z_3$.

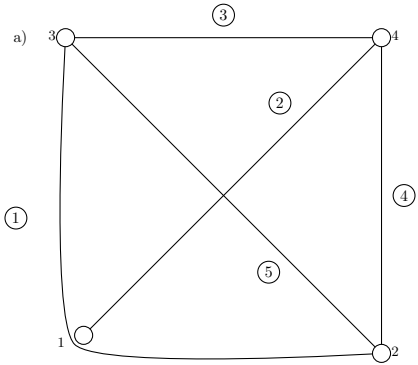
\begin{figure}[!htb]
	\centering
	\begin{tikzpicture}[scale=0.6]
		\node (t') at (-1,0) {a)};

		\coordinate (A) at (0,0);
		\coordinate (B) at (7,0);
		\coordinate (C) at (7,-7);
		\coordinate (D) at (0.4,-6.6);
		\coordinate (Dprime) at (0.2,-6.8);
		
		\draw (A) -- (B);
		\draw (B) -- (C);
		\draw (C) -- (A);
		\draw (D) -- (B);
		\draw plot [smooth, tension=0.3] coordinates {(A) (Dprime) (C)}	;
		
		\draw[fill=white] (A) circle (0.2);
		\draw[fill=white] (B) circle (0.2);
		\draw[fill=white] (C) circle (0.2);
		\draw[fill=white] (D) circle (0.2);
		
		\node[label={[shift={(-0.3,-0.3)}]\small{$3$}}] at (A) {};
		\node[label={[shift={(0.3,-0.3)}]\small{$4$}}] at (B) {};
		\node[label={[shift={(0.3,-0.3)}]\small{$2$}}] at (C) {};
		\node[label={[shift={(-0.3,-0.3)}]\small{$1$}}] at (Dprime) {};
		\node[shape=circle,draw,inner sep=2pt] at (-1.1,-4) {1};
		\node[shape=circle,draw,inner sep=2pt] at (3.5,0.5) {3};
		\node[shape=circle,draw,inner sep=2pt] at (7.5,-3.5) {4};
		\node[shape=circle,draw,inner sep=2pt] at (4.75,-1.45) {2};
		\node[shape=circle,draw,inner sep=2pt] at (4.5,-5.2) {5};
	\end{tikzpicture}
\caption{\label{FiveModeStructure}The graph of causal relations for the four region code.  The vertex labels correspond to the causal diamonds, and the edges correspond to the causal connections between diamonds.  The ugly curved line connecting vertices 1, 2, and 3 represents the causal curve passing through all three regions, directed from 2 through 1 to 3.  While this curve destroys the aesthetics of the figure, it greatly simplifies the code. Circled numbers label the five modes used in the optical implementation.  Errors $E_i$ correspond to the erasure of all modes not adjacent to vertex $i$.}
\end{figure}

The graph of causal structure is shown in \cref{FiveModeStructure}, where the curved line represents this multi-diamond causal curve.  As before, we associate one mode with each edge of this graph.
The stabilizer generators of our five-mode code are specified by the rows of the following parity-check matrix
\begin{equation}\label{fivemodegeneratormatrix}
\bordermatrix{
&1  & 2  & 3  & 4 & 5  &\VR 1 & 2 & 3  & 4 & 5 \cr
&-1 & -1 & 1  & 1 & 0  &\VR 0 & 0 & 0  & 0 & 0 \cr
&0  & 0  & -1 & 1 & -2 &\VR 0 & 0 & 0  & 0 & 0 \cr
&0  & 0  & 0  & 0 & 0  &\VR 1 & 1 & 1  & 1 & 0 \cr
&0  & 0  & 0  & 0 & 0  &\VR 0 & 0 & -1 & 1 & 1
}
\end{equation}
where the numbers 1-5 above label the modes as shown in \cref{FiveModeStructure}.  For example, the stabilizer generator specified by the first row is
\begin{equation}
g_1=-X_1-X_2+X_3+X_4,
\end{equation}
where $X_j$ is the~$X$ operator on mode $j$.  Similarly, the generator specified by the fourth row is
\begin{equation}
h_2=-P_3+P_4+P_5,
\end{equation}
where $P_i$ is the observable conjugate to~$X$ on mode $i$.  For concreteness, $h_2$ corresponds to displacing the~$X$ quadrature by $-1$ unit in each of modes 4 and 5, and displacing mode 3 by $+1$ unit.

The stabilizer code associated with these generators can correct arbitrary displacement errors on certain subsets of the modes.  In particular, the error model we can correct is summarized by
\begin{equation}\label{errormodel}
\bordermatrix{
    &1  & 2 & 3 & 4 & 5 &\VR 1 & 2 & 3 & 4 & 5 \cr
	E_1 &   &   & * & * & * &\VR   &   & * & * & * \cr
	E_2 &   & * & * &   &   &\VR   & * & * &   &   \cr
	E_3 &   & * &   & * &   &\VR   & * &   & * &   \cr
	E_4 & * &   &   &   & * &\VR * &   &   &   & * }
\end{equation}
where $*$ is an arbitrary displacement error on that mode. Each error $E_j$ will be a unitary operator that corresponds to erasure of all modes not adjacent to vertex $j$ in \cref{FiveModeStructure}.  For example, we can correct
\begin{equation}
	E_3=\exp\{i\left(aX_2+bX_4+cP_2+dP_4\right)\}
\end{equation}
where $a$, $b$, $c$, and $d$ are arbitrary.  This is equivalent to arbitrary displacements in phase space of modes 2 and 4.  This unique error model arises in the context of information replication, as we want to reconstruct the state using only information in a known subset of modes.  Whereas other CV quantum error correcting codes can protect against a more familiar error model (e.g., arbitrary (small) displacement errors~\cite{Gottesman2001} everywhere), our code allows us to correct against arbitrarily large, \emph{located} errors.

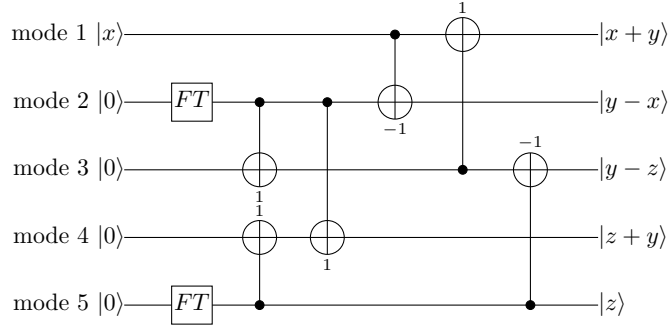
\begin{figure}[!htb]
\centering
\begin{tikzpicture}[scale=0.9, every node/.style={inner sep=0pt}]  
		\node [left](mode1start) at (0,4) {mode 1 $\ket{x}$};
		\node [left](mode2start) at (0,3) {mode 2 $\ket{0}$};
		\node [left](mode3start) at (0,2) {mode 3 $\ket{0}$};
		\node [left](mode4start) at (0,1) {mode 4 $\ket{0}$};
		\node [left](mode5start) at (0,0) {mode 5 $\ket{0}$};
		\node [right](mode5end) at (7,0) {$\ket{z}$};
		\node [right](mode4end) at (7,1) {$\ket{z+y}$};
		\node [right](mode3end) at (7,2) {$\ket{y-z}$};
		\node [right](mode2end) at (7,3) {$\ket{y-x}$};
		\node [right](mode1end) at (7,4) {$\ket{x+y}$};
		
		\drawwires(5,7);
	
		\node [gate] (FT) at (1,0) {$FT$};
		\node [gate] (FT) at (1,3) {$FT$};
		
		\cplus(2,0,2,1,1);	
		\cplus(2,3,2,2,1);
		\cplus(3,3,3,1,1);
		\cplus(4,4,4,3,-1);
		\cplus(5,2,5,4,1);
		\cplus(6,0,6,2,-1)
	\end{tikzpicture}
\caption{\label{EncodingCircuit}An encoding circuit for the encoded state in \cref{encodedstate}.  The $FT$ gates are Fourier transforms, whereas the QND gates are controlled-sum gates, as defined in \cref{cplusdefn}.  Optically, these correspond to quantum non-demolition gates.  The subscripts on the QND gates correspond to a gain in the controlled sum.  When working with continuous variables, the CNOT gate is no longer self-inverse, and a controlled-sum gate with a gain of $-1$ represents controlled-subtraction (i.e., $\text{QND}_{-1}=\text{QND}_1^\dagger$).}
\end{figure}

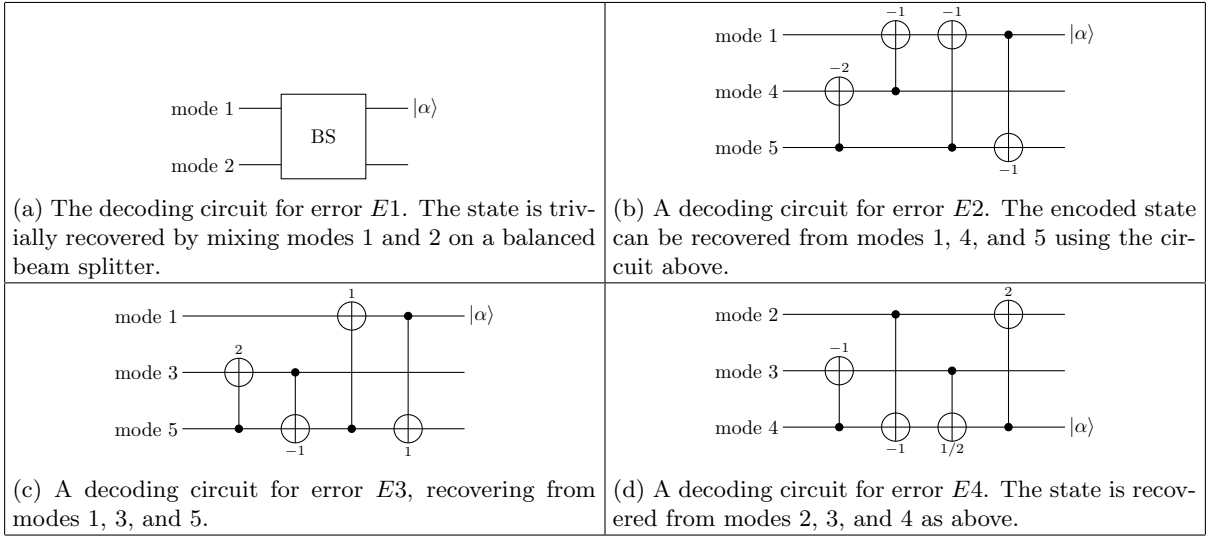
\begin{figure}
\centering
\def\arraystretch{2}
\begin{tabular}{|@{\;}l@{ }|@{\;}r@{ }|}
\hline
\begin{subfigure}[t]{0.47\textwidth}\centering
\begin{tikzpicture}[scale=0.75, every node/.style={inner sep=2pt}]  
		\node [left](mode1start) at (0,1) {mode 1};
		\node [left](mode2start) at (0,0) {mode 2};
		\node [right](mode2end) at (3,0) {};
		\node [right](mode1end) at (3,1) {$\ket{\alpha}$};
		
		\drawwires(2,3);
		
		\draw[fill=white, draw=black] (0.75,-0.25) rectangle (2.25,1.25) node[midway] {BS};
	\end{tikzpicture}
\caption{The decoding circuit for error $E1$. The state is trivially recovered by mixing modes 1 and 2 on a balanced beam splitter.\label{E1circuit}}
\end{subfigure}
&
\begin{subfigure}[t]{0.47\textwidth}\centering
\begin{tikzpicture}[scale=0.75, every node/.style={inner sep=2pt}]  
		\node [left](mode1start) at (0,2) {mode 1};
		\node [left](mode4start) at (0,1) {mode 4};
		\node [left](mode5start) at (0,0) {mode 5};
		\node [right](mode5end) at (5,0) {};
		\node [right](mode4end) at (5,1) {};
		\node [right](mode1end) at (5,2) {$\ket{\alpha}$};
		
		\drawwires(3,5);
		
		\cplus(1,0,1,1,-2);
		\cplus(2,1,2,2,-1);
		\cplus(3,0,3,2,-1);
		\cplus(4,2,4,0,-1);
	\end{tikzpicture}
\caption{A decoding circuit for error $E2$.  The encoded state can be recovered from modes 1, 4, and 5 using the circuit above.\label{E2circuit}}
\end{subfigure}
\\ \hline
\begin{subfigure}[t]{0.47\textwidth}\centering
\begin{tikzpicture}[scale=0.75, every node/.style={inner sep=2pt}]  
		\node [left](mode1start) at (0,2) {mode 1};
		\node [left](mode3start) at (0,1) {mode 3};
		\node [left](mode5start) at (0,0) {mode 5};
		\node [right](mode5end) at (5,0) {};
		\node [right](mode3end) at (5,1) {};
		\node [right](mode1end) at (5,2) {$\ket{\alpha}$};
		
		\drawwires(3,5);
		
		\cplus(1,0,1,1,2);	
		\cplus(2,1,2,0,-1);
		\cplus(3,0,3,2,1);
		\cplus(4,2,4,0,1);
	\end{tikzpicture}
\caption{A decoding circuit for error $E3$, recovering from modes 1, 3, and 5.\label{E3circuit}}
\end{subfigure}
& 
\begin{subfigure}[t]{0.47\textwidth}\begin{center}
\begin{tikzpicture}[scale=0.75, every node/.style={inner sep=2pt}]  
		\node [left](mode2start) at (0,2) {mode 2};
		\node [left](mode3start) at (0,1) {mode 3};
		\node [left](mode4start) at (0,0) {mode 4};
		\node [right](mode4end) at (5,0) {$\ket{\alpha}$};
		\node [right](mode3end) at (5,1) {};
		\node [right](mode2end) at (5,2) {};
		
		\drawwires(3,5);
		
		\cplus(1,0,1,1,-1);	
		\cplus(2,2,2,0,-1);
		\cplus(3,1,3,0,1/2);
		\cplus(4,0,4,2,2);
	\end{tikzpicture}
\caption{A decoding circuit for error $E4$.  The state is recovered from modes 2, 3, and 4 as above.\label{E4circuit}}\end{center}
\end{subfigure}
\\ \hline
\end{tabular}
\caption{\label{IdealDecoders}Ideal decoding circuits using controlled-sum gates. The gates shown correspond to a controlled addition (\emph{i.e.,} quantum non-demolition gate) where the subscript indicates the required gain.  A dagger represents controlled-difference, which is just the adjoint of the controlled-sum.}
\end{figure}

The stabilizer generators define the following basis for the codespace
\begin{equation}\label{encodedstate}
\ket[\text{enc}]{x}=\int dydz \ket{x+y,y-x,y-z,z+y,z}.
\end{equation}
A circuit that prepares the encoded state is shown in \cref{EncodingCircuit}.  Note that this circuit is shown for completeness, but a simple optical implementation to this circuit is provided in \cref{sec:proposed}.
Once the information has been encoded, distributing the shares in spacetime allows one to complete the information replication task.  To demonstrate that the task has been completed, the encoded state is subjected to erasure of a known subset of modes, as in \cref{errormodel}.  We can recover the information from any of $E_1$ to $E_4$ using the decoding circuits shown in \cref{IdealDecoders}.

\section{Proposed experimental implementation}
\label{sec:proposed}

\begin{figure}[!t]
	\centering
\begin{tikzpicture}[scale=0.9, every node/.style={inner sep=0pt}]  
			\node[label={[label distance=0.5em]below:\small{$y_4$}}] (y4) at (0,0) {};
			\node[label={[label distance=0.5em]below:\small{$y_2$}}] (y2) at (4,0) {};
			\node[label={[label distance=0.5em]above:\small{$y_1$}}] (y1) at (3.7,4) {};
			\node[label={[label distance=0.5em]left:\small{$y_3$}}] (y3) at (-0.15,4) {};
			\node[label={[label distance=0.5em]above:\small{$z_4$}}] (z4) at (0.15,4) {};
			\node[label={[label distance=0.5em]below:\small{$z_2$}}] (z2) at (2,0) {};
			\node[label={[label distance=0.5em]right:\small{$z_1$}}] (z1) at (4.075,2) {};
			\node[label={[label distance=0.5em]right:\small{$z_3$}}] (z3) at (4,4) {};
			\node[label={[label distance=1em]right:\scriptsize{mirror}}] (mirror) at (0,2) {};
			\draw (-0.25,2) -- (0.25,2);
			\draw (0.15,1.9) -- (0.25,2);
			\draw (0.05,1.9) -- (0.15,2);
			\draw (-0.05,1.9) -- (0.05,2);
			\draw (-0.15,1.9) -- (-0.05,2);
			\draw (-0.25,1.9) -- (-0.15,2);
			
			\draw[fill=black] (y1) circle (0.1);
			\draw[fill=black] (y2) circle (0.1);
			\draw[fill=black] (y3) circle (0.1);
			\draw[fill=black] (y4) circle (0.1);
			\draw[fill=blue] (z1) circle (0.1);
			\draw[fill=blue] (z2) circle (0.1);
			\draw[fill=blue] (z3) circle (0.1);
			\draw[fill=blue] (z4) circle (0.1);
			
			\draw[thick, -latex] ($(y2)-(0.075,0)$) -- ($(z3)-(0.075,0)$);
			\draw[thick, -latex] ($(y2)+(0.075,0)$) -- ($(z3)+(0.075,0)$);
			\draw[thick, -latex] (y4) -- (z2);
			\draw[thick, -latex] (y1) -- (z4);
			\draw[thick, -latex] (y3) -- (mirror) -- (z4);
			
			\node[shape=circle,draw,inner sep=1pt] at (4.3,1) {\scriptsize $1$};
			\node[shape=circle,draw,inner sep=1pt] at (-0.3,3) {\scriptsize $3$};
			\node[shape=circle,draw,inner sep=1pt] at (1,-0.3) {\scriptsize $4$};
			\node[shape=circle,draw,inner sep=1pt] at (2,4.3) {\scriptsize $2$};
			\node[shape=circle,draw,inner sep=1pt] at (3.7,3) {\scriptsize $5$};
\end{tikzpicture}
\caption{%
	A bird's-eye-view (i.e., top down) of an optical setup for realizing the information replication task shown in \cref{4regions}.  Modes are represented by solid lines with their direction of propagation indicated by arrows. Points $y_3$ and $z_4$ are at the same spatial location, but they are shown separated for clarity.  Similarly, $y_1$ and $z_3$ are at the same location.   Points $y_2$, $z_1$, and $z_3$ are collinear.  In our ideal scheme share number 3 was shown to remain spatially stationary, but since that was not required, in the experimental setup a mirror is used instead to return share number 3 to its initial position. Modes are labelled by the circled numbers 1 through 5, with the same numbering as in \cref{FiveModeStructure}.\label{birdseye}}
\end{figure}
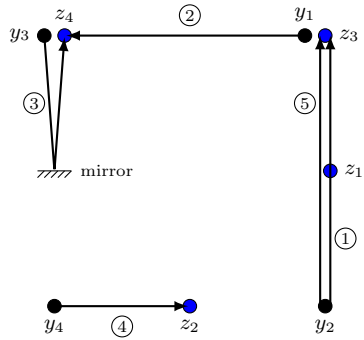

In this section we consider how to realize a spacetime replication protocol.
We focus on the simplest nontrivial case:
four spacetime regions as discussed in \cref{FiveModeCodeSection}.
According to that idealized protocol, we need a five-mode code,
which means in this case that we need a five-channel interferometer
and perhaps also a six-channel interferometer. Since we'd like to realize information replication optically, the five modes in the code will need to be physically oriented according to the spacetime configuration shown in \cref{4regions}.  An example of such an orientation is shown in \cref{birdseye}, which is a bird's-eye-view of a suggested optical layout.

This section begins with a discussion in \cref{subsec:opasgame}
concerning an operational interpretation of information replication as a two-party game.  In \cref{subsec:requisite}
we discuss the optical devices needed to simulate such a game experimentally.  Note that non-optical media that involve coupled harmonic oscillators, squeezing,
and quantum nondemolition (QND) measurements~\cite{YMH+08}
could be viable instead of optical elements, provided they can be made to follow spacetime trajectories compatible with our motivating example.  The encoding and decoding processes are elaborated in \cref{subsec:coding}.  In \cref{subsec:model} we address a realization involving finite squeezing and faulty optical elements.

\subsection{Replication as an optical game}
\label{subsec:opasgame}

In this subsection we revisit the spacetime replication protocol,
which we present as a game. 
The context of a two-player game is important for analyzing experimental limitations and thresholds for declaring success.
In \cref{subsubsec:operationalizing},
we cast the spacetime replication task as an operational two-party game amenable to optical implementation.

\subsubsection{Operationalizing the replication task}
\label{subsubsec:operationalizing}

We reprise and operationalize the summoning task discussed earlier.
The referee creates a state~$\rho$ in a localized neighborhood centered at spacetime point~$s$.
The state~$\rho$ acts on some Hilbert space that is not exclusive to the Hilbert space Alice accesses.
Whereas earlier we described the referee's state as the pure state~$\ket{\phi}\in\mathcal{H}_\text{A}$,
here we allow the referee to wittingly or unwittingly create a state that can be mixed or be entangled with degrees of freedom outside~$\mathcal{H}_\text{A}$.
Alice's share of the state is~$\operatorname{tr}_\text{E}\rho$ for~$E$ referring to the environment,
which is the world beyond Alice's control.

One of the referee's agents,
stationed at spacetime location~$y_j$,
requests that one of Alice's agents reveal the supplied quantum state at reveal point~$z_j$ in the $j^\text{th}$ causal diamond.
If the bipartite state~$\rho$ over~$\mathcal{H}_\text{A}\otimes\mathcal{H}_\text{E}$ is pure,
the Referee can, in principle,
perform a single-shot determination of whether Alice has successfully played the summoning game by applying a binary measurement that projects onto~$\rho$ or its complement.
For example, the Referee can create two copies of the pure state~$\rho$ and then perform a controlled-SWAP operation~\cite{Wan01}
to determine whether the resultant state is identical to the original state or not.
The measurement outcome is either success or failure, written into a single bit of information.

Experimentally demonstrating the full game as described above would require techniques that fall well outside our current abilities. As such, we instead describe in \cref{subsubsec:certifying} a secondary, subordinate protocol in which a Certifier simulates gameplay with Alice in order to determine (statistically) whether or not she would win a real game with a full-power Referee.  In the simulated gameplay, the Certifier assumes the role of a Referee, albeit with bounded experimental capabilities, and cannot prepare arbitrarily pure states or use perfect single-shot measurements.  We stress here that these experimental limitations apply not to Alice (since her protocol is equipped to handle arbitrary input), but rather to the Certifier.  Our hope is that this distinction provides a hierarchy of experimental milestones that can be achieved over time; the first such milestone (certification) is achievable using current technology, while the next milestone (gameplay involving a full Referee) may be quite far off.  We elaborate on the distinction between Certifier and Referee, and describe the certification protocol in \cref{subsubsec:certifying}.

\subsection{Requisite optical devices}
\label{subsec:requisite}

This subsection is about the requisite optical elements for the information replication game.
In \cref{subsubsec:sources} we discuss the light sources, optical parametric oscillators (OPOs),
and linear optical elements.
The quantum nondemolition (QND) gate and its adjoint  are 
described in \cref{subsubsec:qnd}.
The detectors, the optical homodyne tomography procedure, and feedforward for control are discussed in 
\cref{subsubsec:detectors}.

\subsubsection{Sources, optical parametric oscillation and linear optics}
\label{subsubsec:sources}

In this subsection, we consider the sources that are needed for certification and for playing the replication game.
The game allows the Referee to send an arbitrary state,
which can be single-mode or multi-mode, and which can be in an entangled state between Alice's mode and other modes outside Alice's domain.
In practice, only a few types of states can be generated as propagating fields (as opposed to cavity states),
such as coherent states, squeezed states, Fock states, and Schr\"{o}dinger cat states~\cite{DM03}.

The single-mode coherent state can be expressed as the displaced vacuum state
$\ket{\alpha}=D(\alpha)\ket{0}$ for
\begin{equation}
\label{eq:displacement}
	D(\alpha)
		=\exp\left\{\alpha\hat{a}^\dagger-\alpha^*\hat{a}\right\},
\end{equation}
for~$D(\alpha)$ the displacement operator and~$\ket{0}$ the vacuum state.

The coherent state is considered to be an excellent description of a laser output state
provided that the laser is stable, operates in a single mode,
and is highly coherent,
which can be thought as phase stability with respect to a local oscillator.
Thus, stable lasers are used as coherent-state sources,
which can serve as states for Alice's replication or as local oscillator states serving as phase references for
all states including coherent and squeezed states.

Alice employs classically controlled displacements given by
\begin{equation}
\label{eq:Disp}
	D_c:(x,y_\text{classical})\mapsto(x+cy_\text{classical},y_\text{classical}).
\end{equation}
These unitaries correspond to a displacement (using the operator in \cref{eq:displacement}) by an amount $c$ times the classical value $y_\text{classical}$.  Displacements (shown in our figures as Disp) without an argument correspond to $c=1$ in our notation.

Using the one-mode squeezing operator~\cite{LK87}
\begin{equation}
\label{eq:onemodesqueezing}
	S_a(\xi)=\exp\left\{\frac{1}{2}\left(\xi\hat{a}^{\dagger 2}-\xi^*\hat{a}^2\right)\right\},
\end{equation}
where $\xi=r_1 e^{i\phi_1}$, and the two-mode squeezing operator~\cite{LK87}
\begin{equation}
\label{eq:twomodesqueezing}
	S_{ab}(\eta)=\exp\left\{\eta\hat{a}^{\dagger}\hat{b}^{\dagger}-\eta^*\hat{a}\hat{b}\right\},
\end{equation}
where $\eta=r_2 e^{i\phi_2}$, the one- and two-mode squeezed states are
\begin{equation}
	\ket{\xi} =\frac{1}{\sqrt{\cosh r_1}}\sum_{n=0}^{\infty}e^{i n\phi_1}(\tanh r_1)^n \frac{\sqrt{(2n)!}}{n! 2^n}\ket{2n}
\end{equation}
and
\begin{equation}
	\ket{\eta}=\frac{1}{\cosh r_2}\sum_{n=0}^{\infty}e^{i n\phi_2}(\tanh r_2)^n \ket{n,n},
\end{equation}
respectively.
Physically the one- and two-mode squeezing operators~(\ref{eq:onemodesqueezing})
and~(\ref{eq:twomodesqueezing})
are realized by degenerate and nondegenerate OPO, respectively~\cite{LK87}.
Alternatively, nondegenerate parametric amplification suffices if the two-mode squeezed
state is directed into a balanced beam splitter,
which yields a product state of a one-mode squeezed state~$\ket{\xi}$
and its anti-squeezed counterpart~$\ket{-\xi}$ in the other mode.

Coherent states, squeezed states and other states can be manipulated by linear optics.
Linear optical elements do not have any additional electromagnetic power source such as electrical or an optical pump.
If a linear optical element is lossless, it preserves flux (and hence photon number) in the quantum field case.

The beam splitter is a four-port linear optical element~\cite{CST89}:
it has two input ports and two output ports,
and we treat the beam splitter as a flux-preserving transformation independent of the frequency of the mode.
The beam splitter is described by the transformation
\begin{equation}
\label{eq:beamsplitter}
	B(\theta)
		=\exp\left\{\frac{\theta}{2}\left(\hat{a}\hat{b}^\dagger-\hat{a}^\dagger\hat{b}\right)\right\},
\end{equation}
where~$\sin\theta$ is the beam splitter reflectivity.
A mirror is just a beam splitter with~$\theta=\pi/2$ in which case reflectivity is~$1$.

A phase shifter is given by $\exp\left(\text{i}\varphi\hat{a}^\dagger\hat{a}\right)$,
which shifts the phase of the mode by~$\varphi$.
A phase shift of $\varphi=\pi/2$ effects a transformation of the in-phase~$x$ quadrature to the out-of-phase~$p$ quadrature
and the~$p$ quadrature to~$-x$.
This special case of $\varphi=\pi/2$ is the Fourier Transform (FT) of the field~\cite{LB99}.
The phase shifter with~$\varphi=\pi$ is the $\pi$ gate,
which maps $(x,y)\mapsto(-x,-y)$.

Finally, discarding, or dumping, a mode is achieved via linear optics.
This identification can be seen by considering using a mirror to direct a mode out of the region of interest so the mode is neither detected nor influences the rest of the apparatus.
As a mirror is linear optical, so is the mode dumping procedure.

\subsubsection{Quantum nondemolition gates}
\label{subsubsec:qnd}

The CV SUM gate~\cite{BSBN02}, or QND gate~\cite{BvL05}, effects the mapping
\begin{equation}
\label{eq:QNDx}
	\operatorname{QND}_c:(x,y)\mapsto(x+cy,y)
\end{equation}
and its adjoint effects
\begin{equation}
\label{eq:QNDxdagger}
	\operatorname{QND}_c^\dagger:(x,y)\mapsto(x-cy,y).
\end{equation}
This is the coherent version of the classically controlled displacement \cref{eq:Disp}, and is precisely the transformation in \cref{cplusdefn}.  One way to realize this gate is in the so-called ``off-line'' scheme~\cite{FMA05,YMH+08}, which we explain below.

The ``off-line'' QND gate requires two incoming beams of light
corresponding to the two-mode basis~$\{x,y\}$
as in \cref{eq:QNDx} and two OPO-generated squeezed vacuum states.
The two incoming beams are directed into a special type of Mach-Zehnder interferometer with squeezing gates in each arm.
The first mirror of the interferometer has beam splitter reflectivity~$\frac{R}{1+R}$,
and the second beam splitter has reflectivity~$\frac{1}{1+R}$.
Tunability of all beam splitters is achieved by constructing the beam splitters from two polarizing beam splitters and a half-wave plate.

The two arms of the interferometer each mix the beam in the respective arm with an OPO-generated one-mode squeezed-vacuum state
at a beam splitter of reflectivity~$R$.
This beam splitter mixes the signal beam with the squeezed vacuum,
and one of the two beam-splitter outputs is subjected to homodyne detection whose signal controls an electro-optic modulator~\cite{LR09},
which, in turn, modulates an auxiliary beam that mixes at a 99:1 beam splitter with the signal field.
The squeezing operation is similar in both arms of the interferometer and depends on the relative phases of the two paths.
The transformation from input to output is given by \cref{eq:QNDx} for
$a=1/\sqrt{R}-\sqrt{R}$
and has been demonstrated experimentally for $a=1$ and $a=3/2$.
The adjoint operation~(\ref{eq:QNDxdagger}) is achievable by controlling the relative phase shift between the two arms of the interferometer. 

Finite squeezing reduces gate performance by adding noise~\cite{YMH+08}.
Expressions~(\ref{eq:QNDx}) and~(\ref{eq:QNDxdagger}) hold only in the infinite squeezing limit.

\subsubsection{Detectors, tomography and feedforward}
\label{subsubsec:detectors}

Homodyne detection~\cite{LR09} is central to CV quantum information processing.
A homodyne detector mixes the signal field with a local oscillator field in a coherent state;
this mixing takes place at a balanced beam splitter~\cite{CST89}.
The two output fields are directed to photodetectors,
and the difference in the two signals corresponds to a measurement in the quadrature basis in the limit that the local oscillator field strength is large.
Mathematically the quadrature measurement is $\proj{x}$ for the local oscillator in phase with the signal
or $\proj{p}$ (for~$p$ the canonical conjugate to~$x$) if the local oscillator is out-of-phase,
i.e., phase shifted by~$\pi/2$.
Measuring in a continuum of bases~$x\cos\varphi+p\sin\varphi$ is accessible by tuning the local oscillator phase~$\varphi$.

A state can be fully characterized through preparing many copies of the state and 
subjecting it to many measurements for each of a large number of randomly chosen local-oscillator phases~$\varphi$.
The gathered information,
obtained through this sampling procedure,
can be processed to estimate the state of the field.
This process is known as optical homodyne tomography.

In a realistic experiment, optical homodyne tomography can be employed by a Certifier to analyze the full process itself~\cite{LKK+08},
such as Alice's spacetime replication scheme.
Alice is supplied with many copies of a wide range of coherent-state amplitudes and phases and performs her task on each one.
For sufficiently many coherent states supplied for this process,
the Certifier can be confident in the estimate of the process provided that the process does not have significant support outside the domain of coherent states being used for this test.  This is described further in \cref{subsubsec:certifying}

\subsection{Encoding and decoding}
\label{subsec:coding}

Equivalent versions of the encoding and decoding circuits in~\cref{EncodingCircuit,IdealDecoders}
can be efficiently implemented using quantum optics.  Moreover, the optical circuits required to implement the encoding and decoding circuits are feasible using current technology.

\subsubsection{Encoding}
\label{subsubsec:encoding}

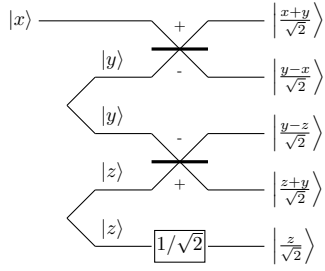
\begin{figure}[!t]
	\centering
		\begin{tikzpicture}[scale=0.75]
		\node [left](mode1start) at (0,0)  {$\ket{x}$};
		\node (mode2start) at (0,-1) {};
		\node (mode3start) at (0,-2) {};
		\node (mode4start) at (0,-3) {};
		\node (mode5start) at (0,-4) {};
		
		\draw(mode1start)--(2,0);
		\draw($(mode2start)+(0.5,-0.5)$)--(1,-1);
		\draw($(mode3start)+(0.5,0.5)$)--(1,-2);
		\draw($(mode4start)+(0.5,-0.5)$)--(1,-3);
		\draw($(mode5start)+(0.5,0.5)$)--(1,-4);
		
		\foreach \x in {1,...,4}
		{
			\draw(1,-\x)--(2,-\x);
		}
		
		\node (sq1) at (1.3,-1+0.3){$\ket{y}$};
		\node (sq2) at (1.3,-2+0.3){$\ket{y}$};
		\node (sq3) at (1.3,-3+0.3){$\ket{z}$};
		\node (sq4) at (1.3,-4+0.3){$\ket{z}$};
		
		\beamsplitterpm(2,0);
		\beamsplittermp(2,-2)
		
		\node [right](mode1end) at (4,0)  {$\Ket{\frac{x+y}{\sqrt{2}}}$};
		\node [right](mode2end) at (4,-1) {$\Ket{\frac{y-x}{\sqrt{2}}}$};
		\node [right](mode3end) at (4,-2) {$\Ket{\frac{y-z}{\sqrt{2}}}$};
		\node [right](mode4end) at (4,-3) {$\Ket{\frac{z+y}{\sqrt{2}}}$};
		\node [right](mode5end) at (4,-4) {$\Ket{\frac{z}{\sqrt{2}}}$};
		
		\foreach \x in {0,...,4}
		{
			\draw(3,-\x)--(4,-\x);
		}
		
		\node[gate] (mode5squeeze) at (2.5,-4) {$1/\sqrt{2}$};
		
	\end{tikzpicture}
\caption{%
	An optical circuit for encoding a position eigenstate state~$\ket{x}$ into five optical modes.
	The input state to be encoded enters in mode~1.
	We require two two-mode squeezed states (or four single mode squeezed states followed by a round of beam splitters).
	These squeezed states are labelled by the position quadratures $y$ and $z$, which are integration variables in the state \cref{opticalencodedstate}.
	At a beam splitter, the mode labelled with a small ``-'' picks up a minus sign when adding the ``x'' quadrature of the mode with the ``+''.  In order to match the form of our encoded state, we require an additional squeezer on mode 5 in order to get the correct factor of $\sqrt{2}$.  However, this squeezing operation can be deferred to the decoding.\label{OpticalEncodingCircuit}}
\end{figure}

Notice that the encoded state \cref{encodedstate} can be rewritten (up to normalization) as
\begin{equation}
\label{opticalencodedstate}
	\ket[\text{enc}]{x}
		=\int\text{d}y\text{d}z \Ket{\frac{x+y}{\sqrt{2}},\frac{y-x}{\sqrt{2}},\frac{y-z}{\sqrt{2}},\frac{z+y}{\sqrt{2}},\frac{z}{\sqrt{2}}}.
\end{equation}
In this form, it is easy to see that the encoded state can be prepared using a very simple optical circuit: mixing two two-mode squeezed states with the input state on a series of beam splitters, as illustrated in \cref{OpticalEncodingCircuit}. 
Note the presence of the single-mode squeezer on mode 5 to get the correct factor of $\sqrt{2}$.  This squeezer can effectively be deferred to the decoding stage whenever mode 5 is not lost and ignored whenever mode 5 is lost.

The referee starts with an arbitrary state, and Alice and her four agents start with vacuum states.
We treat the referee's arbitrary state as a coherent state as the coherent states form a basis and our protocol is linear.
The referee passes the arbitrary state to Alice.

Alice's agents have to generate two pairs of two-mode entangled states.
This generation is performed by spontaneous parametric down coversion or other mathematically equivalent means.
The squeezing parameter is~$r$.

Two pairs of Alice's agents mix modes at a balanced beam splitter~\cite{CST89}.
The agent controlling mode 5 sends his beam through a single-mode squeezer indicated by~$1/\sqrt2$
so $\text{e}^r=\sqrt2$, which implies $r=\frac{1}{2}\ln2$.

\subsubsection{Decoding}
\label{subsubsec:decoding}

\begin{figure}
\centering
\def\arraystretch{8}
\begin{tabular}{|l|r|}
\hline
\begin{subfigure}[t]{0.5\textwidth}\centering
	\centering
		\begin{tikzpicture}[scale=0.75]
		\node [left](mode1start) at (0,0) {mode 1};
		\node [left](mode2start) at (0,-1) {mode 2};
		\beamsplitterpm(1,0);
		\node [right](mode1end) at (3,0) {$\ket{\alpha}$};
		\draw(mode1start)--(1,0);
		\draw(mode2start)--(1,-1);
		\draw(2,0)--(3,0);
		\draw(2,-1)--(3,-1);
	\end{tikzpicture}	
\caption{The encoded state is trivially recovered from modes~1 and 2 by mixing the modes on a balanced beam splitter. \label{E1optical}}
\end{subfigure}
&
\begin{subfigure}[t]{0.44\textwidth}\raggedleft
	\input	145recovery
\caption{%
	The encoded state can be recovered from modes~1, 4, and 5 using the circuit above.\label{E2optical}}
\end{subfigure}
\\ \hline
\begin{subfigure}[t]{0.5\textwidth}\centering
	\centering
	\input	135recovery
\caption{The encoded state can be recovered from modes~1, 3, and 5 using the optical circuit above.\label{E3optical}}
\end{subfigure}
& 
\begin{subfigure}[t]{0.44\textwidth}\centering
	\centering
	\input	234recovery
\caption{The encoded state can be recovered from modes~2, 3, and~4 using the optical circuit above.\label{E4optical}}
\end{subfigure}
\\ \hline
\end{tabular}
\caption{\label{OpticalDecoders}Optical decoding circuits using quantum non-demolition gates.
The QND gates correspond to non-demolition gates, where the subscript indicates the required gain and a dagger represents Hermitian conjugate.
Beam splitters are represented by crossed lines, where the small $(\pm)$ indicates that the mode with the $(-)$ acquires a phase of $\pi$ on the beam splitter.  Gates labelled with a $\pi$ are phase shifters with a phase of $\pi$. Gates with numbers represent single-mode squeezing by the number in the box.
Measurements are shown as either $x$ or $p$ curved gates -- these correspond to homodyne detection in either the computational basis or its conjugate.  Classical wires correspond to feedforward of the measurement outcomes.  The feedforward is used here in conjunction with displacements, i.e. gates labelled ``Disp''.  If ``Disp'' is followed by a numeric value, the displacement is to be done using a gain of that value (i.e. the measurement outcome is multiplied by this value before being added to the displaced mode.)}
\end{figure}
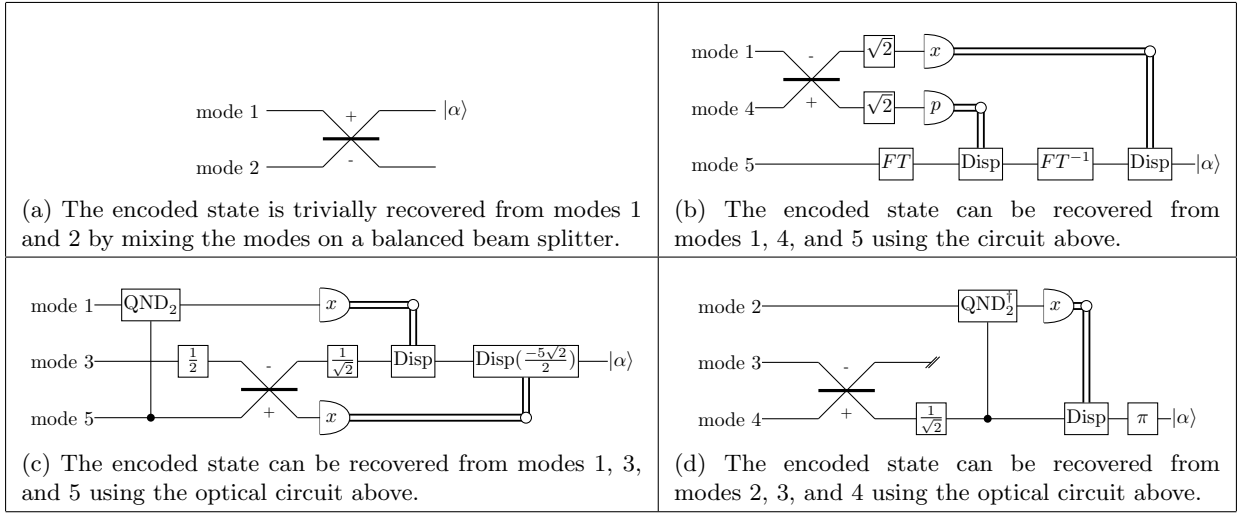

To \emph{decode} the information given access to an appropriate subset of the five modes, Alice's agents could attempt to realize the ideal decoding circuits in \cref{IdealDecoders} using the optical circuits shown in \cref{OpticalDecoders}.  Using the language of quantum error correction, having access to only a known subset of modes may be called an ``erasure error'' on the complementary set of modes; we will continue to use this terminology henceforth.

Note that, while we used homodyne detection and feedforward in these circuits, we could instead use Bloch-Messiah reduction \cite{Braunstein2005b} to rewrite the circuits using only a round of beam splitters,
followed by single-mode squeezing and another round of beam splitters.

If error $E_1$ in \cref{errormodel} occurs, modes~3, 4, and~5 are erased.
To decode the error, the players simply mix the remaining modes~1 and 2 on a balanced beam splitter, recovering the encoded coherent state~$\ket\alpha$ on mode~1.

If error $E_2$ in \cref{errormodel} occurs, modes~2 and 3 are erased.  To decode the error, the players first mix modes 1 and 4 on a balanced beamsplitter, and perform a Fourier transform on mode 5 using a phase shifter with phase equal to $\pi/2$.  Players then measure modes~1 and~4 using homodyne measurement (measuring $x$ on mode~1 and the conjugate basis on mode~4).  Players feed the classical data from mode~4 forward and perform a displacement of mode 5 by an amount fed forward from mode~4 using a gain of $1$.  Mode 5 then undergoes an inverse Fourier transform using a phase shift of $-\pi/2$, and the classical data from the measurement of mode~1 is fed forward to mode 5.  Another displacement of mode 5 by the classical data from mode~1 yields the encoded coherent state $\ket{\alpha}$ on mode 5.

If error $E_3$ in \cref{errormodel} occurs, modes~2 and~4 are erased.  To decode the error, the players first use the a QND gate with gain of 2 from mode~3 to mode~1.  Players then squeeze mode~3 using a single-mode squeezer with a squeezing amount of $r=\ln 2$.  modes~3 and 5 are then mixed on a balanced beam splitter.  Players then squeeze mode~3 by an amount $r=\ln\sqrt 2$ and perform homodyne measurement of modes~1 and 5 to obtain the classical value of the $x$ quadrature.
The classical data from mode~1 is then fed forward to mode~3 and mode~3 is displaced by this amount with a gain of $1$.  The classical data from mode 5 is then fed forward from mode 5 and mode 5 is displaced by this amount using a gain of $\frac{-5\sqrt{2}}{2}$.  The encoded coherent state $\ket{\alpha}$ is thus recovered on mode~3.

If error~$E_4$ in \cref{errormodel} occurs, modes~1 and 5 are erased.  To decode the error, the players first mix modes~3 and~4 on a balanced beam splitter.  mode~3 is then discarded as it is no longer needed.  Players then use single-mode squeezing on mode~4 by an amount of $r=\ln\sqrt 2$ and then perform the adjoint of a QND gate with gain of 2 from mode~4 to mode~2.  mode~2 is then measured with a homodyne detection to determine the value of the $x$ quadrature, which is then fed forward classically to mode~4.  Players displace mode~4 by an amount corresponding to the outcome of the homodyne measurement on mode~2 using a gain of $1$.  Players then perform a phase shift with a phase of $\pi$ on mode~4 to recover the encoded coherent state $\ket{\alpha}$ on mode~4.

Note that in all four of the above decoding schemes we can offload several of the operations to classical post-processing if tomography is to be used to verify the decoded state.  For example, in \cref{E3optical} we can push the two classically controlled displacements and the single mode squeezing of mode~3 into post-processing by performing tomography on mode~3 after the beam splitter operation and transforming the measured outcome appropriately. 

\subsection{A realistic model involving faulty gates and finite squeezing}
\label{subsec:model}

\subsubsection{An achievable proposal: certification of Alice}
\label{subsubsec:certifying}
As described in \cref{subsec:opasgame}, we distinguish between a Referee playing the two-party replication game with Alice, and a Certifier simulating gameplay with Alice to determine if she would be capable of winning a game with a Referee.  This distinction is made because implementing gameplay between Alice and a Referee in an experiment could require techniques beyond our current capabilities.  As such, we focus on the certification task, as it simulates gameplay using experimentally feasible techniques.

In the ideal game, the state provided to Alice by a Referee can be perfectly pure, and the Referee might employ a witness to test for success or failure, using a single-shot test.  In reality, an experimenter can not prepare input states that are perfectly pure, and single-shot tests will not be entirely conclusive.  As such, an experimenter can instead model the role of a Certifier who might resort to multiple shots and employ tomography~\cite{LR09} to ascertain statistically that Alice is succeeding. Furthermore, an experimenter's apparatus is imperfect,
which means the Certifier allows errors to be made, which an adversarial Alice can exploit to cheat.

If Alice is certified to play, she can still cheat as much as the Certifier's loopholes admit, but at least the certificate shows that Alice could play honestly at the certified level if she chose to do so.  Following current experimental standards, Alice is certified according to the average fidelity of her demonstrated spacetime-replication apparatus. This average fidelity depends on the chosen prior of input states~\cite{JBS02}.

Now let us consider this prior distribution of input states.  In order to test Alice's capability for playing the game, the Certifier only needs to check that Alice can transmit and reveal the overcomplete basis of coherent states.  This follows because an arbitrary CV quantum state can be represented in the \emph{diagonal} coherent state representation as $\int d^2\alpha f(\alpha)\proj{\alpha}$, where the function $f$ is possibly singular.  Note that this is not true for orthonormal bases, as this method would be insensitive to dephasing.  However, if the Certifier verifies that Alice can transmit and reveal all of the coherent states $\ket{\alpha}$, then by viewing the process as a quantum channel, we can use linearity of quantum channels to ensure that Alice could also replicate arbitrary states $\int d^2\alpha f(\alpha)\proj{\alpha}$. If the Hilbert space dimension is large, as is the case for CV quantum information, then sampling Alice's ability to transmit and reveal over a prior distribution of basis states suffices for certification. For example, the first CV quantum teleportation experiment~\cite{Furusawa1998} demonstrated a high average fidelity for a Gaussian prior of coherent-state complex amplitudes~$\{\alpha\}$.  In the next subsubsection we consider fidelity thresholds based on Alice's resource requirements to play.

To complete the certification task, a Certifier simulates gameplay with Alice and decides whether or not she would be capable of winning the idealized game with a Referee of full power.
Unlike the ideal game in which the Referee supplies Alice with a pure state and determines her success by measuring whether or not she supplied the original state at a reveal point,
the Certifier must decide Alice's success on a statistical basis.  The Certifier supplies Alice with the same state many times, which is mixed but can be arbitrarily close to being a pure state, or he can provide Alice with a share of a larger entangled state.  He does not tell Alice anything about the state except that the state has support only on the basis that Alice can manage. For example, Alice will have a maximum number of photons she can handle subject to limitations of optical components and detectors.

The certification task is realized via multi-photon active interferometry (passive optical elements plus parametric optical processes and feedforward)~\cite{LB99,Braunstein2005b} with five input ports and five output ports (hence a ``ten-port interferometer''). In a practical simulation of the replication game based on current technology, the Certifier will send either a coherent state from some prior distribution of complex amplitudes or a two-mode squeezed state with one share sent to Alice and the other reserved for recombination with Alice's output state for reconciliation purposes.

For the case that the Certifier provides a coherent state to Alice, he could perform optical homodyne tomography on the output state.  From tomography, he can compare the output state to the input state and infer the fidelity.  The Certifier then decides success if the fidelity exceeds a sufficiently high quantity.  We discuss fidelity thresholds in \cref{subsubsec:fidelity}.

In addition to asking Alice to replicate several coherent states, each instance being replicated multiple times for tomography purposes, the Certifier can also test Alice's playing ability by creating a two-mode squeezed state and sending Alice one share.  The Certifier combines the two squeezed state modes at a beam splitter to yield a product of single-mode squeezed
states at the two output ports~\cite{KS96} and then can perform state tomography on one mode to infer entanglement fidelity for Alice's replication procedure. The Certifier decides to pass Alice if she has demonstrated sufficient fidelity.

\subsubsection{Finite squeezing}
The encoding in~\cref{OpticalEncodingCircuit} uses two sets of two-mode squeezed states to prepare the encoded state in \cref{opticalencodedstate}.  Experimentally, these two-mode squeezed states can only be produced using some finite amount of squeezing.  As such, we can only perform \emph{approximate} error correction and recover the encoded state with some fidelity $\mathcal{F}\leq1$.

\begin{figure}[!htb]
\centering
	\includegraphics[width=0.6\textwidth]{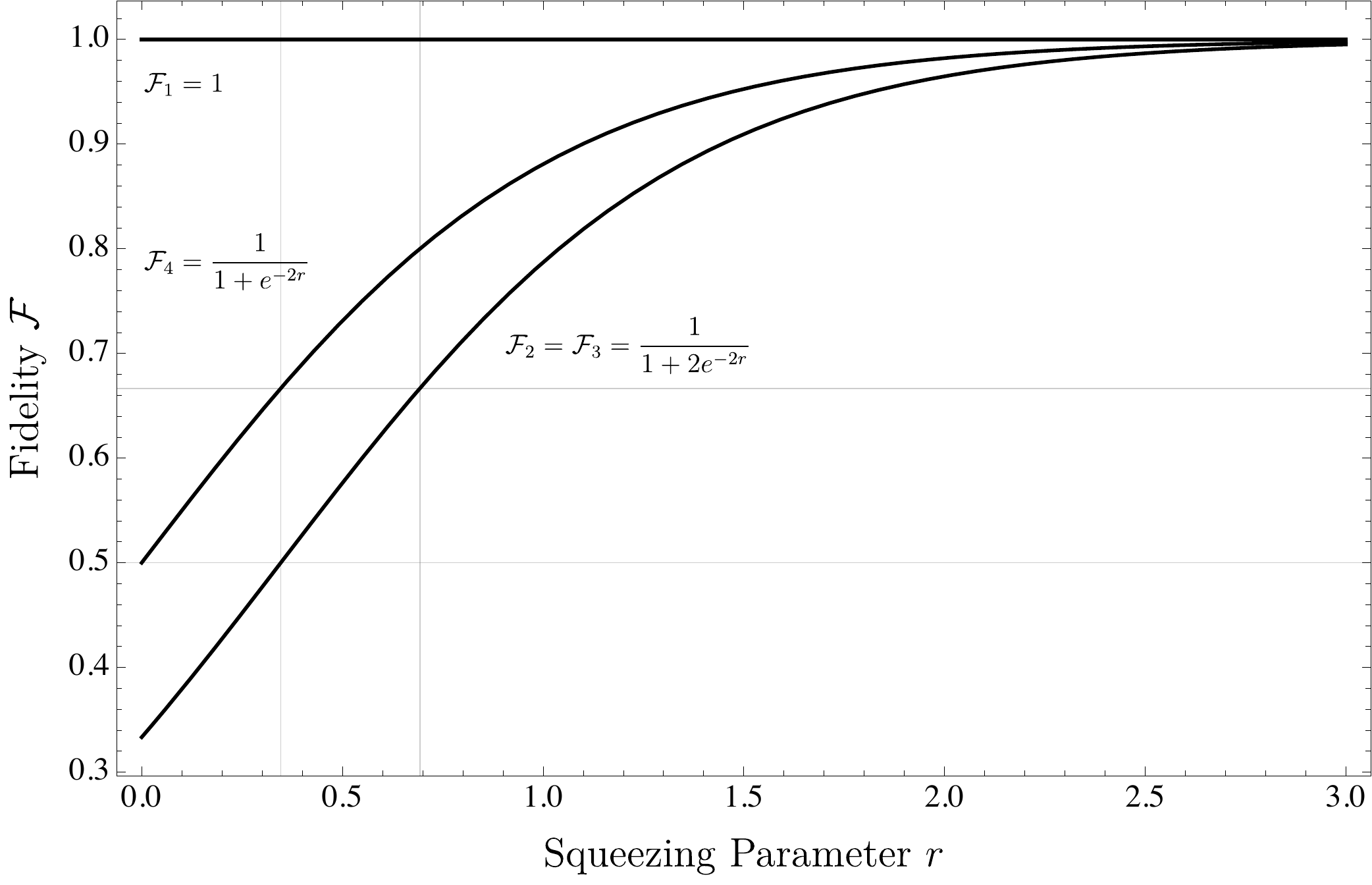}
\caption{%
	Recovery fidelity as a function of input squeezing parameter for each of the perfect, optical decoding circuits.\label{fidelityplot}%
	}
\end{figure}

Suppose we can prepare the encoded state using two-mode squeezing with a squeezing parameter, $r$.  Then, if the optical decoding circuits can be implemented without the introduction of additional noise, we find recovery fidelities of
\begin{align}
	\mathcal{F}_1(r)&=1\\
	\mathcal{F}_2(r)&=\frac{1}{1+2e^{-2r}}\\
	\mathcal{F}_3(r)&=\frac{1}{1+2e^{-2r}}\\
	\mathcal{F}_4(r)&=\frac{1}{1+e^{-2r}},
\end{align}
where the subscripts label the error from which we recover.
These fidelities are shown in \cref{fidelityplot}.

The utility of \cref{fidelityplot} is in showing the requisite amount of fidelity required 
for playing the game as discussed in Subsubsection~\ref{subsubsec:fidelity}.
Specifically sufficient squeezing should be present for the worst case of the four fidelities~$\mathcal{F}_i$
to exceed half or two-thirds or otherwise depending on the rigor of the threshold.

\subsubsection{Fidelity thresholds}
\label{subsubsec:fidelity}

The Certifier provides Alice with coherent state inputs~$\{\ket{\alpha}\}$
according to some prior distribution~$P(\alpha)$.
If Alice's apparatus works perfectly, she returns the supplied state~$\ket{\alpha}$,
which the Certifier verifies.
In practice this certification would be accomplished via state tomography
using optical homodyne detection~\cite{LR09}.
Alternatively, full CV quantum process tomography could be used~\cite{LKK+08}.

After multiple shots and allowing for many local oscillator phases,
average fidelity~$\bar{F}$ of the process is ascertained.
The Certifier could simply issue a certificate stating the fidelity achieved.
On the other hand the Certifier could issue certificates whenever Alice demonstrates that she has surpassed a landmark fidelity indicating that she is employing crucial resources for the game.

The lowest average-fidelity bar is $\bar{F}=1/2$,
which corresponds to Alice using squeezing in her experiment rather than employing phase-correlated pair coherent states in two modes as a cheap substitute for two-mode squeezed states.
The $\bar{F}=1/2$ threshold was employed in demonstrating the first instance of experimental CV quantum teleportation in the sense that the teleportation protocol employs genuine squeezing \cite{Furusawa1998,Braunstein2000,BFKvL01}.

A higher bar is established by requiring that Alice cannot succeed at replicating quantum information in an impossible configuration of causal diamonds.  For example, if the fidelity bar is set below $2/3$, then Alice could succeed at replicating information in the configuration shown in panel (c) of \cref{replicationexamples} using an optimal $1\rightarrow 2$ cloning technique \cite{GG01,Cerf2000b,Cerf2000a,BFKvL01}.  If the fidelity bar is set even lower, Alice could succeed at replicating information in configurations that require more than one violation of the no cloning theorem.  As such, we require a minimum fidelity of $2/3$.
Note that the same fidelity threshold can be argued for if we require that Alice be ``oblivious'' to the input state.  In other words, if Alice is prohibited from retaining an approximate copy of the input state that is closer to the input than the state she presents for verification, the optimal cloning argument places a restriction on the fidelity threshold of $2/3$.  This argument was also used in the context of CV quantum teleportation by \citet{GG01}. 


As average gate fidelity is a poor indicator of Alice's capability for teleporting entanglement~\cite{JBS02},
a different certification process is required if Alice is to be certified to play an information replication game
that requires Alice to deliver a share of a larger entangled state to the reveal point.
In this case the Certifier provides one share of a bipartite entangled state such as 
a two-mode squeezed state to Alice.  Alice performs her replication protocol, and the Certifier could check, for example, if the returned bipartite state can be used to violate a Bell inequality.

\section{Conclusions and outlook}\label{ConclusionsSection}
The information replication theorem of \citet{Hayden2012} states that the \emph{only} conditions on the passage of quantum information through spacetime are no-cloning and no-signalling.  This theorem is both simple and powerful, as it indicates that surprising collections of causal diamonds can simultaneously hold the same quantum information, provided the causal structure of the problem satisfies a natural set of consistency conditions.  The theorem applies broadly to quantum information in all forms, regardless of the physical system in which the information is encoded.  In particular, we have shown that the theorem applies to continuous-variable quantum information.  To realize information replication for allowed configurations of causal diamonds, one applies the principles of quantum error correction as it allows for the delocalization of information in space.

To succeed at information replication for continuous variable systems, we developed a new class of CV stabilizer codes using a novel approach to code construction based on ideas from simplicial homology.  Our codes are continous-variable CSS codes that encode one bosonic mode into many. The codes are capable of replicating quantum information for any $N>3$ spacetime regions, since they correct against errors corresponding to pure loss of a known subset of modes.  The error model is specific to the information replication problem, and the codes we construct are unusual as we can recover an encoded state using a vanishing fraction of the shares of the code.  In particular, if we have $n=\binom{N}{2}$ modes, we recover using only $2/N$ shares (i.e., $O(1/\sqrt{n})$ of the physical modes).  The formalism we developed applies generally to any allowed replication problem.

We also designed an independent code (using an \emph{ad hoc} construction on five modes) for a particularly interesting configuration of four spacetime regions.  The configuration represents the simplest non-trivial example of information replication that does not reduce to previously demonstrated error correcting codes.  We described the code in terms of quantum optical circuits, taking great care to ensure that both the encoding and decoding circuits are experimentally feasible using existing optical technology.  We also outlined a proposal for a quantum optics experiment to demonstrate the replication of quantum information in spacetime. We hope that our five mode proposal will be realized in an experimental quantum optics laboratory.  

In addition to the possibility of future experimental work, many theoretical questions remain.  For instance, it would be interesting to completely characterize the extent to which redundancy in the graph of causal structure can reduce the resources required for information replication.  We used redundancy to bring the general six mode code down to our \emph{ad hoc} five mode code, but we did not explore these redundancies further.  Similarly, one could try to characterize the most efficient \emph{codes} that arise whenever such a redundancy is present.   Another interesting question concerns the development of encoding and decoding circuits that implement our general code.  We presented encoding and decoding circuits for our five mode code, but not for the general codes, and it is conceivable that an algorithm for determining these circuits exists.  If such an algorithm exists, it is not known clear whether or not it would be efficient.
\clearpage
\section*{Acknowledgments}
We thank Emily Davis, Akira Furusawa, Alex Lvovsky, Nicolas Menicucci, Fernando Pastawski, Michael Walter, and Beni Yoshida for helpful discussions. This research is supported by AFOSR (FA9550-16-1-0082), CIFAR, FQXi, and the Simons Foundation. SN is supported by a Stanford Graduate Fellowship, and GS is supported by an NSERC postgraduate scholarship and the AFOSR funding. BCS acknowledges financial support from AITF and NSERC. 

\appendix
\section{Appendix}
\subsection{Proof of error correcting property of the general code}\label{ErrorCorrectionProof}
Let ${\mathcal{C}}_v = \left \langle \vec{v}_{23},\vec{v}_{34}, \dots ,\vec{v}_{N-1,N}\right \rangle$ and ${\cal{C}_A}=\left \langle \vec{A}_1, \vec{A}_2, \dots ,\vec{A}_N \right \rangle$ be the linear subspaces spanned by $\{\vec{v}_{ij}\}$ and $\{\vec{A}_k\}$. Before we consider the error correcting properties of our general code, note the following simple facts:

\textit{Fact 1.} \label{fact:fact1} All triangles $\vec{T}_{ijk}=\vec{e}_{ij}+\vec{e}_{jk}+\vec{e}_{ki}$ are in ${\cal{C}}_v$.

\textit{Fact 2.} \label{fact:fact2} From \textit{Fact 1} it can be seen that ${\cal{C}}_v$ contains all directed loops $\vec{e}_{i_1i_2}+\vec{e}_{i_2i_3}+...+\vec{e}_{i_{k-1}i_{k}}+\vec{e}_{i_{k}i_{1}}$.

\textit{Fact 3.} \label{fact:fact3} The space orthogonal to ${\cal{C}}_A$ is a superposition of directed loops. Therefore, from \textit{Fact 2}, ${\cal{C}}_{v} \oplus {\cal{C}_A}$ is equal to the space of all~$\binom{N}{2}$-dimensional real vectors and ${\cal{C}}_v$ is orthogonal to ${\cal{C}_A}$.

\textit{Fact 4.} \label{fact:fact4} $\sum_{i=1}^{N}{\vec{A}_i}=0$.

Suppose we want to correct the error
\begin{equation}
	\EP=E_{i}^{\dagger} E_j = e^{i \theta} e^{i\vec{\EP}_{X}\cdot\vec{X}} e^{i \vec{\EP}_{P}\cdot\vec{P}},
\end{equation}
where~$\vec{\EP}_{X}$ and~$\vec{\EP}_{P}$ are real vectors describing the error $\EP$, and $\theta$ is some phase.  In our error model, we know that for a given vertex~$r$ on the graph of causal structure there are no errors on edges adjacent to~$r$.  That is,
\begin{equation}\label{orthcon2}
	\vec{e}_{rj}.\vec{\EP}_{X} = \vec{e}_{rj}.\vec{\EP}_{P}=0, \quad\forall j\neq r. 
\end{equation}
We can now prove our error correction theorem.

\emph{Proof of Theorem 1: } Suppose that the causal diamond in which we want to reconstruct the state corresponds to vertex~$r$. To show that the stabilizer can correct errors, it is sufficient to show that if $\EP$ commutes with all of the stabilizer generators then it is in the stabilizer group (up to an unimportant global phase). However it can be easily checked that
\begin{equation}
[g_{jk},\EP]=0 \iff \vec{v}_{jk}\cdot\vec{\EP}_P=0
\end{equation}
and
\begin{equation}[h_k,\EP]=0 \iff \vec{w}_k\cdot\vec{\EP}_X=0.
\end{equation}
Therefore, it is sufficient to prove the following statements:
\begin{enumerate}
\item If for all $2 \leq j<k\leq N$ , $\vec{v}_{jk}\cdot\vec{\EP}_{P}=0$ then $\vec{\EP}_P$ is a linear combination of $\{\vec{w}_l\}$. (Hence $e^{i \vec{\EP}_P\cdot\vec{P}} \in S$.)
\item If for all $2 \leq l \leq N-1$ , $\vec{w}_{l}\cdot\vec{\EP}_{X}=0$ then $\vec{\EP}_X$ is a linear combination of $\{\vec{v}_{jk}\}$. (Hence $e^{i \vec{\EP}_X\cdot\vec{X}} \in S$.)
\end{enumerate}
For condition 1, using \textit{Fact 1} we have
\begin{equation}\label{TdotE}
\vec{T}_{rij}\cdot\vec{\EP}_{P}=\vec{e}_{ri}\cdot\vec{\EP}_{P}+\vec{e}_{ij}\cdot\vec{\EP}_{P}+\vec{e}_{jr}\cdot\vec{\EP}_{P}=0.
\end{equation}
Using \cref{orthcon2},  \cref{TdotE} is equivalent to $\vec{e}_{ij}\cdot \vec{\EP}_{P}=0$.  Thus, $\vec{\EP}_{P} =0$ and condition 1 is trivially satisfied.

For condition 2, first let $\alpha= \vec{A}_1\cdot\vec{\EP}_X$ ($\alpha\in\RR$). Now, as $\vec{w}_j\cdot\vec{\EP}_X=0$ (for $2\leq j \leq N-1$) we have
\begin{equation}
(\vec{A}_1 +\vec{A}_j)\cdot\vec{\EP}_X=0\implies \vec{A}_j\cdot\vec{\EP}_{X}=-\alpha.
\end{equation}
Using \textit{Fact 4}, 
\begin{equation}
\sum_{j=1}^{N}{\vec{A}_i\cdot\vec{\EP}_X}=0.
\end{equation}
so that $\alpha - (N-2)\alpha + \vec{A}_N\cdot\vec{\EP}_X=0$, or
\begin{equation}
\vec{A}_N\cdot\vec{\EP}_X=\alpha(N-3).
\end{equation}
Now, if $N\geq 4$,
\begin{equation}
\vec{A}_j\cdot \vec{\EP}_X\propto\alpha
\end{equation}
for all $j$ including $j=r$. But we know from the orthogonality relation \cref{orthcon2} that $\vec{A}_r\cdot\vec{\EP}_X=0$.  This fixes $\alpha=0$ so that $\vec{A}_j\cdot\vec{\EP}_X=0$ for all $j$.  From fact 3 we can then conclude that $\vec{\EP}_X \in {\cal{C}}_v$ and $\vec{\EP}_X$ can be written as a linear combination of $\{\vec{v}_{jk}\}$.
$\blacksquare$
%
\subsection{Homological construction of the error correcting code}\label{Homology}
In this subsection we describe a general method for constructing quantum error correcting codes using simplicial homology.  Instead of defining simplicial homology in full generality, we will illustrate it with an example that is of particular importance for the error correction scheme used in our spacetime replication procedure.  Existing homological code construction techniques provide codes that protect a number of modes/qudits equal to the dimension of the first homology group of some underlying manifold~\cite{bombin2007homological,bravyi2014homological}.  In our case, the first homology group will turn out to be trivial, and existing techniques define encoded \emph{states} (rather than subspaces).  Our technique is similar to, but differs from, conventional methods, and it provides us with an encoded subspace that protects against suitably defined errors.  In this subsection, we first provide a lightning overview of the necessary ingredients from simplicial homology (motivated by our example), and then we construct a family of homological quantum error correcting codes.  We conclude this subsection with a brief discussion of the similarities and differences between our technique and existing methods.

\begin{figure}[!htb]
\centering
\includegraphics[width=0.6\textwidth]{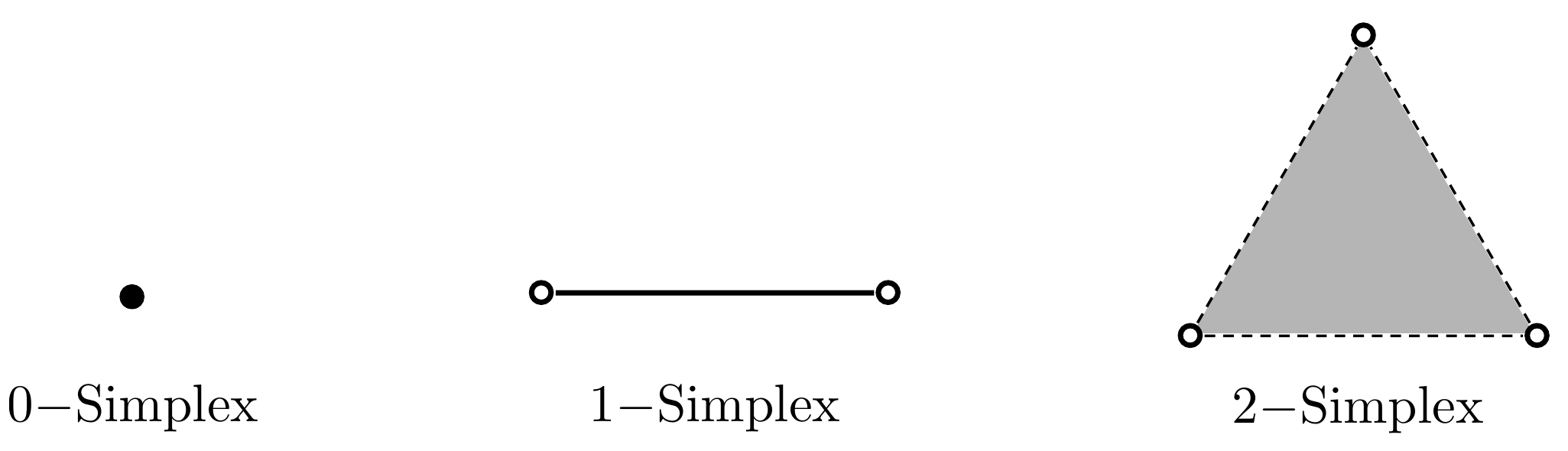}
\caption{Low-dimensional simplices.\label{fig:012sim}}
\end{figure}

Consider an ($N-1$)-dimensional tetrahedron. This tetrahedron has~$N$ vertices~$\{e_1,e_2,\dots,e_N\}$,~$\binom{N}{2}$ edges $\{e_{jk}\,|\,1\leq j<k\leq N\}$, $\binom{N}{3}$ triangles $\{e_{j,k,l}\,|\,1\leq j < k< l\leq N\}$ and generically $\binom{N}{k}$ $(k-1)$-dimensional sub-tetrahedra ($e_{i_1,\dots,i_k}\,|\,1\leq i_1<i_2<\dots<i_k\leq N$). We will henceforth use the term k-simplex to denote a $k$-dimensional tetrahedron. For instance, vertices are $0$-dimensional simplices, edges are $1$-dimensional simplices and triangles are $2$-dimensional simplices, etc. See \cref{fig:012sim} for graphical examples.

We proceed by introducing a vector space $C_k$ called a $k-$chain. $C_k$ is defined to be the vector space containing formal linear combinations of all $k$-dimensional simplices. Specifically, $C_k$ is the span of all of the basis vectors $\hat e_{i_1,\dots,i_k}$.  Elements of $C_k$ have the form of $c_k=\sum{\alpha_{i_1,\dots,i_k}\hat e_{i_1,\dots,i_k}}$, where the sum runs over $1\leq i_1<i_2<\dots<i_k\leq N$, and the coefficience $\alpha_{i_j}$ are real numbers.


In the quantum information replication picture described in this manuscript, we are mainly interested in the first three $C_k$'s , namely $C_0$, $C_1$ and $C_2$. It will be soon clear that it is helpful to think of the $(-1)	$-chain $C_{-1}$ to be the space of real numbers, a $1$-dimensional vector space.

\begin{figure}[!htb]
\centering
\includegraphics[width=0.8\textwidth]{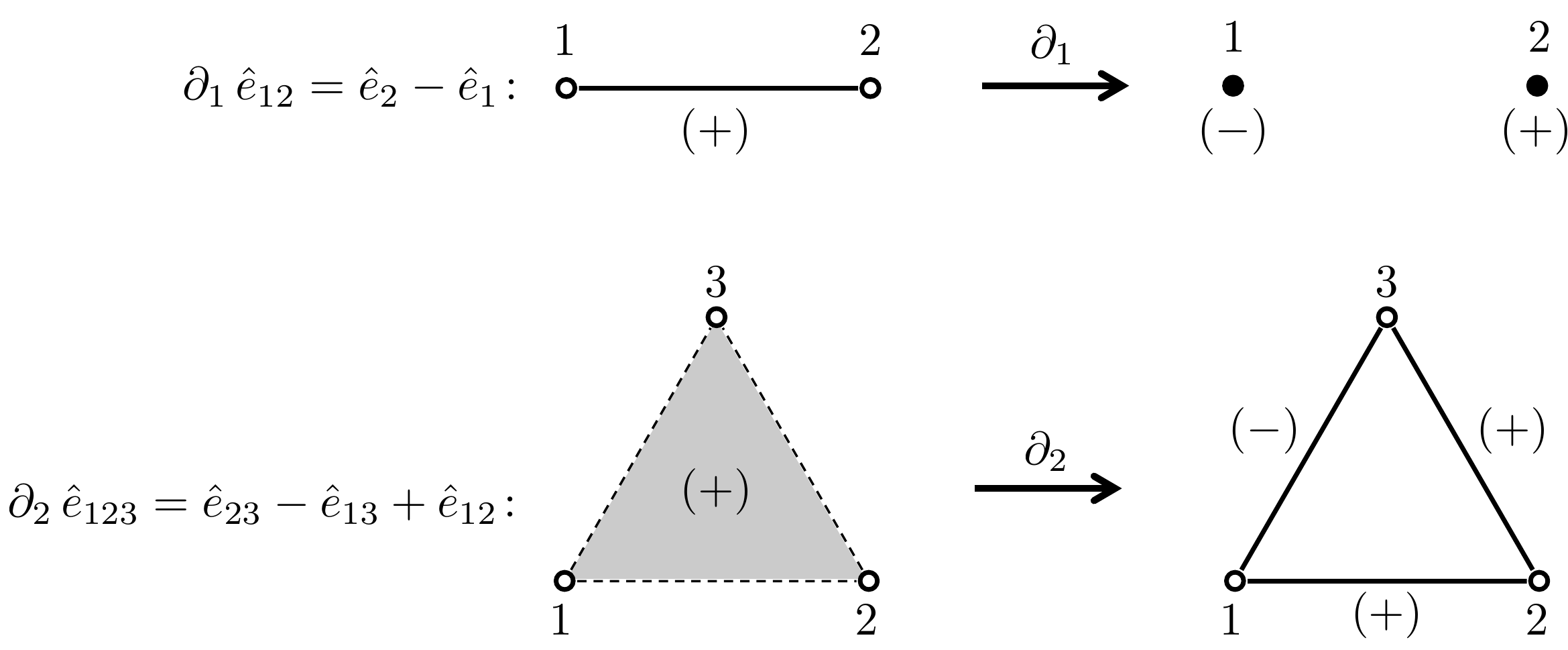}
\caption{Examples of the action of the boundary operator on low-dimensional simplices.\label{fig:lowdimboundary}}
\end{figure}

The next element we introduce is the boundary operator $\partial_k \,\colon\, C_k\to {C_{k-1}}$. The boundary operator maps elements of a $k$-chain to elements of a $k-1$-chain. The action of the boundary operator on a basis element is defined as follows:
\begin{equation}
\partial_k \, \hat e_{i_1,\dots\dots,i_k} = \sum_{m=1}^{k} { (-1)^{m+1} \,\hat e_{i_1,\dots,i_{m-1},i_{m+1},\dots,i_k}}\quad \textrm{where} \quad 1\leq i_1<i_2<\dots<i_k\leq N.
\end{equation}
The index notation in the above equation simply means that we omit the vertex labelled by index $i_m$.   \Cref{fig:lowdimboundary} shows the action of the boundary operator on low-dimensional simplices.  There is also an operator $\partial_{k+1}^T: C_k\rightarrow C_{k+1}$, which is given by the transpose of the associated boundary operator. It is known as the ``boundary transpose'' or ``coboundary'' operator.

Using the above definition, one can prove the following important property for the boundary operator:
\begin{equation}
\partial_k\partial_{k+1}=0.
\end{equation} 
It is important to note that $\partial_0$ sends $C_0$ to $\RR$ (which we call $C_{-1}$) and that its action is to simply add the~$N$ coordinates of the element of $C_0$ in its standard basis:
\begin{equation}
\partial_0 \left(\sum_{m=1}^N {\alpha_m \hat e_m} \right) = \sum_{m=1}^{N}{\alpha_m} \;\;\in C_{-1}\quad \textrm{where }\; \alpha_m \in \RR.
\end{equation}

We can use the topological properties of the $N-1$-dimensional simplex to gain intuition about the structure of $C_{k}$. For instance, if an element of $C_k$ ($0 \leq k \leq N-2$) has null boundary, then it is the boundary of an element of $C_{k+1}$.  In other words,
\begin{equation}
\textrm{If}\,\, \partial_k x=0\,\, \textrm{then}\,\,  x=\partial_{k+1} y \,\, \textrm{for some}\,\, y \in C_{k+1}.
\end{equation}
It is possible to decompose each $C_k$ ($0\leq k \leq N-2$) into the direct sum of two vector spaces. Consider the subspace $R_k=\text{Im}\,\partial_k^T$ and $L_k=(R_k)^{\bot}$. Obviously, $C_k=R_k \oplus L_k$.

\emph{Claim:} $L_k =\text{Im}\,\partial_{k+1}$

\emph{Proof:} We prove the statement in two steps:
\newline
\emph{1.} $L_k \subseteq \text{Im}\,\partial_{k+1}$.

Suppose that $x\in L_k= R_k^\bot$. Then for any $z\in C_{k-1}$ we have that $x\cdot(\partial_k^T z)=0$ or equivalently $ (\partial_k x)\cdot z=0$. This means that $\partial_k x=0$. Therefore, we conclude that $x=\partial_{k+1} y$ for some $y\in C_{k+1}$, and thus $x \in \text{Im}\,\partial_{k+1}$.
\newline
\emph{2.} $\text{Im}\,\partial_{k+1} \subseteq L_k$.

\begin{figure}[!htb]
\centering
\includegraphics[width=0.6\textwidth]{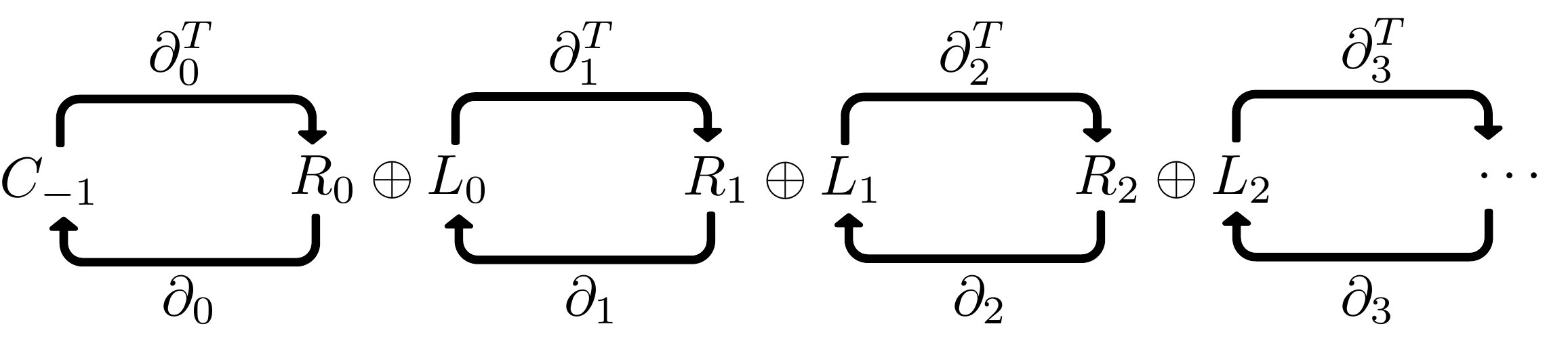}
\caption{The boundary operator and its action on $C_k=R_k \oplus L_k$. Note that when $\partial_k$ ($\partial_{k+1}^T$) acts on $L_k$ ($R_k$), it sends all elements in $L_k$ ($R_k$) to $\vec 0$. It is easy to see that $\partial_k$ ($\partial_{k+1}^T$) is a one-to-one function on $R_k$ ($L_k$). It can also be proved that $\dim(R_k)= \binom{N-1}{k}$ and $\dim(L_k)=\binom{N-1}{k+1}$.\label{fig:ChainDecomposed}}
\end{figure}

Suppose that $x\in \text{Im}\,\partial_{k+1}$. Then there exists a $y \in C_{k+1}$ such that $x= \partial_{k+1} y$. In order to show that $x\in L_k=(R_k)^\bot$, we need to show that~$X$ is orthogonal to all of the elements of $R_k$. Consider an arbitrary element of $R_k$, such as $\partial_k^T z \in R_k$. Then we have that $x\cdot (\partial_k^T z)= (\partial_{k+1} y)\cdot (\partial_k^T z)= (\partial_k \partial_{k+1} y)\cdot z=0$, which is the desired result.  See \cref{fig:ChainDecomposed} for a diagrammatic description of the structure of $C_k=R_k \oplus L_k$.
$\blacksquare$

Returning to the spacetime replication picture, consider a scenario with~$N$ causal diamonds. We identify each causal diamond with one vertex of an $(N-1)$-simplex. Here, $C_0$ is the space of causal diamonds and $C_1$ is the space of links between the diamonds (i.e., the space of CV modes, as one mode lives on each edge).

Our goal is to design a CSS code. As such, we would like two subspaces of $C_1$ labeled by $C_X$ and $C_P$ that we will use to define the~$X$ part and~$P$ part of the stabilizer code. The basic condition that all stabilizer elements commute can be translated to the condition that any element of $C_X$ is orthogonal to any element of $C_P$:
\begin{equation}\label{eq:homortho}
C_X \,\bot\, C_P.
\end{equation}
This fact, along with $R_k\, \bot\, L_k$, suggests the following construction for $C_X$ and $C_P$:
\begin{equation}\label{eq:homcx}
C_X =\partial_2 C_2 = L_1,
\end{equation}
and
\begin{equation}\label{eq:homcp}
C_P=\partial_1^T Q \subseteq R_1 \quad \textrm{for some subspace}\,\, Q \,\,\textrm{of}\,\, C_0.
\end{equation}

With this choice of construction, \cref{eq:homortho} is automatically satisfied. Our job is then to find appropriate $Q$ subspaces satisfying the desired error correction conditions, and encoding one CV mode.
Note that in order to encode one mode, we need $\dim(C_X)+\dim(C_P) = \binom{N}{2}-1$. Therefore, $\dim (C_P) = \dim (R_1)-1=N-2$.

The generator matrix for the stabilizer group is then of the following form:
\begin{center}
\includegraphics[width=0.4\textwidth]{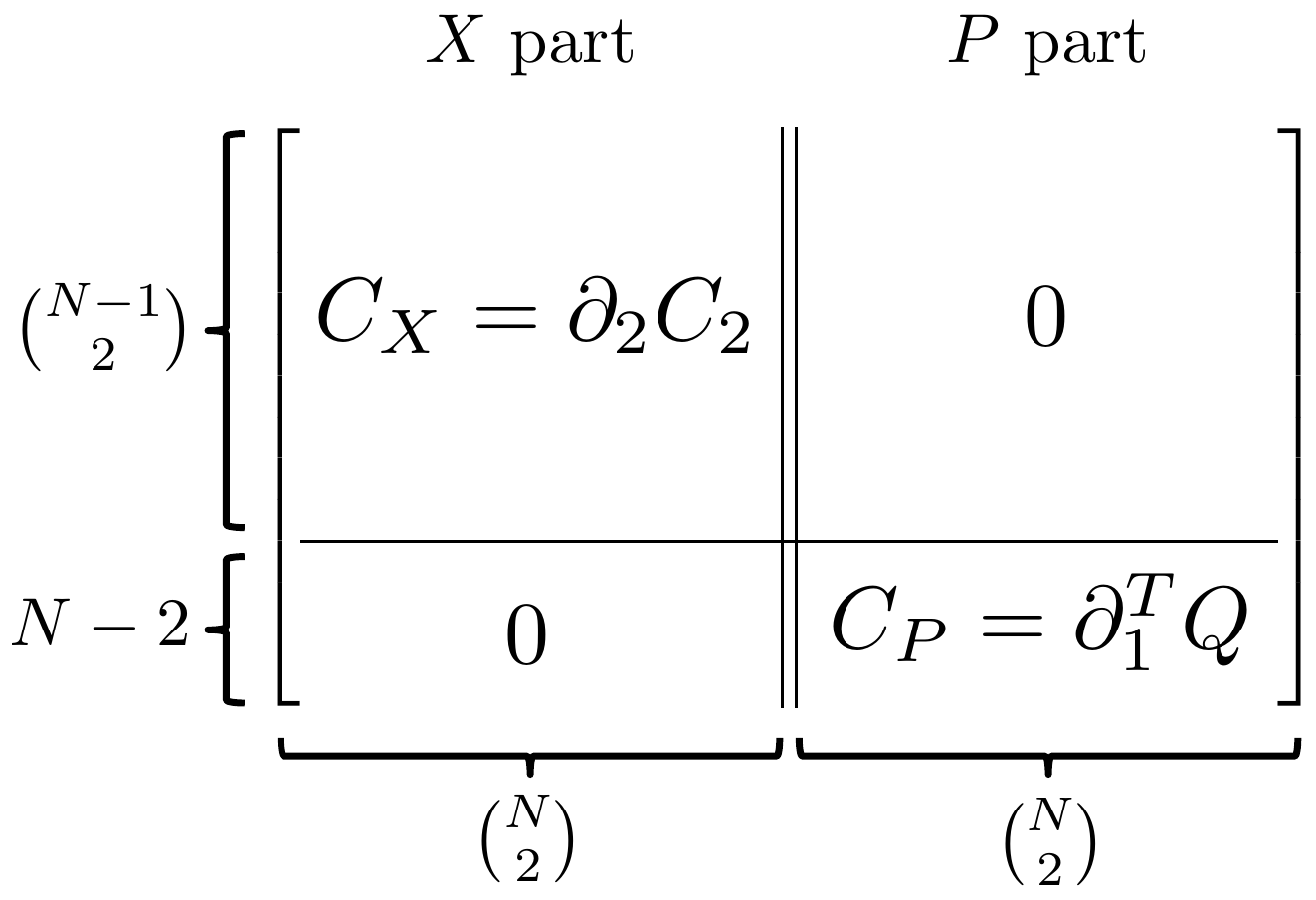}
\end{center}

Just as we did in \cref{ErrorCorrectionProof}, we define the error operators as
\begin{equation}
\EP=E_{i}^{\dagger} E_j = e^{i \theta} e^{i\vec{\EP}_{X}\cdot\vec{X}} e^{i \vec{\EP}_{P}\cdot\vec{P}}.
\end{equation}
The familiar error correction condition for stabilizer codes can be stated in plain language as follows.  If (and only if) an error $\EP$ commutes with all of the stabilizer generators then must be an element of stabilizer group to be correctable. Therefore, in order to verify the error correcting properties of the code it suffices to prove the following two conditions:
\begin{align}
\EP_P\cdot C_X=0 &\Rightarrow \EP_P \in C_P\label{eq:homcon1}\\
\EP_X\cdot C_P=0 &\Rightarrow \EP_X \in C_X. \label{eq:homcon2}
\end{align}
Note that $\EP_P\cdot C_X=0$ is used as a short hand notation for $\forall c_X \in C_X \,:\, \EP_P\cdot c_X=0$.

To prove \cref{eq:homcon1}, we will actually prove the stronger statement that $\EP_P\cdot C_X=0 \Rightarrow \EP_P=0$. Suppose that $\EP_P\cdot C_X=0$.
As $C_X=L_1$, it follows that $\EP_P \in R_1$.
Therefore, $\EP_P=\partial_1^T \alpha$ for some $\alpha \in C_1$.

Now suppose that the causal diamond in which we wish to replication quantum information is diamond $j$. This means there cannot be any errors on edges adjacent to the vertex $j$. In other words, for any index $k$ not equal to $j$, $\hat e_{jk}\cdot(\partial_1^T \alpha)=0$ or equivalently $(\partial \hat e_{jk})\cdot\alpha=0$ or $(\hat e_j - \hat e_k)\cdot\alpha=0$. If $\alpha=\sum_l{\alpha_l \hat e_l}$, then this means that for all $l$, $\alpha_j=\alpha_l$. Therefore, all of the $\alpha_j$ are equal. By the definition of $\partial_0^T$ and $C_{-1}$ we have that $\alpha \in \partial_0^T C_{-1}$. Thus $\EP_P=\partial_1^T \alpha \in \partial_1^T \partial_0^T C_{-1}=(\partial_0 \partial_1)^T C_{-1} =0$ as claimed.

For \cref{eq:homcon2}, consider a set of~$N$-dimensional orthonormal vectors $\vec{q}_1,\dots,\vec{q}_{N-2}$, which constitute a basis for $Q$. If $\EP_X\cdot C_P=0$, we have $\EP_X\cdot(\partial_1^T \vec q_j)= ( \partial_1 \EP_X)\cdot\vec q_j=0$ where $1\leq j \leq N-2$. Also, we can define an element $\vec q_0$ as $\vec q_0 =\sum{\hat e_i}$. Note that $\partial_0^T$ has a column vector representation which is in fact equal to $\vec q_0$.  Thus, we can conlude that $\vec q_0$ is orthogonal to $\partial_1\EP_X$.   Therefore, we have:
\begin{equation}
(\partial_1 \EP_X)\cdot \vec{q}_m=0 \quad \forall 0\leq m \leq N-2
\end{equation}

Notice once more that if the diamond in which we wish to replicate the information is diamond $j$, then there are no errors on the edges adjacent to vertex $j$. Hence, $\partial_1 \EP_X$ is zero on its $j$th coordinate. If we put all of the vectors $\vec q_m$ in a $\left(\binom{N}{2}-1\right)\times \binom{N}{2}$ matrix $M$ and define the vector $\vec{v}=\sum{v_j \hat e_j}=\partial_1 \EP_X$, we obtain the matrix equation
\begin{equation}
	M\cdot\vec{v}=
\left[ \begin{array}{c|c|c|c|c}
    \vec{q}_0 & \vec{q}_{1} &\vec{q}_{2} & \cdots & \vec{q}_{N-2}
\end{array}\right ]^T\cdot
\left[ 
\begin{array}{c}
        v_1\\
        v_2\\
        \vdots\\
        v_n
\end{array}\right ]=0 \quad \textrm{where } \; v_j=0.
\end{equation}
We would like to conclude that $\vec{v}=0$. If one defines the square matrix $M_j$ to be the matrix derived from M by removing its $j$th column, we know that
\begin{equation}
M\cdot \vec v=M_j\cdot \left[\begin{array}{c} v_1\\\vdots\\v_{j-1}\\v_{j+1}\\\vdots\\v_n\end{array}\right]=0.
\end{equation}
Then in order for~$\vec{v}$ to be $\vec 0$, it is sufficient that $M_j$ be invertible.

However, as we do not know a priori in which diamond we wish to reproduce the state (i.e., which vertex is the recovery vertex), we require that all of the minors $M_j$ be invertible. Thus, we should choose $\vec{q}_1,\dots,\vec q_{N-2}$ in such a way that all of the $M_j$ are invertible (remember that $\vec{q}_0$ is fixed to be the vector of all ones). This property is generically correct and one specific example of such a matrix is the following ``butterfly'' matrix $M_B$
\begin{equation}
M_B=
\begin{bmatrix}
    1&1&1&\dots&1&1&1 \\ 
    1&1&0&\dots&0&0&0 \\ 
    1&0&1&\dots&0&0&0 \\
    \vdots&\vdots&\vdots&\ddots&\vdots&\vdots&\vdots\\ 
    1&0&0&\dots&1&0&0 \\ 
    1&0&0&\dots&0&1&0 \\ 
\end{bmatrix}.
\end{equation}
We will conclude by showing that this specific choice of $M_B$ (or equivalently the choice of $Q$) leads to the same general code constructed in \cref{CVSection}.

First consider the~$X$ part of the stabilizer group described in \cref{CVSection}, which is generated by the vectors $\vec v_{jk}$. It was noted in \cref{ErrorCorrectionProof} (\emph{Fact 1}) that all triangles~$\vec T_{ijk}$ are also in the~$X$ part of the stabilizer group. However, the triangle vector $\vec T_{ijk}$ is equal to $\partial_2 \hat e_{ijk}$ in the language of homology. Therefore, the~$X$ part of the two constructions (homological and graph theoretic) coincide.

For the~$P$ part, we first need to study $\partial_1^T \hat e_j$. We can write it in the standard basis of $C_2$:
\begin{equation}
	\partial_1^T \hat e_j= \sum {r_{kl}^j \hat e_{kl}} \quad \textrm{where} \quad 1\leq k < l\leq N.
\end{equation}
Thus the inner product of $\partial_1^T \hat e_j$ and $\hat e_{kl}$ can be calculated in order to find the coefficients:
\begin{equation}
	r_{kl}^j=\hat e_{kl} \cdot (\partial_1^T \hat e_j) = (\partial_1 \hat e_{kl})\cdot \hat e_j= (\hat e_l - \hat e_k)\cdot\hat e_j=\delta_{jl}-\delta_{jk}\quad \Rightarrow\quad \partial_1^T \hat e_j=\sum_{j=1}^{j-1}{\hat e_{kj}} - \sum_{k=j+1}^{N}{\hat e_{jk}}.
\end{equation}
One can easily see that $\partial_1^T \hat e_j$ is the same as the star vector~$\vec A_j$ defined in \cref{CVSection} and that $\partial_1^T \vec q_j= \partial_1^T(\hat e_1+\hat e_j)=\vec A_1+\vec A_j=\vec w_j$. Then, as $\{\partial_1^T \vec q_j\}$ is the set of generators of the~$P$ part of the stabilizer code in our homological construction and $\{ \vec w_j\}$ is the set of~$P$-type generators in the graph theoretic construction in \cref{CVSection}, we conclude that the~$P$ parts also coincide.  Therefore, $M_B$ produces the code described in the \cref{CVSection}. 

In conventional homological code constructions~\cite{bombin2007homological,bravyi2014homological}, the number of encoded modes (or qudits) is equal to the dimension of the first homology group of the underlying manifold.  For the example described in this subsection, the first homology group of our simplicial complex is trivial, so conventional methods provide only a stabilized state.  To circumvent this problem and arrive at a fully-fledged quantum error correcting \emph{code}, we needed to deviate from conventional methods.  To see where our trajectory took a turn, note that existing homological techniques prepare a CSS code for which the $X$ and $P$ type stabilizer generators are images of the boundary and the coboundary operators acting on their full domain. More precisely, if $C_X$ and $C_P$ are the $X$ and $P$ parts of the stabilizer group, existing techniques define $C_X= \partial_2 C_2$ and $C_P=\partial_1^T C_0$. In our construction, we see from \cref{eq:homcp} that, unlike conventional methods, we do not use the image of the entire $0$-chain $C_0$ to define $C_P$, but rather we use a subspace of $Q\subset C_0$.  In doing so, we prepare one fewer generators than conventional methods, and we are left with a stabilizer code.   Schematically, one could arrive at our code using conventional methods by first constructing (one too many) stabilizer generators, taking appropriate linear combinations of the generators, and then throwing one away.
%
\subsection{Derivation of ideal and optical decoding circuits}\label{IdealToOptical}
In this appendix we first show how one can obtain ideal encoding and decoding circuits in \cref{EncodingCircuit,IdealDecoders}, and then we derive the optical encoding and decoding circuits in \cref{OpticalEncodingCircuit,OpticalDecoders} from the ideal circuits.

Using the stabilizer generators in \cref{fivemodegeneratormatrix}, we found that a basis for our code space is given in \cref{encodedstate}.  The spacetime information replication task involves reconstructing the full input state from a known subset of the encoded shares.  An arbitrary input state can be written as
\begin{equation}
\ket{\psi}=\int \psi(x) \ket{x}dx,
\end{equation}
where $\psi$ is some wave function.  For our purposes $\psi(x)$ will be a Gaussian, although we will not need this fact in the following derivations.  Assuming perfect encoding (i.e., infinite squeezing), the encoded state is given by
\begin{equation}
\ket[\text{enc}]{\psi}=\int \psi(x) \ket{x+y,y-x,y-z,z+y,z}dxdydz.
\end{equation}
Thus, the density matrix is given by
\begin{equation}
\rho_\text{enc}=\int \psi(x)\psi^*(x') \ketbra[1]{x+y}{x'+y'}\ox\ketbra[2]{y-x}{y'-x'}\ox\ketbra[3]{y-z}{y'-z'}\ox\ketbra[4]{z+y}{z'+y'}\ox\ketbra[5]{z}{z'}\substack{dxdydz\\dx'dy'dz'} .
\end{equation}

\subsubsection{Ideal decoders}
To reconstruct the input state from some subset of the marginals, we need to compute the reduced density matrix on the remaining subset of the shares.  The careful reader is cautioned that in the following derivation, we will not be precise with respect to normalization or other (potentially infinite) factors that arise due to the nature of infinite dimensional Hilbert spaces.  Note that we use $\ket{\alpha}$ for the recovered state to make contact with the coherent states described in \cref{subsubsec:encoding}.

The simplest reconstruction is done using only modes 1 and 2 (erasure of modes 3, 4, and 5), for which it is easy to see (from the encoding) that we can decode using a single beam splitter as in \cref{E1circuit}:
\begin{equation}
\begin{tikzpicture}[scale=0.75, every node/.style={inner sep=2pt}]
		\node [left](mode1start) at (0,1) {mode 1};
		\node [left](mode2start) at (0,0) {mode 2};
		\node [right](mode2end) at (3,0) {};
		\node [right](mode1end) at (3,1) {$\ket{\alpha}$};
		\drawwires(2,3);
		\draw[fill=white, draw=black] (0.75,-0.25) rectangle (2.25,1.25) node[midway] {BS};
	\end{tikzpicture}
	\raisebox{1.5em}{\quad.}
\end{equation}
As one might expect, in the case of finite squeezing, the reduced density matrix on modes 1 and 2 \emph{after} passing them through a balanced beamsplitter is precisely $\proj{\psi}\ox\rho_\text{th}$, where $\ket{\psi}$ is the input state, and $\rho_\text{th}$ is a thermal state with a temperature determined by the squeezing parameter $r$, namely
\begin{equation}
T=\frac{\hbar\omega}{\log\left(\frac{1+\cosh(2r)}{1-\cosh(2r)}\right)}.
\end{equation}

Reconstructing the input state from a different subset of the shares is slightly trickier.  Suppose, for example, that we want to recover from $E_2$ and reconstruct using modes 1, 4, and 5 (i.e., erasure of modes 2, and 3).  We can compute the reduced density matrix as follows
\begin{equation}
\rho_{145}=\int \bra[2]{u}\bra[3]{v}\rho\ket[2]{u}\ket[3]{v} dudv.
\end{equation}
Contracting with the tensor factors for modes 2 and 3 in $\rho$ gives two delta functions $\delta\left[(y-x)-(y'-x')\right]$ and $\delta\left[(y-z)-(y'-z')\right]$.  Integrating over $z'$ and then $y'$ leaves us with the reduced density matrix
\begin{equation}
\rho_{145}=\int \psi(x)\psi^*(x') \ketbra[1]{x+y}{2x'+y-x}\ox\ketbra[4]{z+y}{z+y-2x+2x'}\ox\ketbra[5]{z}{z-x+x'} dx dx' dy dz.
\end{equation}

In order to decode the information from this reduced density matrix and recover the input state $\proj{\psi}$, we need to transform the state into a tensor product of the form
\begin{equation}
\proj[1]{\psi}\ox\sigma_{45}=\int \psi(x)\psi^*(x') \ketbra[1]{x}{x'}dxdx'\ox \sigma_{45},
\end{equation}
for some density operator $\sigma$. The operations we have at our disposal are implemented as symplectic transformations of the basis states (e.g., $\ket{x}\mapsto\ket{Ax}$ for some state $\ket{x}$ and symplectic transformation $A$).

We first apply a controlled sum (QND) gate controlled on mode 5 to mode 4 with a gain of $-2$.  Unpacking the definitions, this means we subtract twice the value in the register for mode 5 from the value in the register for mode 4.  Graphically, we apply
\begin{equation}
\begin{tikzpicture}[scale=0.8, every node/.style={inner sep=2pt}]
		\node [left](mode1start) at (0,2) {mode 1};
		\node [left](mode4start) at (0,1) {mode 4};
		\node [left](mode5start) at (0,0) {mode 5};
		\drawwires(3,2);
		\cplus(1,0,1,1,-2);
\end{tikzpicture}
\raisebox{2.5em}{\quad.}
\end{equation}
The new state becomes
\begin{equation}
\rho_{145} \xmapsto{\bm{4}-2\times\bm{5}} \rho_{145}'=\int \psi(x)\psi^*(x') \ketbra[1]{x+y}{2x'+y-x}\ox\ketbra[4]{y-z}{y-z}\ox\ketbra[5]{z}{z-x+x'} dx dx' dy dz.
\end{equation}
We then apply a QND controlled on mode 4 to mode 1 with a gain of $-1$ to get
\begin{equation}
\rho_{145}' \xmapsto{\bm{1}-\bm{4}} \rho_{145}''=\int \psi(x)\psi^*(x') \ketbra[1]{x+z}{2x'-x+z}\ox\ketbra[4]{y-z}{y-z}\ox\ketbra[5]{z}{z-x+x'} dx dx' dy dz.
\end{equation}
Next we apply a QND controlled on mode 5 to mode 1 with a gain of $-1$ to get
\begin{equation}
\rho_{145}'' \xmapsto{\bm{1}-\bm{5}} \rho_{145}'''=\int \psi(x)\psi^*(x') \ketbra[1]{x}{x'}\ox\ketbra[4]{y-z}{y-z}\ox\ketbra[5]{z}{z-x+x'} dx dx' dy dz.
\end{equation}
Finally we apply a QND controlled on mode 1 to mode 5 with a gain of $-1$ to get
\begin{equation}
\rho_{145}''' \xmapsto{\bm{1}-\bm{5}} \rho_{145}''''=\int \psi(x)\psi^*(x') \ketbra[1]{x}{x'}\ox\ketbra[4]{y-z}{y-z}\ox\ketbra[5]{z-x}{z-x} dx dx' dy dz.
\end{equation}
While this state may initially look entangled (between modes 1 and 5), note that we can change variables for mode 5 so that the integral fully factors into our desired result:
\begin{equation}
\rho_{145}^{\text{decoded}}=\proj{\psi}\ox \id_{45},
\end{equation}
where the identity on modes 4 and 5 arises because we have not taken into account the necessary normalization factors.  Combining all of these steps into a single circuit gives us the decoding circuit for $E_2$
\begin{equation}
\begin{tikzpicture}[scale=0.8, every node/.style={inner sep=2pt}]
		\node [left](mode1start) at (0,2) {mode 1};
		\node [left](mode4start) at (0,1) {mode 4};
		\node [left](mode5start) at (0,0) {mode 5};
		\node [right](mode1end) at (5,2) {$\ket{\alpha}$};
		\drawwires(3,5);
		\cplus(1,0,1,1,-2);
		\cplus(2,1,2,2,-1);
		\cplus(3,0,3,2,-1);
		\cplus(4,2,4,0,-1);
\end{tikzpicture}
\raisebox{2.5em}{\quad,}
\end{equation}
just has we have in \cref{E2circuit}.

Similarly for $E_3$, if we lose modes 2 and 4, we recover using modes 1, 3, and 5.  The reduced density matrix in this case is
\begin{equation}
\rho_{135}=\int \psi(x)\psi^*(x') \ketbra[1]{x+y}{2x'+y-x}\ox\ketbra[3]{y-z}{y-z+2x'-2x}\ox\ketbra[5]{z}{z+x-x'} dx dx' dy dz.
\end{equation}
The transformations to isolate the encoded state proceeds as follows.
\begin{align}
\rho_{135}=&\int \psi(x)\psi^*(x') \ketbra[1]{x+y}{2x'+y-x}\ox\ketbra[3]{y-z}{y-z+2x'-2x}\ox\ketbra[5]{z}{z+x-x'} dx dx' dy dz\notag\\
\bm{3}+2\times\bm{5}	\mapsto &\int \psi(x)\psi^*(x') \ketbra[1]{x+y}{2x'+y-x}\ox\ketbra[3]{y+z}{y+z}\ox\ketbra[5]{z}{z+x-x'} dx dx' dy dz\notag\\
\bm{5}-\bm{3}			\mapsto &\int \psi(x)\psi^*(x') \ketbra[1]{x+y}{2x'+y-x}\ox\ketbra[3]{y+z}{y+z}\ox\ketbra[5]{-y}{-y+x-x'} dx dx' dy dz\notag\\
\bm{1}+\bm{5}			\mapsto &\int \psi(x)\psi^*(x') \ketbra[1]{x}{x'}\ox\ketbra[3]{y+z}{y+z}\ox\ketbra[5]{-y}{-y+x-x'} dx dx' dy dz\notag\\
\bm{5}+\bm{1}			\mapsto &\int \psi(x)\psi^*(x') \ketbra[1]{x}{x'}\ox\ketbra[3]{y+z}{y+z}\ox\ketbra[5]{x-y}{x-y} dx dx' dy dz,
\end{align}
where again we see that the encoded state is present on mode 1.  These transformations are implemented by the decoding circuit in \cref{E3circuit}
\begin{equation}
\begin{tikzpicture}[scale=0.8, every node/.style={inner sep=2pt}]
		\node [left](mode1start) at (0,2) {mode 1};
		\node [left](mode3start) at (0,1) {mode 3};
		\node [left](mode5start) at (0,0) {mode 5};
		\node [right](mode1end) at (5,2) {$\ket{\alpha}$};
		\drawwires(3,5);
		\cplus(1,0,1,1,2);	
		\cplus(2,1,2,0,-1);
		\cplus(3,0,3,2,1);
		\cplus(4,2,4,0,1);
\end{tikzpicture}
\raisebox{2.5em}{\quad.}
\end{equation}

Lastly, we can recover from $E_4$ and decode using only modes 2, 3, and 4 as follows.  The reduced density matrix in this case is given by
\begin{equation}
\rho_{234}=\int \psi(x)\psi^*(x') \ketbra[2]{y-x}{y+x-2x'}\ox\ketbra[3]{y-z}{y-z+x-x'}\ox\ketbra[4]{z+y}{z+y+x-x'} dx dx' dy dz,
\end{equation}
and the recovery progression is
\begin{align}
\rho_{234}=&\int \psi(x)\psi^*(x') \ketbra[2]{y-x}{y+x-2x'}\ox\ketbra[3]{y-z}{y-z+x-x'}\ox\ketbra[4]{z+y}{z+y+x-x'} dx dx' dy dz\notag\\
\bm{3}-\bm{4}				\mapsto &\int \psi(x)\psi^*(x') \ketbra[2]{y-x}{y+x-2x'}\ox\ketbra[3]{-2z}{-2z}\ox\ketbra[4]{z+y}{z+y+x-x'} dx dx' dy dz\notag\\
\bm{4}-\bm{2}				\mapsto &\int \psi(x)\psi^*(x') \ketbra[2]{y-x}{y+x-2x'}\ox\ketbra[3]{-2z}{-2z}\ox\ketbra[4]{z+x}{z+x'} dx dx' dy dz\notag\\
\bm{4}+\frac{1}{2}\times\bm{3}	\mapsto &\int \psi(x)\psi^*(x') \ketbra[2]{y-x}{y+x-2x'}\ox\ketbra[3]{-2z}{-2z}\ox\ketbra[4]{x}{x'} dx dx' dy dz\notag\\
\bm{2}+2\times\bm{4}		\mapsto &\int \psi(x)\psi^*(x') \ketbra[2]{y+x}{y+x}\ox\ketbra[3]{-2z}{-2z}\ox\ketbra[4]{x}{x'} dx dx' dy dz,
\end{align}
giving us the decoding circuit in \cref{E4circuit}
\begin{equation}
\begin{tikzpicture}[scale=0.8, every node/.style={inner sep=2pt}]
		\node [left](mode2start) at (0,2) {mode 2};
		\node [left](mode3start) at (0,1) {mode 3};
		\node [left](mode4start) at (0,0) {mode 4};
		\node [right](mode4end) at (5,0) {$\ket{\alpha}$};		
		\drawwires(3,5);
		\cplus(1,0,1,1,-1);	
		\cplus(2,2,2,0,-1);
		\cplus(3,1,3,0,1/2);
		\cplus(4,0,4,2,2);
\end{tikzpicture}
\raisebox{2.5em}{\quad.}
\end{equation}

\subsubsection{Optical decoders}
Our goal now is to find efficient optical implementations of the decoding circuits derived above. We begin by cataloging a selection of helpful circuit identities, summarized in  \cref{identity1,identity2,identity3,identity4,identity5,identity6,identity7,identity8}.
\begin{align}
\raisebox{-2.3em}{\begin{tikzpicture}[scale=0.8, every node/.style={inner sep=2pt}]
		\drawwires(2,3);
		\cplus(1,0,1,1,a);
		\cplus(2,1,2,0,b);
\end{tikzpicture}}
&\qquad=\qquad
\raisebox{-2.3em}{\begin{tikzpicture}[scale=0.8, every node/.style={inner sep=2pt}]
		\drawwires(2,5);
		\cplus(2,1,2,0,b);
		\cplus(3,0,3,1,a);
		\node [gate] at (1,0) {$1+ab$};
		\node [gate] at (4,1) {$\frac{1}{1+ab}$};
		\node [label={\scriptsize$1+ab\neq 0$}]at (4.3,0.01) {};
\end{tikzpicture}}\label{identity1}
\\[1.5em]
\raisebox{-1.8em}{\begin{tikzpicture}[scale=0.8, every node/.style={inner sep=2pt}]
		\drawwires(2,3);
		\node[gate] at (1,0) {$a$};
		\cplus(2,0,2,1,b);
\end{tikzpicture}}
&\qquad=\qquad
\raisebox{-1.8em}{\begin{tikzpicture}[scale=0.8, every node/.style={inner sep=2pt}]
		\drawwires(2,3);
		\node[gate] at (2,0) {$a$};
		\cplus(1,0,1,1,ab);
\end{tikzpicture}}\label{identity2}
\\[1.5em]
\raisebox{-2.3em}{\begin{tikzpicture}[scale=0.8, every node/.style={inner sep=2pt}]
		\drawwires(2,3);
		\node[gate] at (1,0) {$a$};
		\cplus(2,1,2,0,b);
\end{tikzpicture}}
&\qquad=\qquad
\raisebox{-2.3em}{\begin{tikzpicture}[scale=0.8, every node/.style={inner sep=2pt}]
		\drawwires(2,3);
		\node[gate] at (2,0) {$a$};
		\cplus(1,1,1,0,b/a);
\end{tikzpicture}}\label{identity3}
\\[1.5em]
\raisebox{-3.3em}{\begin{tikzpicture}[scale=0.8, every node/.style={inner sep=2pt}]
		\drawwires(3,3);
		\cplus(1,1,1,2,a);
		\cplus(2,2,2,0,b);
\end{tikzpicture}}
&\qquad=\qquad
\raisebox{-3.3em}{\begin{tikzpicture}[scale=0.8, every node/.style={inner sep=2pt}]	
		\drawwires(3,4);
		\cplus(1,2,1,0,b);
		\cplus(2,1,2,0,ab);
		\cplus(3,1,3,2,a);
\end{tikzpicture}}\label{identity4}
\\[1.5em]
\raisebox{-3.3em}{\begin{tikzpicture}[scale=0.8, every node/.style={inner sep=2pt}]
		\drawwires(3,3);
		\cplus(1,0,1,1,a);
		\cplus(2,2,2,0,b);
\end{tikzpicture}}
&\qquad=\qquad
\raisebox{-3.3em}{\begin{tikzpicture}[scale=0.8, every node/.style={inner sep=2pt}]	
		\drawwires(3,4);
		\cplus(1,2,1,0,b);
		\cplus(2,0,2,1,a);
		\cplus(3,2,3,1,-ab);
\end{tikzpicture}}\label{identity5}
\\[1.5em]
\raisebox{-1.3em}{\begin{tikzpicture}[scale=0.8, every node/.style={inner sep=2pt}]
		\drawwires(2,2);
		\cplus(1,0,1,1,a);
\end{tikzpicture}}
&\qquad=\qquad
\raisebox{-2.3em}{\begin{tikzpicture}[scale=0.8, every node/.style={inner sep=2pt}]
		\drawwires(2,4);
		\node[gate] at (1,0) {$FT$};
		\node[gate] at (1,1) {$FT$};
		\cplus(2,1,2,0,-a);
		\node[gate] at (3,0) {$FT^{-1}$};
		\node[gate] at (3,1) {$FT^{-1}$};
\end{tikzpicture}}\label{identity6}
\\[1.5em]
\raisebox{-1.0em}{\begin{tikzpicture}[scale=0.8, every node/.style={inner sep=2pt}]
		\draw (0,0) -- (1,0);
		\draw (0,1) -- (1,1);
		\beamsplitterpm(1,1);
		\draw (2,0) -- (3,0);
		\draw (2,1) -- (3,1);
\end{tikzpicture}}
&\qquad=\qquad
\raisebox{-2.3em}{\begin{tikzpicture}[scale=0.8, every node/.style={inner sep=2pt}]
		\drawwires(2,4);
		\cplus(1,1,1,0,-1);
		\cplus(2,0,2,1,1/2);
		\node[gate] at (3,1) {$\sqrt{2}$};
		\node[gate] at (3,0) {$1/\sqrt{2}$};
\end{tikzpicture}}\label{identity7}
\\[1.5em]
\raisebox{-2.3em}{\begin{tikzpicture}[scale=0.8, every node/.style={inner sep=2pt}]
		\drawwires(2,3);
		\cplus(1,1,1,0,1);
		\cplus(2,0,2,1,-1)
\end{tikzpicture}}
&\qquad=\qquad
\raisebox{-1.8em}{\begin{tikzpicture}[scale=0.8, every node/.style={inner sep=2pt}]
		\draw (-3,0) -- (-2,0);
		\draw (-3,1) -- (-2,1);
		\beamsplittermp(-2,1);
		\draw (-1,0) -- (0,0);
		\draw (-1,1) -- (0,1);
		\drawwires(2,3);
		\node[gate] at (0,0) {$\sqrt{2}$};
		\node[gate] at (0,1) {$\sqrt{2}$};
		\cplus(1,0,1,1,-1);
		\node[gate] at (2,1){$1/2$};
\end{tikzpicture}}\label{identity8}\quad .
\end{align}

We will also make extensive use of the well-known principle ``measurement commutes with control''.  In particular
\begin{equation}\label{measurecontrol}
\raisebox{-1.5em}{\begin{tikzpicture}[scale=0.8, every node/.style={inner sep=2pt}]		
		\drawwires(2,2);
		\cplus(1,0,1,1,a);	
		\meas(2,0,$x$);
\end{tikzpicture}}
\quad=\quad
\raisebox{-1.5em}{\begin{tikzpicture}[scale=0.8, every node/.style={inner sep=2pt}]		
		\draw (0,1) -- (3,1);
		\draw (0,0) -- (1,0);
		\cplus(2,0,2,1,a);	
		\cwire({1,0},{2,0});
		\cwire({2,0},{2,0.75});
		\meas(1,0,$x$);
		\draw [fill=white] (2,0) circle (2.75pt);
\end{tikzpicture}}\quad .
\end{equation}

We have four decoding circuits to optimize.  Fortunately, the circuit for $E_1$ is just a beam splitter, so there is nothing to do.  As such, let us first focus on the decoder in \cref{E2circuit}.  One way to simplify this circuit is to use the principle in \cref{measurecontrol}.  Unfortunately, the \emph{last} operation required in this circuit is a controlled sum (QND) gate, where the control is done on the mode on which we will recover our state.  As such, we cannot measure this mode, or we will break our decoder.  To get around this, it will prove advantageous to find an equivalent decoding circuit that does not end with a QND gate controlled on the mode on which we recover.  One way to do this is to use the identities summarized above to rewrite the circuit in a more appealing way.  For instance, we can use \cref{identity1} to commute the last QND gate in the decoding circuit toward the start, and then use \cref{identity2} and \cref{identity3,identity4} to simplify, but in fact there is a simpler way to find such a circuit.

Consider the ideal decoding circuit in \cref{E2circuit}, reproduced below
\begin{equation}
\begin{tikzpicture}[scale=0.8, every node/.style={inner sep=2pt}]
		\node [left](mode1start) at (0,2) {$x$};
		\node [left](mode4start) at (0,1) {$y$};
		\node [left](mode5start) at (0,0) {$z$};
		\node [right](mode1end) at (5,2) {$x-y+z$};
		\node [right](mode4end) at (5,1) {$y-2z$};
		\node [right](mode5end) at (5,0) {$y-x$};
		\drawwires(3,5);
		\cplus(1,0,1,1,-2);
		\cplus(2,1,2,2,-1);
		\cplus(3,0,3,2,-1);
		\cplus(4,2,4,0,-1);
\end{tikzpicture}
\raisebox{2.5em}{\quad.}
\end{equation}
The action of this circuit on the basis state $\ket{x,y,z}$ is to produce the state $\ket{x-y+z,y-2z,y-x}$.  As such, we can represent the action of the decoder as
\begin{equation}
\ket{\vec{x}}\mapsto\ket{A_2\cdot\vec{x}}=\ket{x-y+z,y-2z,y-x},
\end{equation}
where $A_2$ is a matrix, the subscript $2$ comes from the label for $E_2$, and $\ket{\vec{x}}=\ket{x,y,z}$.   For this decoder, 
\begin{equation}\label{E2matrix}
A_2=\left(
\begin{array}{ccc}
	1	&	-1	&	1	\\
	0	&	1	&	-2	\\
	-1	&	1	&	0
\end{array}\right),
\end{equation}
where the rows and columns represent the three modes.  The matrix in \cref{E2matrix} is a convenient representation of the decoding circuit, but it is in fact more fundamental as it represents the set of transformations that must be applied to the modes to decode.  In particular, one way to produce this matrix is to start from the identity and multiply by the matrices representing the QND gates in \cref{E2circuit}.  However, this decomposition is not unique, and we are free to choose any decomposition we like.  We might prefer to choose a decomposition that does not end with a QND gate controlled on the mode we wish to use to recover, or we may decompose using fewer operations.  

As an example, the matrix $A$ can be constructed from the circuit in \cref{E2circuit} as follows.  We begin with the identity matrix $\id_3$ and first apply a QND from mode 5 to mode 4 with a gain of $-2$.  This operation corresponds to subtracting twice row 3 from row 2 in the matrix representation:
\begin{equation}\left(
\begin{array}{ccc}
	1	&	0	&	0	\\
	0	&	1	&	0	\\
	0	&	0	&	1
\end{array}\right)
\xmapsto{R_2-2R_3}
\left(\begin{array}{ccc}
	1	&	0	&	0	\\
	0	&	1	&	-2	\\
	0	&	0	&	1
\end{array}\right).
\end{equation}
The next gate is a QND from mode 4 to mode 1 with a gain of $-1$.  This operation corresponds to the row transformation row 1 minus row 2:
\begin{equation}\left(
\begin{array}{ccc}
	1	&	0	&	0	\\
	0	&	1	&	-1	\\
	0	&	0	&	1
\end{array}\right)
\xmapsto{R_1-R_2}
\left(\begin{array}{ccc}
	1	&	-1	&	2	\\
	0	&	1	&	-2	\\
	0	&	0	&	1
\end{array}\right).
\end{equation}
The next two gates in \cref{E2circuit} are similarly represented as row transformations as follows
\begin{equation}\left(
\begin{array}{ccc}
	1	&	-1	&	2	\\
	0	&	1	&	-2	\\
	0	&	0	&	1
\end{array}\right)
\xmapsto{R_1-R_3}
\left(\begin{array}{ccc}
	1	&	-1	&	1	\\
	0	&	1	&	-2	\\
	0	&	0	&	1
\end{array}\right)
\xmapsto{R_3-R_1}
\left(\begin{array}{ccc}
	1	&	-1	&	1	\\
	0	&	1	&	-2	\\
	-1	&	1	&	0
\end{array}\right).
\end{equation}

If we wish to find a different decomposition of the transformation into simple gates, we can reverse the above process and keep track of our steps.  In other words, we start with the transformation matrix $A$ and perform Gaussian elimination (in whatever order we choose) to return the matrix to the identity.  Each of the row-reduction steps becomes a fundamental quantum operation that we need to apply.  For the matrix in \cref{E2matrix} then, we have
\begin{align}\label{E2rowreduction}
\left(\begin{array}{ccc}
	1	&	-1	&	1	\\
	0	&	1	&	-2	\\
	-1	&	1	&	0
\end{array}\right)
&\xmapsto{R_1\leftrightarrow R_3}
\left(\begin{array}{ccc}
	-1	&	1	&	0	\\
	0	&	1	&	-2	\\
	1	&	-1	&	1
\end{array}\right)
\xmapsto{R_1\times (-1)}
\left(\begin{array}{ccc}
	1	&	-1	&	0	\\
	0	&	1	&	-2	\\
	1	&	-1	&	1
\end{array}\right)
\xmapsto{R_3-R_1}
\left(\begin{array}{ccc}
	1	&	-1	&	0	\\
	0	&	1	&	-2	\\
	0	&	0	&	1
\end{array}\right)\notag\\
&\xmapsto{R_2+2R_3}
\left(\begin{array}{ccc}
	1	&	-1	&	0	\\
	0	&	1	&	0	\\
	0	&	0	&	1
\end{array}\right)
\xmapsto{R_1+R_2}
\left(\begin{array}{ccc}
	1	&	0	&	0	\\
	0	&	1	&	0	\\
	0	&	0	&	1
\end{array}\right).
\end{align}
The operators comprising a decoding circuit can be read off from the row-reduction transformations from the above equation.  In particular, the quantum circuit corresponds to reading the set of transformations \emph{backwards}, and noting that we in fact want the \emph{adjoint} of the row-reduction transformations.  For instance, the last row-reduction transformation in \cref{E2rowreduction} was $R_1+R_2$.  The since $R_1$ corresponds to mode 1 and $R_2$ corresponds to mode 5, the first gate in our decoding circuit becomes a QND controlled on mode 5 with a gain of $-1$ (minus because we need the adjoint of the transformations).  Tracing back, we find and equivalent decoding circuit of
\begin{equation}
\begin{tikzpicture}[scale=0.8, every node/.style={inner sep=2pt}]
		\node [left](mode1start) at (0,2) {mode 1};
		\node [left](mode4start) at (0,1) {mode 4};
		\node [left](mode5start) at (0,0) {mode 5};
		\node [right](mode1end) at (6,2) {$\ket{\alpha}$};		
		\drawwires(3,6);
		\cplus(1,1,1,2,-1);	
		\cplus(2,0,2,1,-2);
		\cplus(3,2,3,0,1);
		\node[gate] at (4,2){$-1$};
		\cvswap({5,0},{5,2});
\end{tikzpicture}
\raisebox{2.5em}{\quad,}
\end{equation}
where $
\begin{tikzpicture}[scale=0.5, every node/.style={inner sep=2pt}]	
		\drawwires(2,2);
		\cvswap({1,0},{1,1});
\end{tikzpicture}
$
represents the SWAP operation on the two affected modes.  
We can remove the final SWAP gate by simply recovering the encoded state on mode 5 instead of mode 1 as
\begin{equation}
\begin{tikzpicture}[scale=0.8, every node/.style={inner sep=2pt}]
		\node [left](mode1start) at (0,2) {mode 1};
		\node [left](mode4start) at (0,1) {mode 4};
		\node [left](mode5start) at (0,0) {mode 5};
		\node [right](mode5end) at (5,0) {$\ket{\alpha}$};		
		\drawwires(3,5);
		\cplus(1,1,1,2,-1);	
		\cplus(2,0,2,1,-2);
		\cplus(3,2,3,0,1);
		\node[gate] at (4,2){$-1$};
\end{tikzpicture}
\raisebox{2.5em}{\quad.}
\end{equation}

With this new (equivalent) circuit in hand, we are in a position to use our identities to find an efficient optical implementation.  We first note that, since they contain no information about the encoded state at the end of the computation, modes 1 and 4 can be measured at the end without disturbing the decoding.  As such, we arbitrarily measure the $x$ quadrature of mode 1 and the $p$ quadrature of mode 4 (this choice will simplify the analysis below).  Next, we notice that we can discard the $-1$ gate, since it also does not affect the decoded state.  We then use identities \cref{identity6} and \cref{measurecontrol} to get
\begin{equation}
\begin{tikzpicture}[scale=0.7, every node/.style={inner sep=2pt}]
		\node [right](mode5end) at (5,0) {$\ket{\alpha}$};		
		\drawwires(3,5);
		\cplus(1,1,1,2,-1);	
		\cplus(2,0,2,1,-2);
		\cplus(3,2,3,0,1);
		\node[gate] at (4,2){$-1$};
\end{tikzpicture}
\raisebox{2.95em}{$\quad\mapsto\quad$}
\begin{tikzpicture}[scale=0.7, every node/.style={inner sep=2pt}]
		\node [right](mode5end) at (6,0) {$\ket{\alpha}$};		
		\drawwires(3,6);
		\cplus(1,1,1,2,-1);	
		\node[gate] at (2,0) {FT};
		\node[gate] at (2,1) {FT};
		\cplus(3,1,3,0,2);
		\node[gate] at (4.2,0) {$\text{FT}^{-1}$};
		\node[gate] at (4.2,1) {$\text{FT}^{-1}$};
		\cplus(5.4,2,5.4,0,1);
		\meas(6,1,$p$);
		\meas(6,2,$x$);
\end{tikzpicture}
\raisebox{2.95em}{$\quad\mapsto\quad$}
\begin{tikzpicture}[scale=0.7, every node/.style={inner sep=2pt}]
		\node [right](mode5end) at (6,0) {$\ket{\alpha}$};		
		\drawwires(3,2);
		\draw (2,0) -- (6,0);
		\cplus(1,1,1,2,-1);	
		\cplus(3,1,3,0,2);
		\node[gate] at (2,0) {FT};
		\node[gate] at (4.2,0) {$\text{FT}^{-1}$};
		\cplus(5.4,2,5.4,0,1);
		\cwire({2,1},{3,1});
		\cwire({2,2},{5.4,2});
		\cwire({3,1},{3,0.25});
		\cwire({5.4,2},{5.4,0.25});
		\draw [fill=white] (3,1) circle (2.75pt);
		\draw [fill=white] (5.4,2) circle (2.75pt);
		\meas(2,1,$p$);
		\meas(2,2,$x$);
\end{tikzpicture}
\raisebox{2.5em}{\quad.}
\end{equation}
Notice that the first QND gate in the new circuit is followed by measurement of $x$ on mode 1 and $p$ on mode 2.  The effective measurement performed when the QND gate is taken into account can be implemented without the QND gate as follows
\begin{equation}
\begin{tikzpicture}[scale=0.8, every node/.style={inner sep=2pt}]
		\node [right](mode1start) at (0,1) {$\substack{x_1\\p_1}$};
		\node [right](mode2start) at (0,0) {$\substack{x_2\\p_2}$};
		\node [right](mode1end) at (3,1) {\small$(x_1-x_2)$};	
		\node [right](mode2end) at (3,0) {\small$(p_1+p_2)$};				
		\drawwires(2,2);
		\cplus(1.15,0,1.15,1,-1);	
		\cwire({2,0},{3,0});
		\cwire({2,1},{3,1});
		\meas(2,0,$p$);
		\meas(2,1,$x$);
\end{tikzpicture}
\raisebox{1.8em}{$\quad = \quad$}
\begin{tikzpicture}[scale=0.8, every node/.style={inner sep=2pt}]
		\node [right](mode1start) at (-1,1) {$\substack{x_1\\p_1}$};
		\node [right](mode2start) at (-1,0) {$\substack{x_2\\p_2}$};
		\drawwiresat({-1,0},2,1);
		\node [right](mode1end) at (4,1) {\small$(x_1-x_2)$};	
		\node [right](mode2end) at (4,0) {\small$(p_1+p_2)$};				
		\drawwiresat({1,0},2,2);
		\beamsplittermp(0,1);
		\node[gate] at (2,0) {$\sqrt{2}$};
		\node[gate] at (2,1) {$\sqrt{2}$};
		\cwire({3,0},{4,0});
		\cwire({3,1},{4,1});
		\meas(3,0,$p$);
		\meas(3,1,$x$);
\end{tikzpicture}
\raisebox{1.8em}{\quad.}
\end{equation}
One can compute this circuit equivalence directly, as above, or arrive at the same conclusion by working in the Heisenberg picture and conjugating the homodyne measurements by QND gates.  We have thus arrived at our efficient decoding circuit for $E_2$, as shown in \cref{E2optical} and below
\begin{equation*}
	\begin{tikzpicture}[scale=0.75, every node/.style={inner sep=0pt}]  
		\node [left](mode1start) at (0,2) {mode 1};
		\node [left](mode4start) at (0,1) {mode 4};
		\node [left](mode5start) at (0,0) {mode 5};
		\node [right](mode5end) at (7.8,0) {$\ket{\alpha}$};
		
		\draw (mode5start) -- (mode5end);

		\node [gate] (FT) at (2.5,0) {$FT$};
		\node [gate] (Disp45) at (4,0) {Disp};
		\node [gate] (FTinv) at (5.5,0) {$FT^{-1}$};
		\node [gate] (Disp15) at (7,0) {Disp};
		
		\node [inner xsep=1.1em](meas4) at (3,1) {};
		\coordinate (control45) at (Disp45|-meas4);
		\cwire(control45,Disp45);
		\cwire(meas4,control45);
		\draw [fill=white] (control45) circle (2.75pt);
		\meas(3,1,$p$);
		\draw (mode4start) -- (0.5,1);
		\draw (1.5,1) -- ($(meas4)$);  
		\node [gate] (sqeeze4) at (2.2,1) {$\sqrt{2}$};

		\node [inner xsep=1.1em](meas1) at (3,2) {};
		\coordinate (control15) at (Disp15|-meas1);
		\cwire(control15,Disp15); 
		\cwire(meas1,control15|-meas1);
		\draw [fill=white] (control15) circle (2.75pt);
		\meas(3,2,$x$);
		\draw (mode1start) -- (0.5,2);
		\draw (1.5,2) -- ($(meas1)$);  
		\beamsplittermp(0.5,2);
		\node [gate] (sqeeze1) at (2.2,2) {$\sqrt{2}$};

	\end{tikzpicture}

\raisebox{2.5em}{\quad.}
\end{equation*}

Derivation of the last two efficient circuits follow similarly to that above.  We begin my constructing the transformation matrix according to the ideal decoding circuit, row-reduce to identity, and form a new decoding circuit by following the row-reductions backwards (and converting to the adjoint of the corresponding operator).  We then appeal to the circuit identities summarized earlier.

Let us now analyze the decoding circuit for $E_3$, shown in \cref{E3circuit}:
\begin{equation}
\begin{tikzpicture}[scale=0.8, every node/.style={inner sep=2pt}]
		\node [left](mode1start) at (0,2) {$x$};
		\node [left](mode3start) at (0,1) {$y$};
		\node [left](mode5start) at (0,0) {$z$};
		\node [right](mode1end) at (5,2) {$x-y-z$};
		\node [right](mode3end) at (5,1) {$y+2z$};
		\node [right](mode5end) at (5,0) {$x-2y-2z$};
		\drawwires(3,5);
		\cplus(1,0,1,1,-2);
		\cplus(2,1,2,2,-1);
		\cplus(3,0,3,2,-1);
		\cplus(4,2,4,0,-1);
\end{tikzpicture}
\raisebox{2.5em}{\quad.}
\end{equation}
The transformation matrix for this decoding circuit is then
\begin{equation}\label{E3matrix}
A_3=\left(
\begin{array}{ccc}
	1	&	-1	&	-1	\\
	0	&	1	&	2	\\
	1	&	-2	&	-2
\end{array}\right).
\end{equation}
We can use Gaussian elimination to reduce this to the identity via
\begin{align}\label{E3rowreduction}
&\left(\begin{array}{ccc}
	1	&	-1	&	-1	\\
	0	&	1	&	2	\\
	1	&	-2	&	-2
\end{array}\right)
\xmapsto{R_1\rightarrow R_2\rightarrow R_3\rightarrow R_1}
\left(\begin{array}{ccc}
	1	&	-2	&	-2	\\
	1	&	-1	&	-1	\\
	0	&	1	&	2
\end{array}\right)
\xmapsto{R_2-R_1}
\left(\begin{array}{ccc}
	1	&	-2	&	-2	\\
	0	&	1	&	1	\\
	0	&	1	&	2
\end{array}\right)
\xmapsto{R_1+2R_3}
\left(\begin{array}{ccc}
	1	&	0	&	2	\\
	0	&	1	&	1	\\
	0	&	1	&	2
\end{array}\right)\notag\\
&\xmapsto{R_2-\frac{1}{2}R_3}
\left(\begin{array}{ccc}
	1	&	0	&	2	\\
	0	&	1/2	&	0	\\
	0	&	1	&	2
\end{array}\right)
\xmapsto{R_3-2R_2}
\left(\begin{array}{ccc}
	1	&	0	&	2	\\
	0	&	1/2	&	0	\\
	0	&	0	&	2
\end{array}\right)
\xmapsto{R_1-R_3}
\left(\begin{array}{ccc}
	1	&	0	&	0	\\
	0	&	1/2	&	0	\\
	0	&	0	&	2
\end{array}\right)
\xmapsto{R_2\times 2 \&R_3\times \frac{1}{2}}
\left(\begin{array}{ccc}
	1	&	0	&	0	\\
	0	&	1	&	0	\\
	0	&	0	&	1
\end{array}\right).
\end{align}
Reading these transformations backwards and taking appropriate adjoints gives the following equivalent decoding circuit
\begin{equation}
\begin{tikzpicture}[scale=0.8, every node/.style={inner sep=2pt}]
		\node [left](mode1start) at (0,2) {mode 1};
		\node [left](mode3start) at (0,1) {mode 3};
		\node [left](mode5start) at (0,0) {mode 5};
		\node [right](mode4end) at (7,1) {$\ket{\alpha}$};		
		\drawwires(3,7);
		\node[gate] at (1,0){$2$};
		\node[gate] at (1,1){$1/2$};
		\cplus(2,0,2,2,1);
		\cplus(3,1,3,0,2);	
		\cplus(4,0,4,1,1/2);
		\cplus(5,0,5,2,-2);
		\cplus(6,2,6,1,1);
\end{tikzpicture}
\raisebox{2.5em}{\quad,}
\end{equation}
where we dropped the final cyclic permutation (which can be implemented with two SWAP gates) and instead recovered on mode 3.

We now make use of several circuit identities (mostly \cref{identity2,identity3,identity4} and \cref{identity8}) to improve efficiency.  The goal is to reduce the number of QND gates and avoid controlled gates with the control on the recovered mode.  Furthermore, we would like to use simple linear optical circuit elements (e.g., beam splitters) whenever possible.
\begin{align}
&\begin{tikzpicture}[scale=0.55, every node/.style={inner sep=2pt}]
		\node [right](mode4end) at (7,1) {$\ket{\alpha}$};		
		\drawwires(3,7);
		\node[gate] at (1,0){$2$};
		\node[gate] at (1,1){$1/2$};
		\cplus(2,0,2,2,1);
		\cplus(3,1,3,0,2);	
		\cplus(4,0,4,1,1/2);
		\cplus(5,0,5,2,-2);
		\cplus(6,2,6,1,1);
\end{tikzpicture}
\raisebox{2.1em}{$\quad = \quad$}
\begin{tikzpicture}[scale=0.55, every node/.style={inner sep=2pt}]
		\node [right](mode4end) at (8,1) {$\ket{\alpha}$};		
		\drawwires(3,8);
		\node[gate] at (5,0){$2$};
		\node[gate] at (1,1){$1/2$};
		\cplus(2,0,2,2,2);
		\cplus(3,1,3,0,1);	
		\cplus(4,0,4,1,1);
		\cplus(6,0,6,2,-2);
		\cplus(7,2,7,1,1);
\end{tikzpicture}
\raisebox{2.1em}{$\quad = \quad$}
\begin{tikzpicture}[scale=0.55, every node/.style={inner sep=2pt}]
		\node [right](mode4end) at (9,1) {$\ket{\alpha}$};		
		\drawwires(3,9);
		\node[gate] at (6,0){$2$};
		\node[gate] at (1,1){$1/2$};
		\cplus(2,0,2,2,2);
		\cplus(3,1,3,0,1);	
		\cplus(4,0,4,1,-1);
		\cplus(5,0,5,1,2);
		\cplus(7,0,7,2,-2);
		\cplus(8,2,8,1,1);
\end{tikzpicture}\notag\\
\raisebox{1.8em}{$= \quad$}
&\begin{tikzpicture}[scale=0.55, every node/.style={inner sep=2pt}]
		\node [right](mode4end) at (9,1) {$\ket{\alpha}$};		
		\drawwires(3,9);
		\node[gate] at (5,0){$2$};
		\node[gate] at (1,1){$1/2$};
		\cplus(2,0,2,2,2);
		\cplus(3,1,3,0,1);	
		\cplus(4,0,4,1,-1);
		\cplus(6,0,6,1,1);
		\cplus(7,0,7,2,-2);
		\cplus(8,2,8,1,1);
\end{tikzpicture}
\raisebox{1.8em}{$\quad = \quad$}
\begin{tikzpicture}[scale=0.55, every node/.style={inner sep=2pt}]
		\node [right](mode4end) at (11,1) {$\ket{\alpha}$};		
		\draw (0,2) -- (11,2);
		\drawwires(2,3);
		\cplus(2,0,2,2,2);
		\node[gate] at (1,1){$1/2$};
		\beamsplittermp(3,1);
		\drawwiresat({4,0},2,7);
		\node[gate] at (5,0) {$\sqrt{2}$};
		\node[gate] at (5,1) {$\sqrt{2}$};
		\cplus(6,0,6,1,-1);
		\node[gate] at (7,1){$1/2$};
		\node[gate] at (7,0){$2$};
		\cplus(8,0,8,1,1);
		\cplus(9,0,9,2,-2);
		\cplus(10,2,10,1,1);
\end{tikzpicture}\notag\\
\raisebox{1.8em}{$\quad = \quad$}
&\begin{tikzpicture}[scale=0.55, every node/.style={inner sep=2pt}]
		\node [right](mode4end) at (10,1) {$\ket{\alpha}$};		
		\draw (0,2) -- (10,2);
		\drawwires(2,3);
		\cplus(2,0,2,2,2);
		\node[gate] at (1,1){$1/2$};
		\beamsplittermp(3,1);
		\drawwiresat({4,0},2,6);
		\node[gate] at (5,1) {$1/\sqrt{2}$};
		\node[gate] at (5,0) {$2\sqrt{2}$};
		\cplus(6,0,6,1,-1/4);
		\cplus(7,0,7,1,1);
		\cplus(8,0,8,2,-2);
		\cplus(9,2,9,1,1);
\end{tikzpicture}
\raisebox{1.8em}{$\quad = \quad$}
\begin{tikzpicture}[scale=0.55, every node/.style={inner sep=2pt}]
		\node [right](mode4end) at (10,1) {$\ket{\alpha}$};		
		\draw (0,2) -- (10,2);
		\drawwires(2,3);
		\cplus(2,0,2,2,2);
		\node[gate] at (1,1){$1/2$};
		\beamsplittermp(3,1);
		\drawwiresat({4,0},2,6);
		\node[gate] at (5,1) {$1/\sqrt{2}$};
		\node[gate] at (5,0) {$2\sqrt{2}$};
		\cplus(6,0,6,1,3/4);
		\cplus(7,2,7,1,1);
		\cplus(8,0,8,2,-2);
		\cplus(9,0,9,1,-2);
\end{tikzpicture}\notag\\
\raisebox{1.8em}{$\quad = \quad$}
&\begin{tikzpicture}[scale=0.55, every node/.style={inner sep=2pt}]
		\node [right](mode4end) at (9,1) {$\ket{\alpha}$};		
		\draw (0,2) -- (9,2);
		\drawwires(2,3);
		\cplus(2,0,2,2,2);
		\node[gate] at (1,1){$1/2$};
		\beamsplittermp(3,1);
		\drawwiresat({4,0},2,5);
		\node[gate] at (5,1) {$1/\sqrt{2}$};
		\node[gate] at (5,0) {$2\sqrt{2}$};
		\cplus(7,0,7,1,-5/4);
		\cplus(6,2,6,1,1);
		\cplus(8,0,8,2,-2);
\end{tikzpicture}
\raisebox{2em}{\quad.}
\end{align}
Notice that the last QND gate is of no consequence to the recovered state, so we can safely drop it.  Furthermore, we can measure modes 1 and 5 safely and commute the measurements past controls to get the optical decoding in \cref{E3optical}, reproduced below
\begin{equation*}
	\begin{tikzpicture}[scale=0.75, every node/.style={inner sep=0pt, minimum size=0pt}]  
		\node [left](mode1start) at (0,2) {mode 1};
		\node [left](mode3start) at (0,1) {mode 3};
		\node [left](mode5start) at (0,0) {mode 5};

		\node [right](mode3end) at (9.1,1) {$\ket{\alpha}$};
		\coordinate (BSstart) at (2.6,1);
		\coordinate (BSend) at (3.6,1);
		\draw (mode3start) -- (BSstart);
		\draw (BSend)-- (mode3end);
		\node [gate] (Sq3half) at (1.75,1) {$\frac{1}{2}$};
		\node [gate] (Sq3aqrt2) at (4.4,1) {$\frac{1}{\sqrt{2}}$};
		\node [gate] (Disp31) at (5.65,1) {Disp};
		\node [gate] (Disp32) at (7.65,1) {$\text{Disp}(\frac{-5\sqrt{2}}{2})$};
		
		\beamsplittermp(2.6,1);
		
		\node [inner xsep=1.1em](meas5) at (4,0) {};
		\coordinate (control53) at (Disp32|-meas5);
		\cwire(control53,Disp32);
		\cwire(meas5,control53);
		\draw [fill=white] (control53) circle (2.75pt);
		\meas(4,0,$x$);
		\draw (mode5start)--(BSstart|-mode5start);
		\draw (BSend|-mode5start) -- ($(meas5)$);  

		\node [inner xsep=1.1em](meas1) at (4,2) {};
		\coordinate (control13) at (Disp31|-meas1);
		\cwire(control13,Disp31);
		\cwire(meas1,control13);
		\draw [fill=white] (control13) circle (2.75pt);
		\meas(4,2,$x$);
		\draw (mode1start) -- ($(meas1)$);  
		\node [gate] (QND21) at (1,2) {$\text{QND}_2$};
		\node (control51) at (QND21|-mode5start) {};
		\fill (control51) circle [radius=2pt];
		\draw (control51) -- (QND21);
		
	\end{tikzpicture}

\raisebox{2.5em}{\quad.}
\end{equation*}

The final decoding circuit (to recover the state after $E_4$), as shown in \cref{E4circuit} effects the following transformation
\begin{equation}
\begin{tikzpicture}[scale=0.8, every node/.style={inner sep=2pt}]
		\node [left](mode2start) at (0,2) {$x$};
		\node [left](mode3start) at (0,1) {$y$};
		\node [left](mode4start) at (0,0) {$z$};
		\node [right](mode2end) at (5,2) {$-x+y+z$};
		\node [right](mode3end) at (5,1) {$y-z$};
		\node [right](mode4end) at (5,0) {$-x+\frac{1}{2}y+\frac{1}{2}z$};
		\drawwires(3,5);
		\cplus(1,0,1,1,-1);
		\cplus(2,2,2,0,-1);
		\cplus(3,1,3,0,1/2);
		\cplus(4,0,4,2,+2);
\end{tikzpicture}
\raisebox{2.5em}{\quad.}
\end{equation}
The transformation matrix for this decoding circuit is then
\begin{equation}\label{E4matrix}
A_4=\left(
\begin{array}{ccc}
	-1	&	1	&	1	\\
	0	&	1	&	-1	\\
	-1	&	\frac{1}{2}	&	\frac{1}{2}
\end{array}\right).
\end{equation}
We can use Gaussian elimination to reduce this to the identity via
\begin{align}\label{E4rowreduction}
&\left(\begin{array}{ccc}
	-1	&	1	&	1	\\
	0	&	1	&	-1	\\
	-1	&	\frac{1}{2}	&	\frac{1}{2}
\end{array}\right)
\xmapsto{R_3\times (-1)}
\left(\begin{array}{ccc}
	-1	&	1	&	1	\\
	0	&	1	&	-1	\\
	1	&	-\frac{1}{2}	&	-\frac{1}{2}
\end{array}\right)
\xmapsto{R_3+R_1}
\left(\begin{array}{ccc}
	-1	&	1	&	1	\\
	0	&	1	&	-1	\\
	0	&	\frac{1}{2}	&	\frac{1}{2}
\end{array}\right)
\xmapsto{R_1-2R_3}
\left(\begin{array}{ccc}
	-1	&	0	&	0	\\
	0	&	1	&	-1	\\
	0	&	\frac{1}{2}	&	\frac{1}{2}
\end{array}\right)\notag\\
&\xmapsto{R_3-\frac{1}{2}R_2}
\left(\begin{array}{ccc}
	-1	&	0	&	0	\\
	0	&	1	&	-1	\\
	0	&	0	&	1
\end{array}\right)
\xmapsto{R_2+R_3}
\left(\begin{array}{ccc}
	-1	&	0	&	0	\\
	0	&	1	&	0	\\
	0	&	0	&	1
\end{array}\right)
\xmapsto{R_1\times (-1)}
\left(\begin{array}{ccc}
	1	&	0	&	0	\\
	0	&	1	&	0	\\
	0	&	0	&	1
\end{array}\right).
\end{align}
Reading these transformations backwards and taking appropriate adjoints gives the following equivalent decoding circuit
\begin{equation}
\begin{tikzpicture}[scale=0.8, every node/.style={inner sep=2pt}]
		\node [left](mode2start) at (0,2) {mode 2};
		\node [left](mode3start) at (0,1) {mode 3};
		\node [left](mode4start) at (0,0) {mode 4};
		\node [right](mode4end) at (7,0) {$\ket{\alpha}$};		
		\drawwires(3,7);
		\node[gate] at (1,2){$-1$};
		\cplus(2,0,2,1,-1);
		\cplus(3,1,3,0,1/2);	
		\cplus(4,0,4,2,2);
		\cplus(5,2,5,0,-1);
		\node[gate] at (6,0) {$-1$};
\end{tikzpicture}
\raisebox{2.5em}{\quad.}
\end{equation}
Using \cref{identity8} we can simplify this circuit to
\begin{equation}
\begin{tikzpicture}[scale=0.8, every node/.style={inner sep=2pt}]
		\node [right](mode4end) at (7,0) {$\ket{\alpha}$};		
		\drawwires(3,7);
		\node[gate] at (1,2){$-1$};
		\cplus(2,0,2,1,-1);
		\cplus(3,1,3,0,1/2);	
		\cplus(4,0,4,2,2);
		\cplus(5,2,5,0,-1);
		\node[gate] at (6,0) {$-1$};
\end{tikzpicture}
\raisebox{3.3em}{$\quad = \quad$}
\begin{tikzpicture}[scale=0.8, every node/.style={inner sep=2pt}]
		\node [right](mode4end) at (7,0) {$\ket{\alpha}$};		
		\draw (0,2) -- (7,2);
		\drawwires(2,1);
		\node[gate] at (1,2){$-1$};
		\beamsplittermp(1,1);
		\drawwiresat({2,0},2,5);
		\node[gate] at (3,1){$\sqrt{2}$};
		\node[gate] at (3,0){$1/\sqrt{2}$};
		\cplus(4,0,4,2,2);
		\cplus(5,2,5,0,-1);
		\node[gate] at (6,0) {$-1$};
\end{tikzpicture}
\raisebox{2.5em}{\quad.}
\end{equation}

We now observe that, after the beam splitter, mode 3 becomes unentangled from the recovery mode (mode 4), so we can ignore the single mode squeezing for mode 2 and in fact dump the mode immediately after it is passed through the beam splitter.  Similarly, we can measure mode 2 and commute the measurement past the final QND gate.  Moving the $-1$ gate on mode 2 to the end of the computation and replacing the $-1$ gate on mode 5 by its optical implementation (a phase shift of $\pi$) gives us the circuit in \cref{E4optical}, reproduced below
\begin{equation*}
	\begin{tikzpicture}[scale=0.75, every node/.style={inner sep=0pt}]  
		\node [left](mode2start) at (0,2) {mode 2};
		\node [left](mode3start) at (0,1) {mode 3};
		\node [left](mode4start) at (0,0) {mode 4};
		
		\draw (mode4start) -- (1,0);
		\draw (mode3start) -- (1,1);
		\draw (mode2start) -- (5,2);

		\beamsplittermp(1,1);
		
		\draw (2,0) -- (7.25,0);
		\node [gate] (Disp4) at (5.75,0) {$\text{Disp}$};
		\node [gate] (Pi4) at (6.75,0) {$\pi$};
		\node [gate] (Sq4) at (3,0) {$\frac{1}{\sqrt{2}}$};
		\node [right](mode1end) at (7.25,0) {$\ket{\alpha}$};
		
		\draw (2,1) -- (3,1);
		\draw (2.9,0.9) -- (3.1,1.1);
		\draw (2.95,0.9) -- (3.15,1.1);
		
		\node [gate] (QND31) at (4,2) {$\text{QND}_2^\dagger$};
		\coordinate (control31) at (QND31|-mode4start);
		\fill (control31) circle [radius=2pt];
		\draw (control31) -- (QND31);
		\node [inner xsep=1.1em](meas1) at (5,2) {};
		\coordinate (control13) at (5.75,2);
		\cwire(control13|-meas1,Disp4);
		\cwire(meas1,control13|-meas1);
		\draw [fill=white] (control13) circle (2.75pt);
		\meas(5,2,$x$);
		
	\end{tikzpicture}

\raisebox{2.5em}{\quad.}
\end{equation*}

\bibliographystyle{unsrtnat}
\bibliography{references}
\end{document}